\documentclass[12pt, a4paper]{article}

\usepackage{amsfonts}
\usepackage{amsmath}
\usepackage[english]{babel}
\usepackage{bm}
\usepackage{caption}
\usepackage{dsfont}
\usepackage[T1]{fontenc}
\usepackage[top = 2.0cm, left = 2.0cm, right = 2.0cm, bottom = 2.0cm]{geometry}
\usepackage{indentfirst}
\usepackage[utf8]{inputenc}
\usepackage{placeins}
\usepackage{setspace}
\usepackage{xcolor}
\usepackage{graphicx}
\usepackage{multirow}
\usepackage{booktabs}
\usepackage{makecell}
\usepackage[superscript,biblabel]{cite}
\usepackage{hyperref}
\title{Improved inference for MCP-Mod approach using time-to-event endpoints with small sample sizes

}

\author{M\'arcio A. Diniz\thanks{Biostatistics Research Center, Samuel Oschin Comprehensive Cancer Center, Cedars-Sinai Medical Center, Los Angeles, California, USA, e-mail to \texttt{marcio.diniz@cshs.org}.
}
	\and Diego I. Gallardo\thanks{Department of Mathematics, Engineering School, University of Atacama, Copiap\'o, Chile, e-mail to \texttt{diego.gallardo@uda.cl}.}
	\and 	Tiago M. Magalh\~aes\thanks{Department of Statistics, Institute of Exact Sciences, Federal University of Juiz de Fora, Juiz de Fora, Brazil, e-mail to \texttt{tiago.magalhaes@ufjf.br}.}
}

\date{}

\newcommand{\vg}[1]{\mbox{\boldmath{$#1$}}}

\begin{document}

\maketitle

\thispagestyle{empty}

\doublespacing

\begin{abstract}
    The  Multiple Comparison Procedures with Modeling
Techniques (MCP-Mod) framework has been recently approved by the U.S. Food and Administration and European Medicines Agency  as fit-for-purpose for phase II studies. Nonetheless, this approach relies on the asymptotic properties of Maximum Likelihood (ML) estimators, which might not be reasonable for small sample sizes.
	In this paper, we derived improved ML estimators and correction for their covariance matrices in the censored Weibull regression model based on the corrective and preventive approaches. We performed two simulation studies to evaluate ML and improved ML estimators with their covariance matrices in (i) a regression framework (ii) the Multiple Comparison Procedures with Modeling Techniques framework. 
	We have shown that improved ML estimators are less biased than ML estimators yielding Wald-type statistics that controls type I error without loss of power in both frameworks. Therefore, we recommend the use of improved ML estimators in the MCP-Mod approach to control type I error at nominal value for sample sizes ranging from 5 to 25 subjects per dose.
	\medskip
	
\end{abstract}

\section{Introduction}
\label{sec:intro}

Adequate designs for early-phase trials is essential to a successful clinical drug development. Traditionally, investigators evaluate safety in phase I trials, proof-of-concept (PoC) in phase IIa trials, and efficacy in phase IIb trials. When drugs are promising in these early stages, phase III trials are designed accruing a large number of patients to provide a definitive evidence of efficacy.  Nonetheless, Jardim et al. \cite{jardim2017factors} showed that 20\% of 80 cancer drug programs submitted between 2009 to 2015 to the Food and Drug Administration (FDA) did not have any data for phase II trials such that 31\% obtained FDA approval; 46\% had positive PoC with of 76\% FDA-approval; and 34\% had negative PoC with only 15\% of FDA-approval. Therefore, attaining proof-of-concept is a predictor to a successful phase III trial.

A classic PoC is designed with a highest dose allowable based on Phase I clinical trial results to compare with placebo in a two-arm design or a historical threshold in a one-arm design. Once proof of concept is shown, dose range studies to identify a minimum effective dose (MED) or target dose (TD) are conducted following two possible approaches: multiple comparisons (MCP) and modeling (Mod). The MCP approach corresponds to evaluate contrasts among doses while preserving the family-wise error rate (FWER), which is robust to distribution assumptions but restricted to the set of doses under investigation; the Mod approach assumes a dose-response relationship allowing to estimate the response for a dose even if that dose is not under investigation but it heavily depends on choosing the appropriate functional form.  

Bretz et al. \cite{bretz2005combining} proposed a framework named Multiple Comparison Procedures with Modeling Techniques (MCP-Mod) unifying phase IIa and IIb trials into a seamless design while taking advantage of both traditional approaches for normally distributed endpoints. Later, Pinheiro et al. \cite{pinheiro2014model} extended this methodology to general parametric models  using generalized least squares estimation allowing statisticians to consider more complex designs and other types of endpoints such as binary and time to event. Regulatory agencies have stated their approval of the MCP-Mod framework as an adequate and efficient methodology for design and analysis of phase II dose-finding studies that will guide dose selection for phase III trials.

The MCP-Mod framework is implemented in two steps: (i) MCP-step consists of a trend test to assess the presence of a dose response signal among a set of pre-specified candidate models while preserving FWER; (ii) Mod-step corresponds to estimate dose-response curves in order to identify the optimal dose that achieves a desired level of response in comparison to placebo among the models that were selected in the MCP-step. Therefore, the properties of the trend test defined as a Wald statistic used in the MCP-step and the maximum likelihood (ML) estimators with their covariance matrix used in the Mod-step are essential to the successful implementation of MCP-Mod approach. However, both steps rely on the the asymptotic properties of the ML estimators, which are only valid for large sample sizes.

In cancer mouse studies, time-to-death is an endpoint that could be used to guide the identification of the minimum effective dose (MED) with large expected effect sizes and small sample sizes. Nonetheless, the application of the MCP-Mod framework is limited due the underlying asymptotic assumptions. In this context, parametric survival models such as the censored Weibull regression model (WRM) was showed to be useful given that provides clinical meaningful interpretation based on event time ratios or hazard ratios\cite{carroll2003use}. Moreover, the asymptotic properties of the ML estimators can be studied and refined for small sample sizes. Recently, Magalhães et al. \cite{MagalhaesGallardoGomez:2019} obtained the skewness coefficient of the distribution of the maximum likelihood estimators for WRM, and Magalhães et al. \cite{MagalhaesGallardo:2020} derived improved test statistics for LR, score and gradient tests but not for the Wald statistic.



Our main goal is to derive improved ML estimators for the regression parameters and its second-order covariance matrix for WRM, then use them as input to the  generalized least squares procedure proposed by Pinheiro et al. \cite{pinheiro2014model} yielding a type I error probability closer to the nominal value when testing proof-of-activity and more accurate MED estimates.   Moreover, we particularize the results from Cox and Snell \cite{CoxSnell:1968} and Magalhães et al. \cite{MagalhaesBotterSandoval:2021} that are very general to WRM and they can be used in a much broader context than the MCP-Mod framework. 

The remaining paper is organized as follows: in Section \ref{sec:weibull}, we revisit the Weibull distribution and its properties; in Section \ref{sec:mcpmod}, we review the main concepts of the MCP-Mod framework; in Section \ref{sec:improved}, we introduce the improved estimators; in Section \ref{sec:application}, we conducted a simulation study to evalute the use of improved estimators in the MCP-Mod framework; finally, in Section \ref{sec:conclusion}, we presented some concluding remarks.

\section{Weibull distribution}
\label{sec:weibull}

The Weibull distribution \cite{Weibull:1951} is commonly used to analysis of time-to-event or lifetime data and a continuous random variable $T$ is called Weibull, if its probability density function (pdf) is 
%
%
\begin{align}
\label{pdf:wei}
f(t;\lambda,\sigma) = \frac{1}{\sigma\lambda^{1/\sigma}} t^{1/\sigma-1} \exp\left\{ -\left( t/\lambda \right)^{1/\sigma} \right\}, \ t > 0,
\end{align}

\noindent where $\sigma > 0$ is the shape parameter and $\lambda > 0$ is the scale parameter, it says $T \sim$ WE$(\lambda,\sigma)$. Two particular models under this parametrization are obtained for $\sigma = 1$ and $\sigma = 1/2$, which represents the exponential and the Rayleigh models with means $\lambda$ and $\lambda \sqrt{\pi/2}$, respectively. In this work, we focused on those models. However, if $\sigma$ is unknown, we assume that it can be replaced by consistent estimate. The $\rho\%$ survival time is given by
%
%
\begin{align}
\label{eq:mst}
t_\rho = \lambda [-\log(\rho)]^{\sigma}
\end{align}
%
%
The regression structure can be incorporated in \eqref{pdf:wei} by making
\begin{align}
\label{eq:reg}
\log(\lambda_i) = {\vg x}_i^\top {\vg \beta}
\end{align}

\noindent where ${\vg \beta}$ is a p-vector of unknown parameters and ${\vg x}_i$ is a vector of predictors related to the $i$th observation. 

In lifetime data, there is the censoring restriction, i.e, if $T_1, \ldots, T_n$ are a random sample from \eqref{pdf:wei}, instead of $T_i$, we observe, under right censoring, $t_i = \min(T_i, L_i)$, where $L_i$ is the censoring time, independent of $T_i$ and $\delta_i = 1$, if $T_i \leq L_i$ or $\delta_i = 0$, otherwise, $i = 1, \ldots, n$. Here, we consider an hybrid censoring scheme, where the study is finalized when a pre-fixed number $r \leq n$ out of $n$ observations have failed, as well as when a prefixed time, say $L_1 = \cdots = L_n = L$, has been reached. The type I censoring is a particular case for $r = n$ and the type II censoring appears when $L_1, \ldots, L_n = +\infty$. Additionally, we add the non-informative censoring assumption, i.e., the random variables $L_i$ does not depend on $\lambda$. Usually, the regression modeling considers the distribution of $Y_i = \log(T_i)$ instead of $T_i$, which is an accelerated lifetime model form, see Kalbfleisch and Prentice\cite{KalbfleischPrentice:2002}. The distribution of $Y_i$ is of the extreme value form with pdf given by
%
%
\begin{align}
\label{pdf:evd}
f(y_i; {\vg x}_i) = \frac{1}{\sigma}\exp\left\{\frac{y_i-\mu_i}{\sigma}-\exp\left(\frac{y_i-\mu_i}{\sigma}\right)\right\}, \quad -\infty < y_i < \infty,
\end{align}

\noindent where $\mu_i = \log \lambda_i$. From this moment, we assume that $\sigma$ is known, Then, the log-likelihood function derived from \eqref{pdf:evd} is given by
%
%
\begin{align*}
\ell({\vg \beta}) = \sum_{i=1}^n \left[\delta_i\left(-n\log \sigma+\frac{y_i-\mu_i}{\sigma}\right)-\exp\left(\frac{y_i-\mu_i}{\sigma}\right)\right].
\end{align*}

The total score function and the total Fisher information matrix for ${\vg \beta}$ are, respectively, ${\vg U}_{{\bm \beta}} = \sigma^{-1} {\vg X}^{\top} {\vg W}^{1/2} {\vg v}$ and ${\bm K}_{{\bm \beta} {\bm \beta}} = \sigma^{-2} {\vg X}^{\top} {\vg W} {\vg X}$, where ${\vg X} = ({\vg x}_1, \ldots, {\vg x}_n)^{\top}$, the model matrix, assuming rank$({\vg X}) = p$, ${\vg W} =$ diag$(w_1, \ldots, w_n)$, $w_i = \mathds{E}\left[\exp\left(\frac{y_i - \mu_i}{\sigma}\right)\right]$ and ${\vg v} = (v_1, \ldots, v_n)^{\top}$, $v_i = \left\{ - \delta_i + \exp\left( \frac{y_i - \mu_i}{\sigma}\right) \right\} w_i^{-1/2}$. It can observed that the value of $w_i$ depends on the mechanism of censoring. That means $w_i = q \times \left\{ 1 - \exp\left[ -L_i^{1/\sigma} \exp(-\mu_i/\sigma) \right] \right\} + (1-q)\times \left(r/n\right)$, where $q = \mathds{P}\left(W_{(r)}\leq \log L_i\right)$ and $W_{(r)}$ denotes the $r$th order statistic from $W_1, \ldots, W_n$. Note that $q=1$ and $q=0$ for types I and II censoring, respectively, as showed in Magalhães et al. \cite{MagalhaesGallardoGomez:2019}. The maximum likelihood estimator (MLE) of ${\vg \beta}$, $\widehat{{\vg \beta}}$, is the solution of ${\vg U}_{{\bm \beta}} = {\bf 0}$. The $\widehat{{\vg \beta}}$ can not be expressed in closed-form. It is typically obtained by numerically maximizing the log-likelihood function using a Newton or quasi-Newton nonlinear optimization algorithm. Under mild regularity conditions and in large samples,
%
%
\begin{align}
\label{eq:mledist}
\widehat{{\bm \beta}} \sim \mbox{N}_p \left( {\bm \beta}, {\bm K}_{{\bm \beta} {\bm \beta}}^{-1} \right),
\end{align}

\noindent approximately. The classic Wald test\cite{Wald:1943} statistic is
%
%
\begin{align}\label{eq:w0}
	W_{MLE} = \left( {\bm C} \widehat{\bm \beta} - {\bm C} {\bm \beta}^{(0)} \right)^{\top} \left\{ {\bm C} \widehat{\bm K}_{{\bm \beta} {\bm \beta}}^{-1} {\bm C}^{\top} \right\}^{-1} \left( {\bm C} \widehat{\bm \beta} - {\bm C} {\bm \beta}^{(0)} \right),
\end{align}

\noindent where ${\vg C}$ is a matrix of contrasts $m \times p$. Under the null hypothesis $\mathcal{H}: {\vg C} {\vg \beta} = {\vg C} {\vg \beta}^{(0)}$, $W_{MLE}$ has a $\chi_{p}^2$ distribution up to an error of order $n^{-1}$. The null hypothesis is rejected for a given nominal level, $\alpha$ say, if the test statistic exceeds the upper $100(1 - \alpha)\%$ quantile of the $\chi_{p}^2$ distribution.

\section{MCP-Mod General approach}
\label{sec:mcpmod}

We briefly summarize the two-stage procedure discussed by Pinheiro et al. \cite{pinheiro2014model} following the same notation. We consider that time-to-event responses $t_{ij}$ for doses $x_i$ given to $j$th subject for $i = 0, \ldots, p$ and $j = 1, \ldots, n_i$ can be described by the Weibull distribution with scale parameters $\lambda_i$ and shape parameter $\sigma$ defined in \eqref{pdf:wei} and \eqref{eq:reg}. Then, we consider the parameter $\mu$ as our response for the dose-response model such that it could be defined as the median survival time \eqref{eq:mst} for $\rho = 0.5$ or, alternatively, $\log(\lambda)$. 

Initially, a set of candidate dose-response models $\mu_i = f(x_i, \vg{\theta})$ is considered such that each model can be rewritten as function of standardized model as below

\begin{align}
\label{eq:model.theta}
f_m(x, \vg{\theta}) = \theta_0 + \theta_1 f_m^0(x, \vg{\theta}^0)
\end{align}

\noindent for $m = 1, \ldots, M$, where $(\theta_0, \theta_1)$ are unknown parameters. In this work, we consider $(M = 5)$ standardized models:  (i) linear: $f^0(x, \vg{\theta}^0) = x$; (ii) emax : $f^0(x, \vg{\theta}^0) = x/(x + ED_{50})$ where $ED_{50}$ can be interpreted as the dose that produces the desired response on 50\% of subjects; (iii) exponential: $f^0(x, \vg{\theta}^0) = \exp\{x/ \delta \} - 1$, where $\delta$ is the exponential rate; (iv) logistic: $f^0(x, \vg{\theta}^0) = 1/(1 + \exp\{(ED_{50} - x)/\delta\})$ (v) beta: $f^0(x, \vg{\theta}^0) = \beta(\delta_1, \delta_2)(x/scal)^{\delta_1}(1 - x/scal)^{\delta_2}$.

\subsection{MCP-step}
\label{sec:mcp_step}

In this step, a set of contrasts corresponding to the candidate models will be tested.  Let $\vg{\mu}$ denote the estimated dose-response parameter vector. For each candidate model, an optimal contrast $\vg{c}^{opt}$ that maximizes the probability of rejecting the hypothesis of non-signal dose-response is derived assuming that the candidate model is correct and guess estimates for $\vg{\theta}^0$.
%
%
\begin{align}
\label{eq:guess.estimates}
\vg{c}^{opt} \propto \vg{S}^{-1} \left( \vg{\mu}_{m}^{0} - \frac{ {\vg{\mu}_{m}^{0}}^{\top}\vg{S}^{-1}\vg{1} }{ \vg{1}\vg{S}^{-1}\vg{1}^{\top} } \right),
\end{align}

\noindent where $\vg{\mu}_{m}^{0} = (f_m^0(x_1, \vg{\theta}^0), \ldots, f_m^0(x_p, \vg{\theta}^0))$ and $\vg{S}$ is the covariance matrix of $\vg{\mu}$. Assuming that $\vg{\hat{\mu}}$ follows approximately $\mbox{N}_p(\vg{\mu}, \vg{S})$, the test of hypotheses for proof-of-concept can be translated to $H_0:\vg{c}^{opt}_m \vg{\mu} = 0$ vs. $H_1:\vg{c}^{opt}_m \vg{\mu} > 0$ for candidate model $m$ based on the Wald test statistic
%
%
\begin{align}
W^{(m)} = (\vg{c}^{opt}_m\vg{\hat{\mu}})^{\top}\left\{ \vg{C}^{opt}\vg{\hat{S}} {\vg{C}^{opt}}^{\top} \right\}^{-1}_{m, m}\vg{c}^{opt}_m\vg{\hat{\mu}}
\end{align}

\noindent where \vg{\hat{S}} is the estimated covariance matrix, $\vg{C}^{opt}$ is the matrix $m \times p$ of optimal contrast with $[\vg{A}]_{m,m}$ denoting the $m^{th}$ diagonal element of matrix $\vg{A}$ for $m = 1, \ldots, M$. Critical values for tests are derived based on the joint distribution for $W = (W^{(1)}, \ldots, W^{(M)})$ allowing one to calculate multiplicity adjusted p-values controlling the FWER at a prespecificed nominal type I error $\alpha$.

\subsection{Mod-step}
\label{sec:mod_step}

In this step, the estimation of non-linear dose responses models is performed in two stages. In the first stage, the parameters $\vg{\mu} = [\mu_1, \ldots, \mu_p]$ are estimated using standard software packages with analysis of variance (ANOVA) parametrization for the design matrix resulting into a separate parameter $\mu_i$ for each dose level $x_i$ for $i = 1, \ldots, D$. In particular for $\mu = \log(\lambda)$, we have $\vg{\mu} = \vg{\beta}$ and $\vg{S} = {\bm K}_{{\bm \beta} {\bm \beta}}^{-1}$.

In the second stage, the non-linear dose-response model $f(x, \vg{\theta})$ is fitted by minimizing the generalized linear squares (GLS) criterion:
\begin{align}\
\label{eq:gls.criterion}
\Psi(\vg{\theta}) = (\vg{\hat{\mu}} - \vg{f(x, \theta)})'\vg{\hat{S}}^{-1}(\vg{\hat{\mu}} - \vg{f(x, \theta)})    
\end{align}
with respect to $\vg{\theta}$.

Then, the minimum effective dose can be estimated as $\widehat{MED} = \{x | f(x, \vg{\hat{\theta}}) > f(0, \vg{\hat{\theta}}) + \Delta\}$, where $\Delta$ is a clinical meaningful threshold and $\vg{\hat{\theta}}$ minimizes \eqref{eq:gls.criterion}. 

\section{Improved inference}
\label{sec:improved}

\subsection{Bias correction}

Inferences based on maximum likelihood method depend strongly on asymptotic properties. Among these properties, the MLE is approximately non-biased, in other words, $\mathds{E}(\widehat{{\bm \beta}} - {\bm \beta}) = \mathcal{O}(n^{-1})$, which is essential to define the mean of the normal distribution of $\widehat{{\bm \beta}}$. Therefore, likelihood inferences based on asymptotic approximation may not be reliable when sample sizes are small or moderate, and two approaches are available to correct the MLE.

\paragraph{The corrective approach.} The bias of the MLE can be written as $\mathds{E}(\widehat{{\bm \beta}} - {\bm \beta}) = \mathds{B}({\bm \beta}) + \mathcal{O}(n^{-2})$, where $\mathds{B}({\bm \beta})$ is a term of order $\mathcal{O}(n^{-1})$, a function of the derivatives of the log-likelihood function. Cox and Snell \cite{CoxSnell:1968} proposed a bias-corrected maximum likelihood estimator (BCE), that can be expressed as $\widetilde{{\bm \beta}} = \widehat{{\bm \beta}} - \mathds{B}(\widehat{{\bm \beta}})$, where $\mathds{B}({\bm \beta})$ is the term of order $n^{-1}$, evaluated in $\widehat{{\bm \beta}}$ and $\mathds{E}(\widetilde{{\vg \lambda}} - {\vg \lambda}) = \mathcal{O}(n^{-2})$, i.e., less biased then MLE of ${\bm \beta}$. The Cox and Snell's method is known as a corrective approach because the MLE is calculated and then, the bias correction is applied. For the censored Weibull regression model, the expression of $\mathds{B}(\widehat{\bm \beta})$ has the form
%
%
\begin{align}
\label{eq:bias}
\mathds{B}(\widehat{\bm \beta}) = - \frac{1}{2 \sigma^3} {\bm P} {\bm Z}_d \left({\bm W} + 2 \sigma {\bm W}^{\prime}\right) {\bm 1},
\end{align}

\noindent where ${\bm P} = {\bm K}_{{\bm \beta} {\bm \beta}}^{-1} {\bm X}^{\top}$, ${\bm Z} = {\bm X} {\bm K}_{{\bm \beta} {\bm \beta}}^{-1} {\bm X}^{\top}$, ${\bm Z}_d$ is a diagonal matrix with diagonal given by the diagonal of ${\bm Z}$, ${\bm W}^{\prime} =$ diag$(w_1^{\prime}, \ldots, w_n^{\prime})$, $w_i^{\prime} = - \sigma^{-1} L_i^{1/\sigma} \exp\{ -L_i^{1/\sigma} \exp(-\mu_i/\sigma) - \mu_i/\sigma \}$ and ${\bm 1}$ is a $n$-dimensional vector of ones. 
%
%

\paragraph{The preventive approach.} As alternative to the corrective approach, Firth \cite{Firth:1993} proposed the following modification in the score vector:
%
%
\begin{align}
\label{eq:scoreFirth}
{\bm U}_{{\bm \beta}}^{\star} = {\bm U}_{{\bm \beta}} - {\bm K}_{{\bm \beta} {\bm \beta}} \mathds{B}({\bm \beta}),
\end{align}

\noindent where $\mathds{B}({\bm \beta})$ is given by \eqref{eq:bias}. The estimator $\check{{\bm \beta}}$, solution of ${\bm U}_{{\bm \beta}}^{\star} = {\bm 0}$, has a bias of order $\mathcal{O}(n^{-2})$. This is a preventive approach because the procedure already computes a less biased estimator than the regular MLE. 

\subsection{Covariance correction}

From the general result of Magalhães et al. \cite{MagalhaesBotterSandoval:2021} , we derived the specific matrix expression for the MLE and BCE second-order covariance matrices for the censored Weibull regression model and it is given by  
%
%
\begin{align}
	\label{eq:cov2G}
	\mbox{{\bf Cov}}_{\bm 2}^{\bm \tau}({\bm \beta}^{\star}) = {\bm K}_{{\bm \beta} {\bm \beta}}^{-1} + {\bm K}_{{\bm \beta} {\bm \beta}}^{-1} \left\{ {\bm \Delta} + {\bm \Delta}^{\top} \right\} {\bm K}_{{\bm \beta} {\bm \beta}}^{-1} + \mathcal{O}(n^{-3}),
\end{align}

\noindent where ${\vg \Delta} = -0.5 {\vg \Delta}^{(1)} + 0.25 {\vg \Delta}^{(2)} + 0.5 \tau_2 {\vg \Delta}^{(3)}$ with
%
%
\begin{align*}
\Delta^{(1)} &= \frac{1}{\sigma^4} {\bm X}^{\top} {\bm W}^{\star} {\bm Z}_{d} {\bm X}, \nonumber \\
\Delta^{(2)} &= - \frac{1}{\sigma^6} {\bm X}^{\top} \left[ {\bm W} {\bm Z}^{(2)} {\bm W} - 2 \sigma {\bm W} {\bm Z}^{(2)} {\bm W}^{\prime} - 6 \sigma^2 {\bm W}^{\prime} {\bm Z}^{(2)} {\bm W}^{\prime} \right] {\bm X}, \nonumber\\
\Delta^{(3)} &= \frac{1}{\sigma^5} {\bm X}^{\top} {\bm W}^{\prime} {\bm W}^{\star\star} {\bm X}, \nonumber
\end{align*}

\noindent ${\bm W}^{\star} =$ diag$(w_1^{\star}, \ldots, w_n^{\star})$, $w_i^{\star} = w_i (w_i -2) - 2 \sigma w_i^{\prime} + \sigma \tau_1 (w_i^{\prime} + 2 \sigma w_i^{\prime\prime})$, ${\bm Z}^{(2)} = {\bm Z} \odot {\bm Z}$, with $\odot$ representing a direct product of matrices (Hadamard product), ${\bm W}^{\star\star}$ is a diagonal matrix, with ${\bm Z} ( {\bm W} + 2 \sigma {\bm W}^{\prime}) {\bm Z}_{d} {\bm 1}$ as its diagonal, ${\bm W}^{\prime\prime} =$ diag$(w_1^{\prime\prime}, \ldots, w_n^{\prime\prime})$, $w_i^{\prime\prime} = - \sigma^{-1} w_i^{\prime} \left[ L_i^{1/\sigma} \exp(-\mu_i/\sigma) - 1 \right]$, ${\bm \tau} = (\tau_1, \tau_2) = (1, 1)$ indicating the second-order covariance matrix of the MLE ${\bm \beta}^{\star} = \widehat{{\bm \beta}}$ denoted by $\mbox{{\bf Cov}}_{\bm 2}(\widehat{{\bm \beta}})$ and ${\bm \tau} = (0, -1)$ indicating the second-order covariance matrix of the BCE ${\bm \beta}^{\star} = \widetilde{{\bm \beta}}$ denoted by $\mbox{{\bf Cov}}_{\bm 2}(\widetilde{{\bm \beta}})$.

\subsection{Wald-type test}

Let $\mbox{{\bf Cov}}_{\bm 2}^{-1}(\widehat{{\bm \beta}})$ and $\mbox{{\bf Cov}}_{\bm 2}^{-1}(\widetilde{{\bm \beta}})$ the inverse of $\mbox{{\bf Cov}}_{\bm 2}(\widehat{{\bm \beta}})$ and $\mbox{{\bf Cov}}_{\bm 2}(\widetilde{{\bm \beta}})$, respectively and considering also the partitions and the notation for the Fisher information matrix discussed in the introductory section, we can propose four modifications to the Wald test in \eqref{eq:w0}:
%
%
\begin{align}
\label{eq:w1}
W_{MLE2} &= \left( {\bm C} \widehat{\bm \beta} - {\bm C} {\bm \beta}^{(0)} \right)^{\top} \left\{ {\bm C} {\widehat{\rm \bf C}}\mbox{{\bf ov}}_{\bm 2}^{-1}(\widehat{{\bm \beta}}) {\bm C}^{\top} \right\}^{-1} \left( {\bm C} \widehat{\bm \beta} - {\bm C} {\bm \beta}^{(0)} \right), \\
\label{eq:w2}
W_{BCE} &= \left( {\bm C} \widetilde{{\bm \beta}} - {\bm C} {\bm \beta}^{(0)} \right)^{\top} \left\{ {\bm C} \widetilde{\bm K}_{{\bm \beta} {\bm \beta}}^{-1} {\bm C}^{\top}\right\}^{-1} \left( {\bm C} \widetilde{{\bm \beta}} - {\bm C} {\bm \beta}^{(0)} \right),\\
\label{eq:w3}
W_{BCE2} &= \left( {\bm C} \widetilde{{\bm \beta}} - {\bm C} {\bm \beta}^{(0)} \right)^{\top} \left\{ {\bm C} {\widetilde{\rm \bf C}}\mbox{{\bf ov}}_{\bm 2}^{-1}(\widetilde{{\bm \beta}}) {\bm C}^{\top} \right\}^{-1} \left( {\bm C} \widetilde{{\bm \beta}} - {\bm C} {\bm \beta}^{(0)} \right),\\
\label{eq:w4}
W_{Firth} &= \left( {\bm C} \check{{\bm \beta}} - {\bm C} {\bm \beta}^{(0)} \right)^{\top} \left\{ {\bm C} \check{{\bm K}}_{{\bm \beta} {\bm \beta}}^{-1} {\bm C}^{\top}\right\}^{-1} \left( {\bm C} \check{{\bm \beta}} - {\bm C} {\bm \beta}^{(0)} \right),
\end{align}

\noindent where ${\widehat{\rm \bf C}}\mbox{{\bf ov}}_{\bm 2}(\widehat{{\bm \beta}})$ is the matrix $\mbox{{\bf Cov}}_{\bm 2}({\bm \beta})$ evaluated at $\widehat{{\bm \beta}}$, $\widetilde{{\bm K}}_{{\bm \beta} {\bm \beta}}$ is the Fisher information evaluated at $\widetilde{{\bm \beta}}$, ${\widetilde{\rm \bf C}}\mbox{{\bf ov}}_{\bm 2}(\widetilde{{\bm \beta}})$ is the matrix $\mbox{{\bf Cov}}_{\bm 2}(\widetilde{{\bm \beta}})$ evaluated at $\widetilde{{\bm \beta}}$, $\check{{\bm K}}_{{\bm \beta} {\bm \beta}}$ is the Fisher information evaluated at $\check{{\bm \beta}}$. Under $\mathcal{H}$, $W_{MLE}$, $W_{BCE}$, $W_{BCE2}$, $W_{Firth}$ follow a $\chi_{p}^2$ distribution.

\subsection{Improved Estimator strategies}

In the supplemental material and the next section, we studied the statistical properties of improved estimators for $\bm \beta$ and its covariance matrix with the following strategies: the classical ML estimator with the Fisher information as covariance matrix (MLE); the classical ML estimator with the corrected covariance matrix defined in \eqref{eq:cov2G} (MLE2); the bias corrected estimator (BCE) given in \eqref{eq:bias} with the Fisher information as covariance matrix; the bias corrected estimator with the corrected covariance matrix (BCE2); and the Firth estimator defined in \eqref{eq:scoreFirth} with its Fisher information as covariance matrix.

\section{Simulation study}
\label{sec:application}

In our collaborative work, investigators wanted to establish a dose-response relationship between a new inhibitor agent for pancreatic cancer in combination with a given dose of gemcitabine in mouse models. Based on preliminary data, a survival median time of 4 months was estimated in control-treated KPC mice model such that previous studies showed no survival benefit with only gemcitabine \cite{olive2009inhibition}. Assuming a Weibull distribution with $\sigma = 0.5$, we calculated the placebo effect equal to 1.57 and the maximum effect of 2.26 considering a hazard ratio of 4. Investigators were interested in the minimum effective dose yielding a minimum hazard ratio of 2 corresponding to $\Delta = 0.693$.

\begin{table}[!h]
    \centering
    \setlength{\tabcolsep}{6.0pt} 
	\renewcommand{\arraystretch}{1.4} 
    \begin{tabular}{c|c|c|c|c|c}
    \hline
    \multirow{2}{*}{Model}  &  \multirow{2}{*}{Constraints} &  \multicolumn{3}{c|}{Guess estimates/True parameters} & \multirow{2}{*}{True MED} \\
    \cline{3-5}
    & & $\theta_0$ & $\theta_1$ & $\theta_2$ & \\
    Constant & - &  $E_0 = 1.569$ & - & - & - \\
    \hline
    Linear & - &  $E_0 = 1.569$ & $\delta = 0.0139$ & - & - \\
    \hline
    Emax & 50\% at $x_4$ & $E_0 = 1.569$ & $E_{Max} = 2.079$ & $ED_{50} = 50.000$ & 25.00 \\
    \hline
    Exponential & 10\% at $x_1$ & $E_0 = 1.569$ & $E_1 = 0.017$ & $\delta = 22.756$ & 84.51 \\
    \hline
    Logistic &  \makecell{10\% at $x_3$ \\ 80\% at $x_4$} &  $E_0 = 1.569$ & $E_{Max} = 1.391$ & \makecell{$ED_{50} = 40.329$\\ $\delta = 6.976$} & 40.37 \\
    \hline
    Beta &30\% at $x_2$ & $E_0 = 1.569$ & $E_{Max} = 1.386$ & \makecell{$\delta_1 = 0.749$\\ $\delta_2 = 1.049$} & 10.61 \\
    \hline
    \end{tabular}
    \caption{Scenarios (Cosntant, Emax, Exponential, Logistic and Beta) and candidate models (Linear, Emax, Exponential, Logistic and Beta) defined based on, respectively, true parameters and guess estimates. True parameters/guess estimates were calculated based on placebo effect of 1.57, maximum effect of 2.96 and constraints with scale parameter of 120 for Beta model. True MED was calculated based on $\Delta = 0.693$.}
    \label{tab:scenarios}
\end{table}

Five scenarios (constant, emax, exponential, logistic and beta model) presented in Table \ref{tab:scenarios} were studied. True parameters were defined based on the aforementioned placebo and maximum effects, and the percent of maximum effect that is achieved at given dose as discussed in Bornkamp et al.\cite{bornkamp2009mcpmod}. For each scenario, the doses 0, 5, 25, 50 and 100 (mg/kg) were considered such that true MED was calculated as continuous dose given in Table \ref{tab:scenarios}. Five candidate models (linear, emax, exponential, logistic and beta model) were considered with guess estimates defined as the true parameters to calculate the optimal contrasts in \eqref{eq:guess.estimates}. For each scenario, the null hypothesis of non-signal of the new inhibitor agent was tested at 5\% significance level, and we used Akaike Information Criteria (AIC) to choose the model to estimate MED when more than one model rejected the non-signal hypothesis. Furthermore, we assumed that the target dose is estimated as a continuous dose for any value within the range of the dose grid. In case, the target dose is estimated outside of the dose grid, then the closest dose is selected.

For both steps of the MCP-Mod framework, the five strategies (MLE, MLE2, BCE, BCE2 and Firth) were evaluated based on a Monte Carlo simulation study with 100,000 replicates for sample sizes ranging from 5 to 100 mice per dose and censoring rate of 10\%, 25\% and 50\%. The following operating characteristics were studied: (i) convergence rate of GLS algorithm in the MCP-Mod General framework; (ii) Probability of incorrectly detecting a dose-response signal under the scenario with a constant dose response curve, i.e., type I error; (iii) Probability of correctly detecting a dose-response signal under scenarios with a non-constant dose-response curve (Emax, Exponential, Logistic, Beta), i.e., power; (iv) Probability of selecting the true model given that there is a signal; (v) Bias of MED estimate and (vi) RMSE of MED estimate. 

\subsection{Results}

\begin{figure}[!ht]
    \centering
    \includegraphics[scale = 0.8]{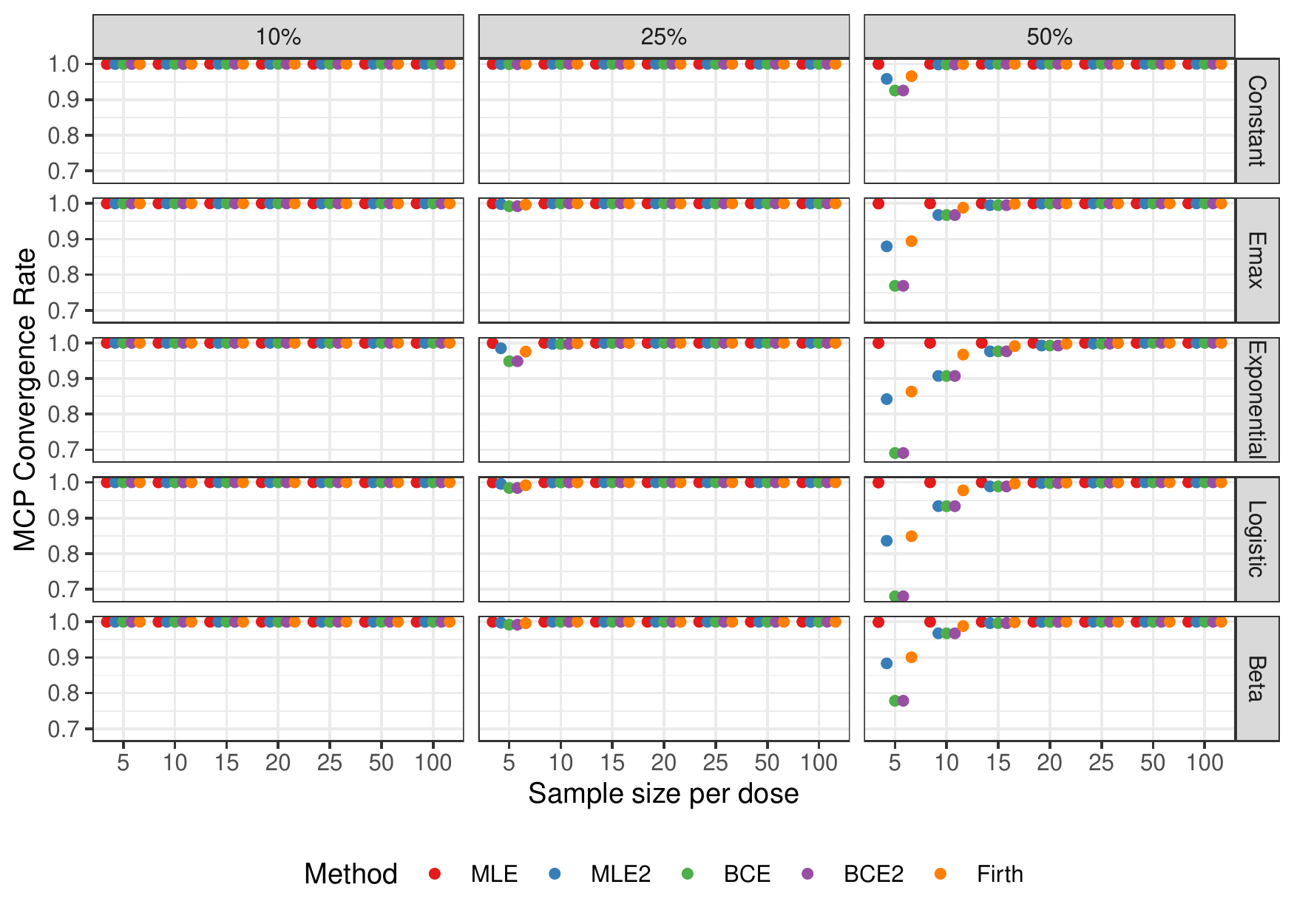}
    \caption{Convergence rate when calculating MLE, MLE2, BCE, BCE2 and Firth estimators and applying them as input for the MCP-Mod framework as function of censoring rate and true model.}
    \label{fig:convergence_rate}
\end{figure}

In Figure \ref{fig:convergence_rate}, convergence rates when calculating estimators and applying them as input for the MCP-Mod framework is presented for different censoring rates and true models. When censoring is 10\%, there is no difference among estimators; when censoring is 25\%, similar conclusion can be drawn but the proposed strategies have lower convergence rates than MLE for the true model Exponential and sample size of 5 subjects per dose: 94\% for BCE and BCE, 97\% for Firth and MLE2 ; when censoring is 50\%, the lower convergence rates of the estimators Firth, MLE2, BCE and BCE2 are also observed in other scenarios with dose-relationship signal for sample sizes of 5 or 10 subjects per dose reaching 69\% for BCE and BCE2 when true model is either Logistic or Exponential, 76\% for BCE and BCE2 when true model is either Beta or Emax. These lower convergence rates are attained because small sample sizes in combination with high censoring rates result into doses with no events (in our case, deaths), therefore, very large estimates for the regression coefficients are obtained and, consequently, singular covariance matrix estimates. 

\begin{figure}[!ht]
    \centering
    \includegraphics[scale = 0.8]{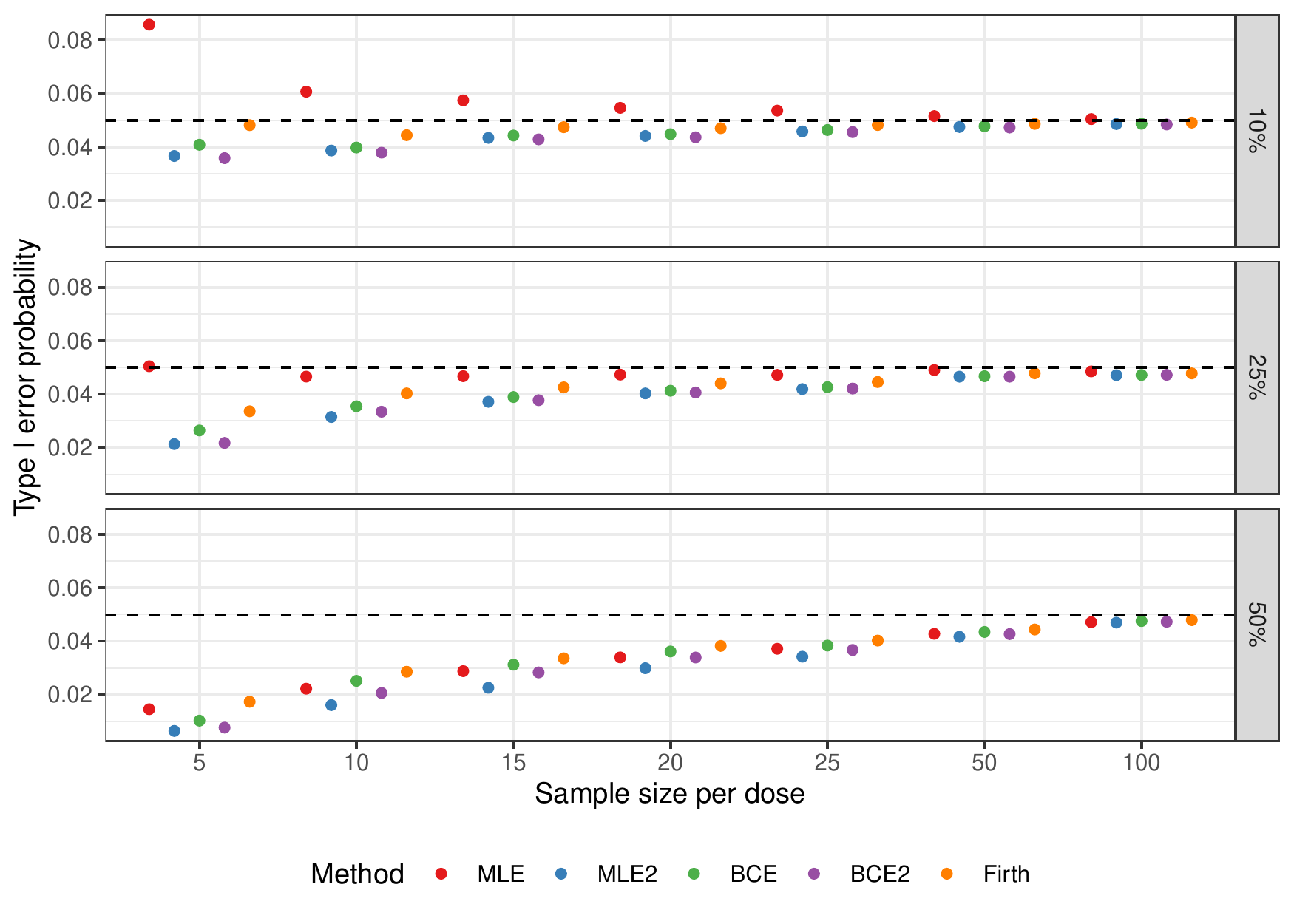}
    \caption{Type I error probability when in the MCP-step using MLE, MLE2, BCE, BCE2 and Firth estimators as function of censoring rate.}
    \label{fig:typeIerror}
\end{figure}

For the MCP-step, type I error probability is displayed for different censoring rates and true models in Figure  \ref{fig:typeIerror}. When censoring is 10\%, the  strategy MLE shows empirical type I error probability inflated  up to 0.086 when sample size is 5 subjects per dose, and reaches the nominal type error probability when sample size is 100 subjects per dose; the proposed strategies are slightly conservative such that the strategy Firth shows type I error probability uniformly closer to 0.05 than the other proposed strategies. When censoring is 25\%, strategies based on refined estimators are more conservative than MLE; among the proposed strategies, the strategy Firth is consistently superior followed by BCE, BCE2 and MLE2 reaching the same performance than MLE strategy for a sample size of 50 subjects per dose.  When censoring is 50\%, all strategies  are overly conservative including MLE;  Firth strategy is uniformly superior followed by BCE, MLE BCE2 and MLE2 up to a sample size of 100 subjects per dose. For all censoring values and strategies,  type I error probability converges to its nominal value for sample size of 100 subjects per dose.

\begin{figure}[!ht]
    \centering
    \includegraphics[scale = 0.9]{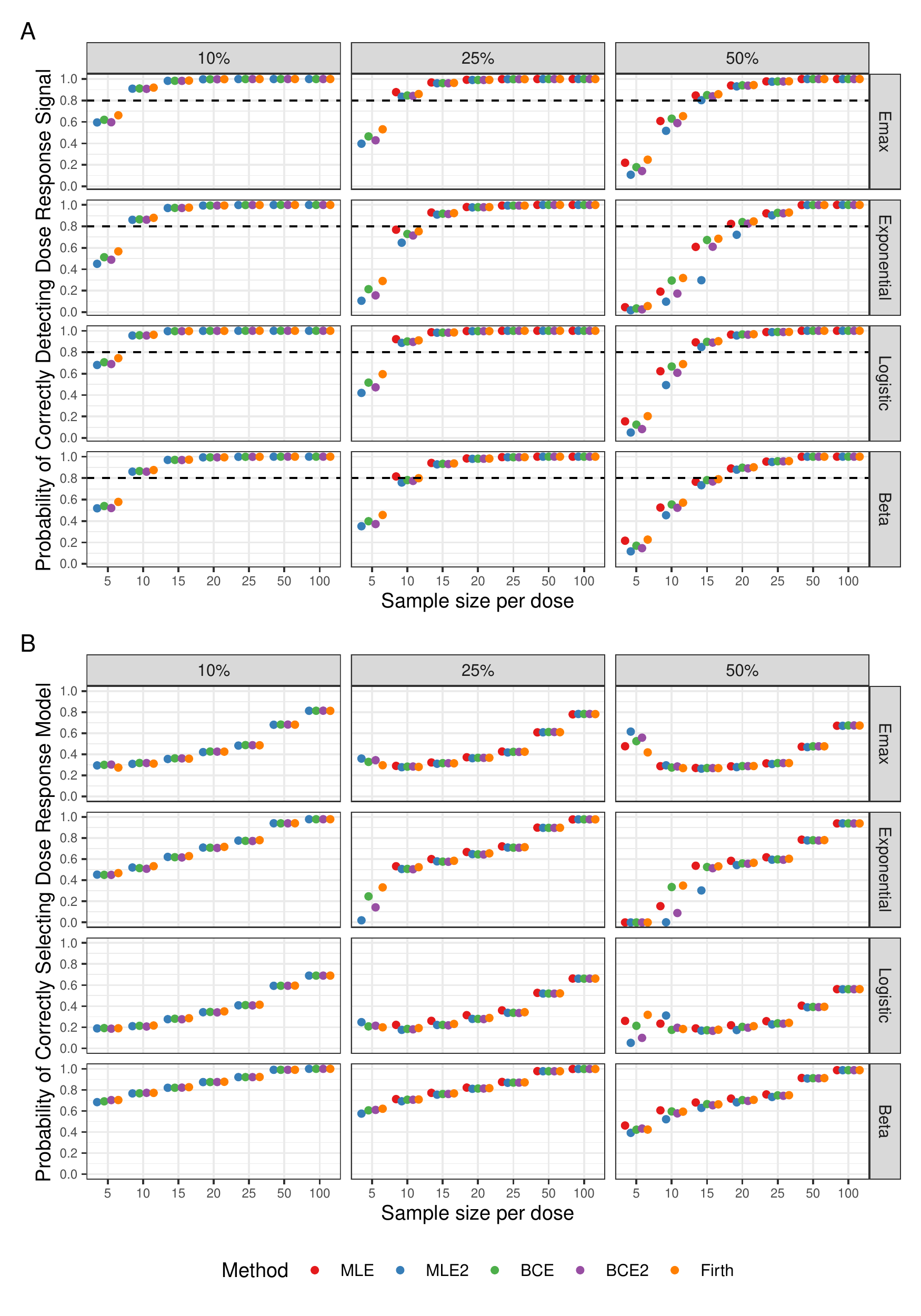}
    \caption{\textbf{A}. Probability of correctly detecting dose-response signal  when in the Mod-step  using MLE, MLE2, BCE, BCE2 and Firth estimators as function of censoring rate and true model. \textbf{B}. Probability of correctly selecting dose-response model using AIC in the Mod-step using MLE, MLE2, BCE, BCE2 and Firth estimators as function of censoring rate and true model.}
    \label{fig:power_ps}
\end{figure}

In Figure \ref{fig:power_ps}A, the probability of correctly detecting dose-response signal in the MCP-step is showed as function of censoring rates and true models. It is expected that the strategy with inflated type I error show higher power than strategies with empirical type I error closer to its nominal value. A fair comparison would require us to re-adjust critical values to reject the null hypothesis such that the empirical type I probability was set at 0.05 for all strategies. Nonetheless, we did not adjust the critical value as this procedure would never be done in practice due computational costs. Therefore, we only discussed strategies that have probability of type I error less or equal to its nominal value of 0.05.

When censoring is 10\%, the probability of correctly detecting dose-response signal is less the target of 0.8 with 5 subjects/dose such that the strategy Firth shows a higher power of at least 0.04 in comparison to the other proposed strategies; for sample size of 10 or larger, differences among strategies are negligible with power above its target for all strategies and true models. When censoring rate is 25\%, the strategy Firth is superior to others by at least 0.06 for 5 subjects/dose; however, all strategies have power less than its target; for sample size of 10 or larger, differences among strategies are negligible including MLE that is uniformly less conservative than the proposed strategies; moreover, power is above its target value for sample size of 10 for all true models except for Exponential as true model. When censoring is 50\%, the strategy Firth shows consistently higher power  while MLE2 strategy shows consistently lower power when compared to others with differences among strategies minor for sample size of 15 or higher subjects/dose, except for the strategy MLE2 and Exponential as true model. 

In Figure \ref{fig:power_ps}B, the probability of correctly selecting dose-response model using AIC is calculated given that we selected at least one model in the MCP-step, i.e., it is a conditional probability. When censoring is 10\%, differences among strategies are negligible. When censoring is 25\%, there is no clear pattern for sample size of 5 subjects/dose; for larger sample sizes, the strategy MLE shows a slight better performance than other strategies up to 0.03 for sample size from 10 to 20 subjects/dose. When censoring is 50\%, there is no clear patterns for sample size of 5 and 10 subjects/dose; for larger sample sizes, the strategy MLE also shows higher probability no more than 0.02 for sample size of 15 to 25 subjects/dose. 

\begin{figure}[!ht]
    \centering
    \includegraphics[scale = 0.9]{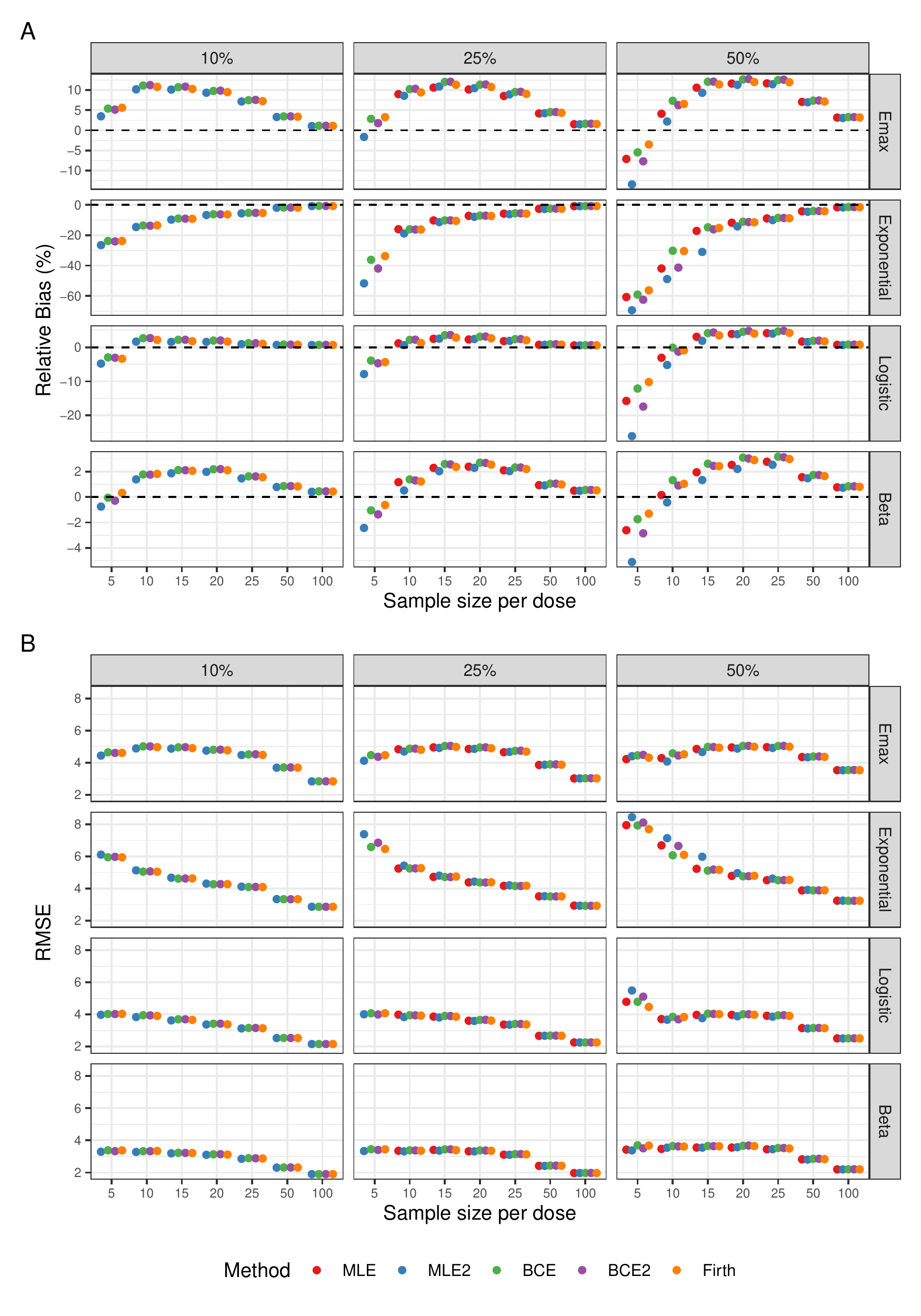}
    \caption{\textbf{A}. Bias of $\widehat{MED}$ derived from the MCP-Mod framework when  using MLE, MLE2, BCE, BCE2 and Firth estimators as function of censoring rate and true model. \textbf{B}. Root-Mean-Square Error of $\widehat{MED}$ derived from the MCP-Mod framework when  using MLE, MLE2, BCE, BCE2 and Firth estimators as function of censoring rate and true model.}
    \label{fig:bias_rmse}
\end{figure}

For Mod-step, we calculated the relative bias and RMSE of $\widehat{MED}$ estimator in Figure \ref{fig:bias_rmse}A and B, respectively, for different censoring rates and true models. When censoring is 10\% and 25\%, MLE and proposed strategies based on improved estimators have negligible differences for bias and RMSE; when censoring is 50\%, the strategies BCE and Firth presented lowest bias followed by MLE and BCE2 with noticeable poorer performance for MLE2 for sample sizes from 5 and 15 while similar conclusions can be drawn for RMSE only the Exponential as true model with sample sizes up to 15 subjects/dose and Logistic for sample size of 5 sujbects/dose. Otherwise, differences are negligible.

\section{Concluding Remarks}
\label{sec:conclusion}

We have derived improved inferences based on the Wald statistic for WRM particularizing general results from Cox and Snell \cite{CoxSnell:1968} and Magalhães et al. \cite{MagalhaesBotterSandoval:2021}, which complements previous results for improved inference based on likelihood ratio, Rao score and gradient statistics discussed in Magalhães and Gallardo \cite{MagalhaesGallardo:2020}. Few authors have presented improved inference for survival models with small sample sizes under the classical approach: Cordeiro and Colosimo \cite{CordeiroColosimo:1997, CordeiroColosimo:1999} and Medeiros \cite{MedeirosLemonte:2021} derived those statistics, respectively, for censored exponential regression models (ERM), a particular case of censored WRM. Also for ERM, Lemonte \cite{Lemonte:2022} presented the second-order covariance matrix of the MLE.

We have also proposed strategies based on  bias-corrected (BCE, BCE2 and Firth) and second-order covariance matrices (MLE2, BCE2) as input for the general MCP-Mod framework introduced by Pinheiros et al.\cite{pinheiro2014model}, which addresses the issue of relying on the asymptotic properties of MLE, which might not be valid for small sample sizes. To the best of our knowledge, this work is the first attempt to apply refined estimators for small sample sizes in the general MCP-Mod framework. Two simulations studies were performed to study the properties of refined estimators in an usual context of regression models and relevant operating characteristics in the MCP-Mod framework. 

In the simulation study presented for general censored Weibull regression model in the supplementary material, we showed numerical evidences that BCE and Firth estimator have lower bias and RMSE than MLE; second-order covariance matrices evaluated at MLE and BCE are closer to their respective empirical covariance matrices in comparison to the first-order covariance matrices; and Wald statistics derived from the combination between bias-corrected estimators and second-order covariance matrices yielded type I error probability closer to the nominal value than the standard Wald statistic with no loss of power. 

In the simulation study for the MCP-Mod framework, we have found that refined estimators and second-order covariance matrices approximate type I error probability to its nominal value in the MCP-step, while there are negligible differences between MLE and refined estimators in the probability of correctly detecting the dose-response signal, probability correctly selecting the dose-response model, bias and RMSE when censoring rates are up to 25\%. For censoring rate of 50\%, we found convergence issues for the corrected estimators and second-order covariance matrices and poorer performance in the assessed operating characteristics.  

In conclusion, we recommend the use of Firth as a strategy using refined estimators for small sample sizes in the MCP-Mod framework. In the context of basic science with limited sample sizes and large effect sizes, we do not expect large censoring rates in mouse experiments. In human trials, smaller effect sizes are pursued such that larger sample sizes are required. In this case, proposed strategies are not needed because all estimators present comparable performance for large sample sizes. Nonetheless, we showed that type I error probability is still inflated even for sample sizes of 25 subjects per dose, therefore, the use of proposed strategies would avoid to dedicate further efforts on non-promising drugs. 

We hope that refined estimators allow statisticians to implement the MCP-Mod framework  with small sample sizes accelerating the pre-clinical and clinical drug development process. An R-package, MCPModBC \cite{mcpmodbc}, is available on CRAN to fit the Weibull model with refined estimators and perform simulations for power considerations. Similar ideas can be applied to other distributions such as Binomial, Negative Binomial and Poisson, and they are currently under investigation. Furthermore, the impact of model misspecification requires further study.

\section{Data Availability}

Data sharing not applicable to this article as no datasets were generated or analysed during the current study.

\clearpage

\clearpage

\clearpage

\clearpage

\bibliographystyle{ama}
\bibliography{Wcdsocm_bib}

\begin{thebibliography}{10}

\bibitem{jardim2017factors}
Jardim Denis~L, Groves Eric~S, Breitfeld Philip~P, Kurzrock Razelle. Factors
  associated with failure of oncology drugs in late-stage clinical development:
  a systematic review  {\it Cancer treatment reviews. } 2017;52:12--21.

\bibitem{bretz2005combining}
Bretz Frank, Pinheiro Jos{\'e}~C, Branson Michael. Combining multiple
  comparisons and modeling techniques in dose-response studies  {\it
  Biometrics. } 2005;61:738--748.

\bibitem{pinheiro2014model}
Pinheiro Jos{\'e}, Bornkamp Bj{\"o}rn, Glimm Ekkehard, Bretz Frank. Model-based
  dose finding under model uncertainty using general parametric models  {\it
  Statistics in medicine. } 2014;33:1646--1661.

\bibitem{carroll2003use}
Carroll Kevin~J. On the use and utility of the Weibull model in the analysis of
  survival data  {\it Controlled clinical trials. } 2003;24:682--701.

\bibitem{MagalhaesGallardoGomez:2019}
Magalh{\~a}es Tiago~M., Gallardo Diego~I., G{\'o}mez H.~W.. Skewness of maximum
  likelihood estimators in the Weibull censored data  {\it Symmetry. }
  2019;11:1351.

\bibitem{MagalhaesGallardo:2020}
Magalh{\~a}es Tiago~M., Gallardo Diego~I.. Bartlett and Bartlett-type
  corrections for censored data from a Weibull distribution  {\it SORT -
  Statistics and Operations Research Transactions. } 2020;44:127--140.

\bibitem{CoxSnell:1968}
Cox David~R., Snell E.~J.. A general definition of residuals  {\it Journal of
  the Royal Statistical Society. Series B (Methodological). } 1968;30:248--275.

\bibitem{MagalhaesBotterSandoval:2021}
Magalh{\~a}es Tiago~M., Botter Denise~A., Sandoval M{\^o}nica~C.. A general
  expression for second-order covariance matrices - an application to
  dispersion models  {\it Brazilian Journal of Probability and Statistics. }
  2021;35:37--49.

\bibitem{Weibull:1951}
Weibull Waloddi. A statistical distribution function of wide applicability
  {\it Journal of Applied Mechanics. } 1951;18:293--297.

\bibitem{KalbfleischPrentice:2002}
Kalbfleisch John~D., Prentice Ross~L.. {\it The statistical analysis of failure
  time data}.
\newblock New Jersey: John Wiley \& Sons2~ed. 2002.

\bibitem{Wald:1943}
Wald Abraham. Test of statistical hypotheses concerning several parameter when
  the number of observations is large  {\it Transactions of the American
  Mathematical Society. } 1943;54:426--482.

\bibitem{Firth:1993}
Firth David. Bias reduction of maximum likelihood estimates  {\it Biometrika. }
  1993;80:27--38.

\bibitem{olive2009inhibition}
Olive Kenneth~P, Jacobetz Michael~A, Davidson Christian~J, et al. Inhibition of
  Hedgehog signaling enhances delivery of chemotherapy in a mouse model of
  pancreatic cancer  {\it Science. } 2009;324:1457--1461.

\bibitem{bornkamp2009mcpmod}
Bornkamp Bj{\"o}rn, Pinheiro Jos{\'e}, Bretz Frank, others . MCPMod: An R
  package for the design and analysis of dose-finding studies  {\it Journal of
  Statistical Software. } 2009;29:1--23.

\bibitem{CordeiroColosimo:1997}
Cordeiro Gaus~M., Colosimo Enrico~A.. Improved likelihood ratio tests for
  exponential censored data  {\it Journal of Statistical Computation and
  Simulation. } 1997;56:303--315.

\bibitem{CordeiroColosimo:1999}
Cordeiro Gaus~M., Colosimo Enrico~A.. Corrected score tests for exponential
  censored data  {\it Statistics \& Probability Letters. } 1999;44:365--373.

\bibitem{MedeirosLemonte:2021}
Medeiros Francisco M.~C., Lemonte Artur~J.. Likelihood-based inference in
  censored exponential regression models  {\it Communications in Statistics -
  Theory and Methods. } 2021;50:3214--3233.

\bibitem{Lemonte:2022}
Lemonte Artur~J.. Covariance matrix of maximum likelihood estimators in
  censored exponential regression models  {\it Communications in Statistics -
  Theory and Methods. } 2022;51:1765--1777.

\bibitem{mcpmodbc}
Diniz M\'arcio~A., Gallardo Diego~I., Magalh{\~a}es Tiago~M.. {\it MCPModBC:
  MCP-Mod with bias corrected estimators} 2023.
\newblock R package version 1.0.

\end{thebibliography}


\begin{thebibliography}{1}

\bibitem{Wald:1943}
Wald Abraham. Test of statistical hypotheses concerning several parameter when
  the number of observations is large  {\it Transactions of the American
  Mathematical Society. } 1943;54:426--482.

\bibitem{Lawley:1956}
Lawley D.. A general method for approximating to the distribution of likelihood
  ratio criteria  {\it Biometrika. } 1956;43:295--303.

\bibitem{CoxSnell:1968}
Cox David~R., Snell E.~J.. A general definition of residuals  {\it Journal of
  the Royal Statistical Society. Series B (Methodological). } 1968;30:248--275.

\bibitem{MagalhaesBotterSandoval:2021}
Magalh{\~a}es Tiago~M., Botter Denise~A., Sandoval M{\^o}nica~C.. A general
  expression for second-order covariance matrices - an application to
  dispersion models  {\it Brazilian Journal of Probability and Statistics. }
  2021;35:37--49.

\end{thebibliography}

\end{document}


\maketitle

\thispagestyle{empty}

\doublespacing

\beginsupplement

We performed a simulation study with 10,000 Monte Carlo replicates where censored Weibull data was generated with the regression structure established in \eqref{eq:reg}. We considered scenarios with $p = 3, 5, 7$ covariates following the standard normal distribution associated with the first $p$ components of the regression coefficient vector ${\bm \beta} = (-2.0, 1.5, -1.0, 2.5, -1.3, 1.8, -0.5)^{\top}$. Censoring was assumed as 10\%, 25\% and 50\%. Bias and RMSE for MLE, Firth and BCE estimators are presented in Supplemental Figures \ref{fig:s1}-\ref{fig:s6}.

Furthermore, we calculated matrix distances between the sampling covariance matrix - {\bf Cov}($\bm \beta^*$) - and the Fisher Information - ${\bm K}_{{\bm \beta} {\bm \beta}}^*$ - and the corrected covariance matrix $\mbox{{\bf Cov}}_{\bm 2}(\bm \beta^*)$ from \eqref{eq:cov2G} evaluated at MLE ($\widehat{\bm \beta}$) and BCE ($\widetilde{\bm \beta}$) in Table \ref{tab:s_cov}. The following distances between matrices ${\bm A}$ and ${\bm B}$ were considered:

\begin{itemize}
\item $d_1 = \max \mbox{diag}\left|{\bm A} - {\bm B}\right|$;
\item $d_2 = \sqrt{ \mbox{tr}\left\{ \left({\bm A} - {\bm B}\right)^{\top} \left({\bm A} - {\bm B}\right) \right\} }$;
\item $d_3 = \sum_{i} \sum_{j} |a_{ij}-b_{ij}|$;
\end{itemize}

Finally, we evaluated type I error testing the composite null  hypothesis $\mathcal{H}: {\vg \beta}_1 = {\vg \beta}_1^{(0)}$ against a composite alternative hypothesis $\mathcal{A}:$ $\mathcal{H}$ is false, where ${\vg \beta}_1^{(0)}$ is a specified vector, ${\vg \beta}_1$ is a $q$-dimensional vector and ${\vg \beta}_2$ contains the remaining $p-q$ parameters. This partition ${\vg \beta} = ({\vg \beta}_1^{\top}, {\vg \beta}_2^{\top})^{\top}$ induces the corresponding partitions
%
%
\begin{align*}
	{\bm K}_{{\bm \beta} {\bm \beta}} = \left(
	\begin{array}{cc}
		{\bm K}_{11} & {\bm K}_{12} \\
		{\bm K}_{21} & {\bm K}_{22} \\
	\end{array}
	\right)
	= \sigma^{-2} \left(
	\begin{array}{cc}
	{\vg X}_1^{\top} {\vg W} {\vg X}_1 & {\vg X}_1^{\top} {\vg W} {\vg X}_2 \\
	{\vg X}_2^{\top} {\vg W} {\vg X}_1 & {\vg X}_2^{\top} {\vg W} {\vg X}_2 \\
	\end{array}
	\right)
	\mbox{ and }
	{\bm K}_{{\bm \beta} {\bm \beta}}^{-1} = \left(
	\begin{array}{cc}
		{\bm K}^{11} & {\bm K}^{12} \\
		{\bm K}^{21} & {\bm K}^{22} \\
	\end{array}
	\right),
\end{align*}

\noindent where ${\vg X} = [{\vg X}_1 \ {\vg X}_2]$, ${\vg X}_1$, ${\vg X}_2$ being $n \times q$ and $n \times (p-q)$, respectively. We consider the following Wald test\cite{Wald:1943} statistics:
%
%
\begin{align}
\label{eq:w0}
W_{MLE} &= \left( \widehat{{\vg \beta}}_1 - {\vg \beta}_1^{(0)} \right)^{\top} \left\{ \widehat{{\bm K}}^{11} \right\}^{-1} \left( \widehat{{\vg \beta}}_1 - {\vg \beta}_1^{(0)} \right),\\
\label{eq:w1}
W_{MLE2} &= \left( \widehat{{\bm \beta}}_1 - {\bm \beta}_1^{(0)} \right)^{\top} \left\{ {\widehat{\rm \bf C}}\mbox{{\bf ov}}_{\bm 2}^{11}(\widehat{{\bm \beta}}) \right\}^{-1} \left( \widehat{{\bm \beta}}_1 - {\bm \beta}_1^{(0)} \right), \\
\label{eq:w2}
W_{BCE} &= \left( \widetilde{{\vg \beta}}_1 - {\vg \beta}_1^{(0)} \right)^{\top} \left\{ \widetilde{{\bm K}}^{11} \right\}^{-1} \left( \widetilde{{\vg\beta}}_1 - {\vg \beta}_1^{(0)} \right),\\
\label{eq:w3}
W_{BCE2} &= \left( \widetilde{{\bm \beta}}_1 - {\bm \beta}_1^{(0)} \right)^{\top} \left\{ {\widetilde{\rm \bf C}}\mbox{{\bf ov}}_{\bm 2}^{11}(\widetilde{{\bm \beta}}) \right\}^{-1} \left( \widetilde{{\bm \beta}}_1 - {\bm \beta}_1^{(0)} \right),\\
\label{eq:w4}
W_{Firth} &= \left( \check{{\bm \beta}}_1 - {\bm \beta}_1^{(0)} \right)^{\top} \left\{ \check{{\bm K}}^{11} \right\}^{-1} \left( \check{{\bm \beta}}_1 - {\bm \beta}_1^{(0)} \right).
\end{align}

In the censored Weibull regression model the statistic \eqref{eq:w0}-\eqref{eq:w4} can be rewritten as 
\begin{align}
W = \left({\vg \beta}_1 - {\vg \beta}_1^{(0)} \right)^{\top} \left( {\vg R}^{\top} {\vg W} {\vg R} \right) \left({\vg \beta}_1 - {\vg \beta}_1^{(0)} \right),
\end{align}

\noindent with ${\vg R} = {\vg X}_1 - {\vg X}_2 {\vg C}$, ${\vg C} = \left( {\vg X}_2^{\top} {\vg W} {\vg X}_2 \right)^{-1} {\vg X}_2^{\top} {\vg W} {\vg X}_1$ represents a $(p-q) \times q$ matrix whose columns are the vectors of regression coefficients obtained in the weighted normal linear regression of the columns of ${\vg X}_1$ on the model matrix ${\vg X}_2$ with ${\vg W}$ as a weight matrix.

Under the null hypothesis $\mathcal{H}$, $W$ has a $\chi_{q}^2$ distribution up to an error of order $n^{-1}$. The null hypothesis is rejected for a given nominal level, $\alpha$ say, if the test statistic exceeds the upper $100(1 - \alpha)\%$ quantile of the $\chi_{q}^2$ distribution.

Similarly, we evaluated power with the composite alternative hypothesis $ \mathcal{A}:{\bm \beta}^\top=(\psi \textbf{1}_q,\textbf{0}_{p-q})^\top$ for $\psi = 0.05, 0.10, 0.25, 0.50, 1.00, 2.00$. Results are presented in Table \ref{tab:s_power}.

\section{Results}

\subsection{Assessing the bias for different estimation procedures}

\begin{figure}[!h]
\begin{center}
\resizebox{\linewidth}{!}{
\begin{tabular}{cccc}
  $p=3$ &
  \begin{minipage}{.28\textwidth}{\includegraphics[width=1\textwidth]{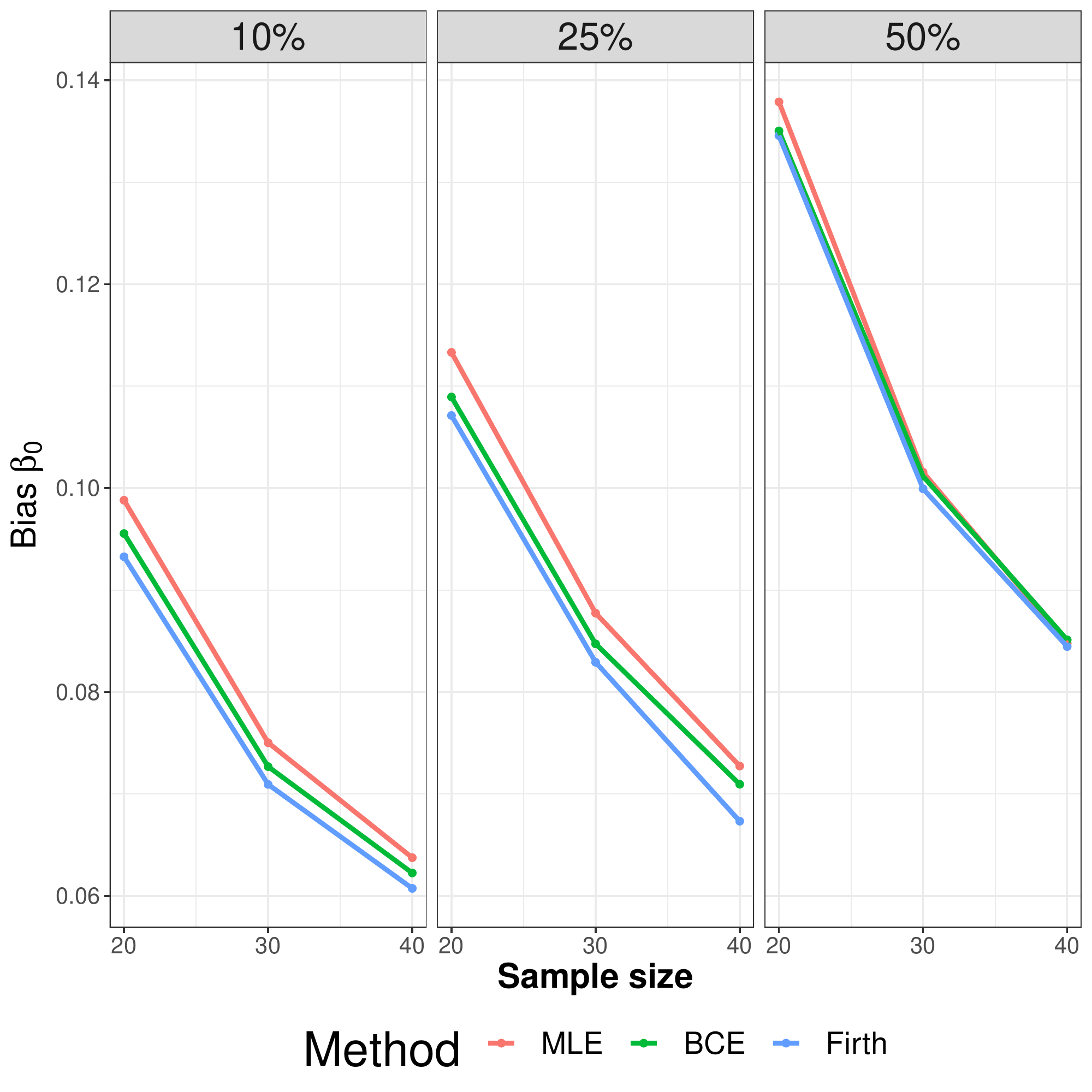}} \end{minipage} &
  \begin{minipage}{.28\textwidth}{\includegraphics[width=1\textwidth]{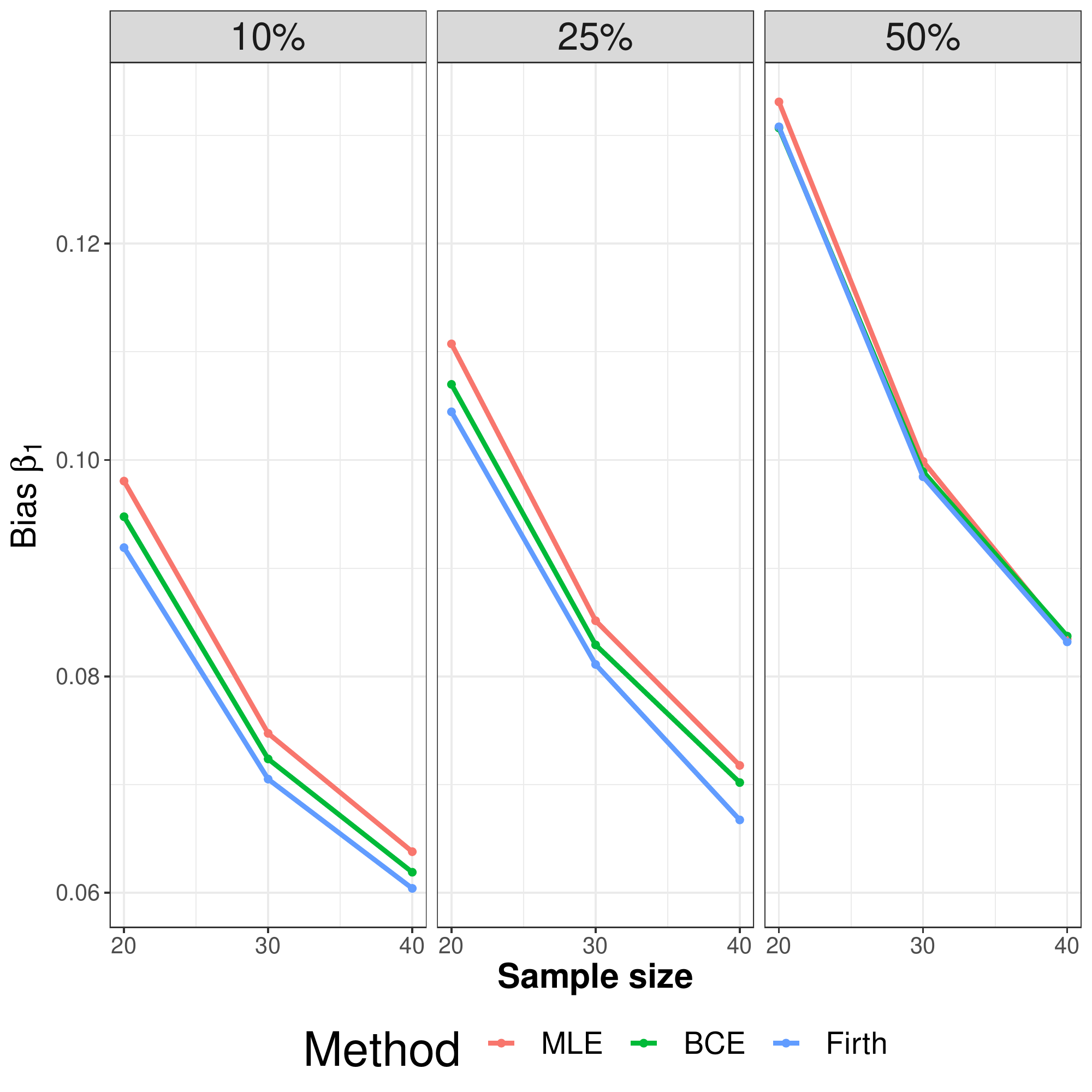}} \end{minipage} &
  \begin{minipage}{.28\textwidth}{\includegraphics[width=1\textwidth]{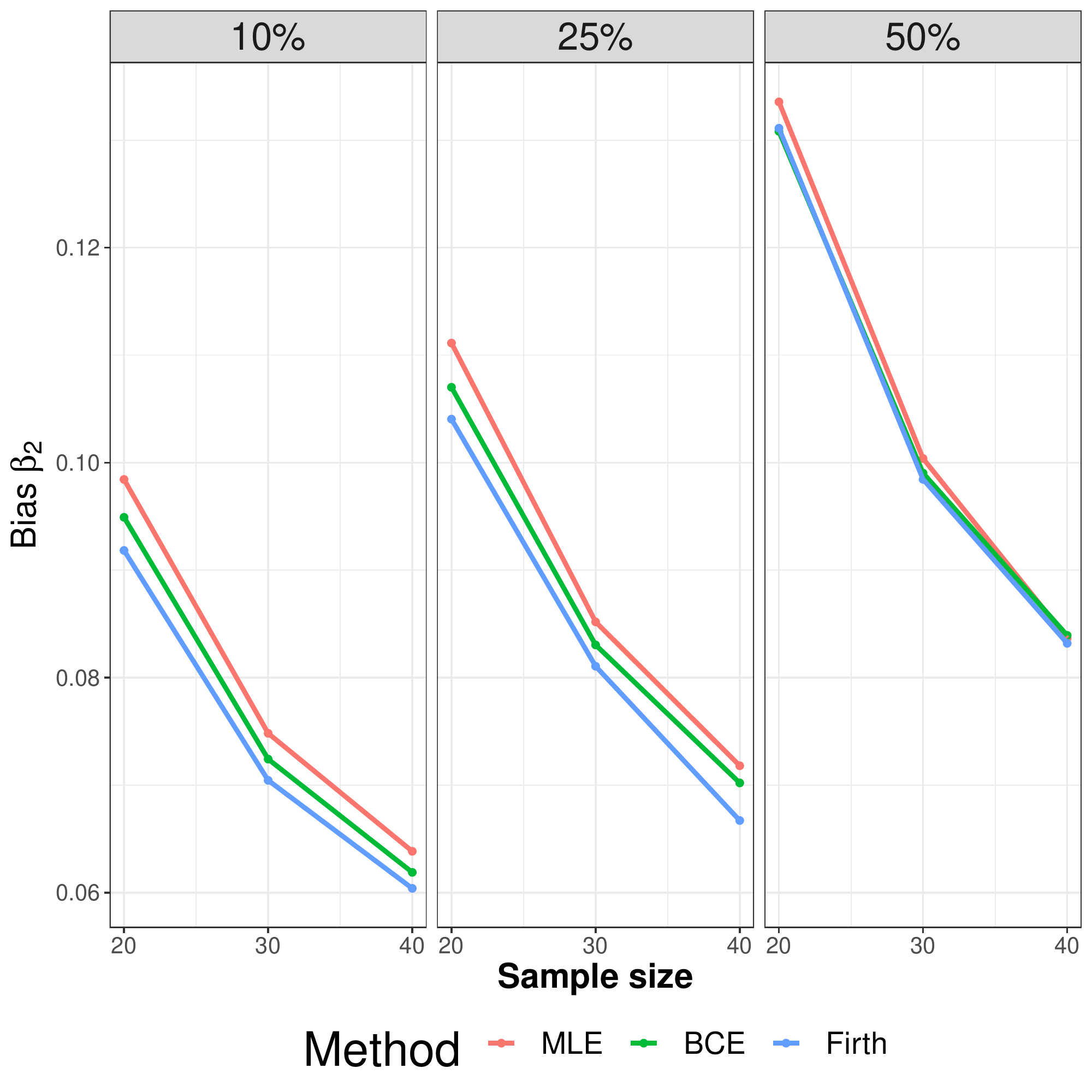}} \end{minipage} \\
  $p=5$ &
  \begin{minipage}{.28\textwidth}{\includegraphics[width=1\textwidth]{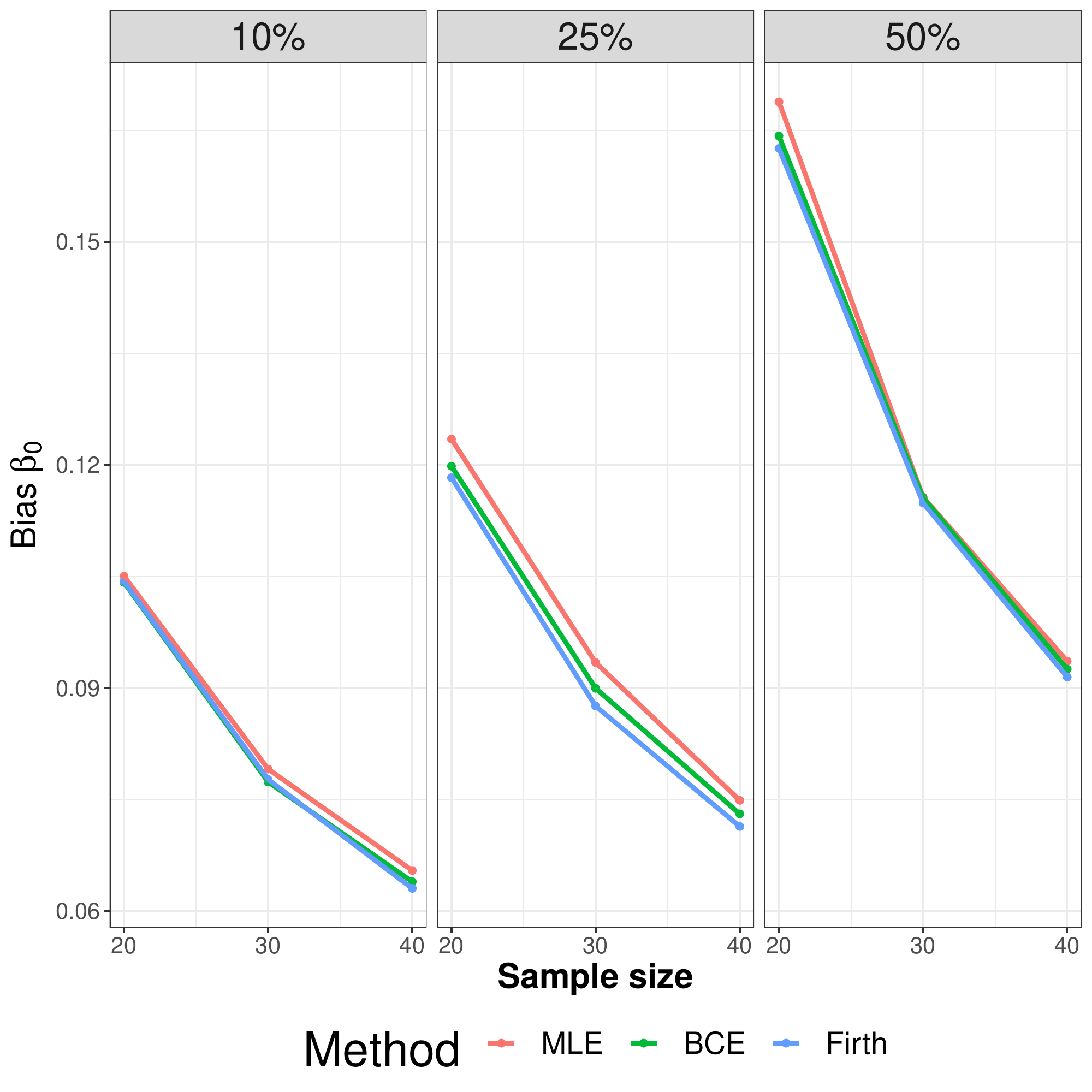}} \end{minipage} &
  \begin{minipage}{.28\textwidth}{\includegraphics[width=1\textwidth]{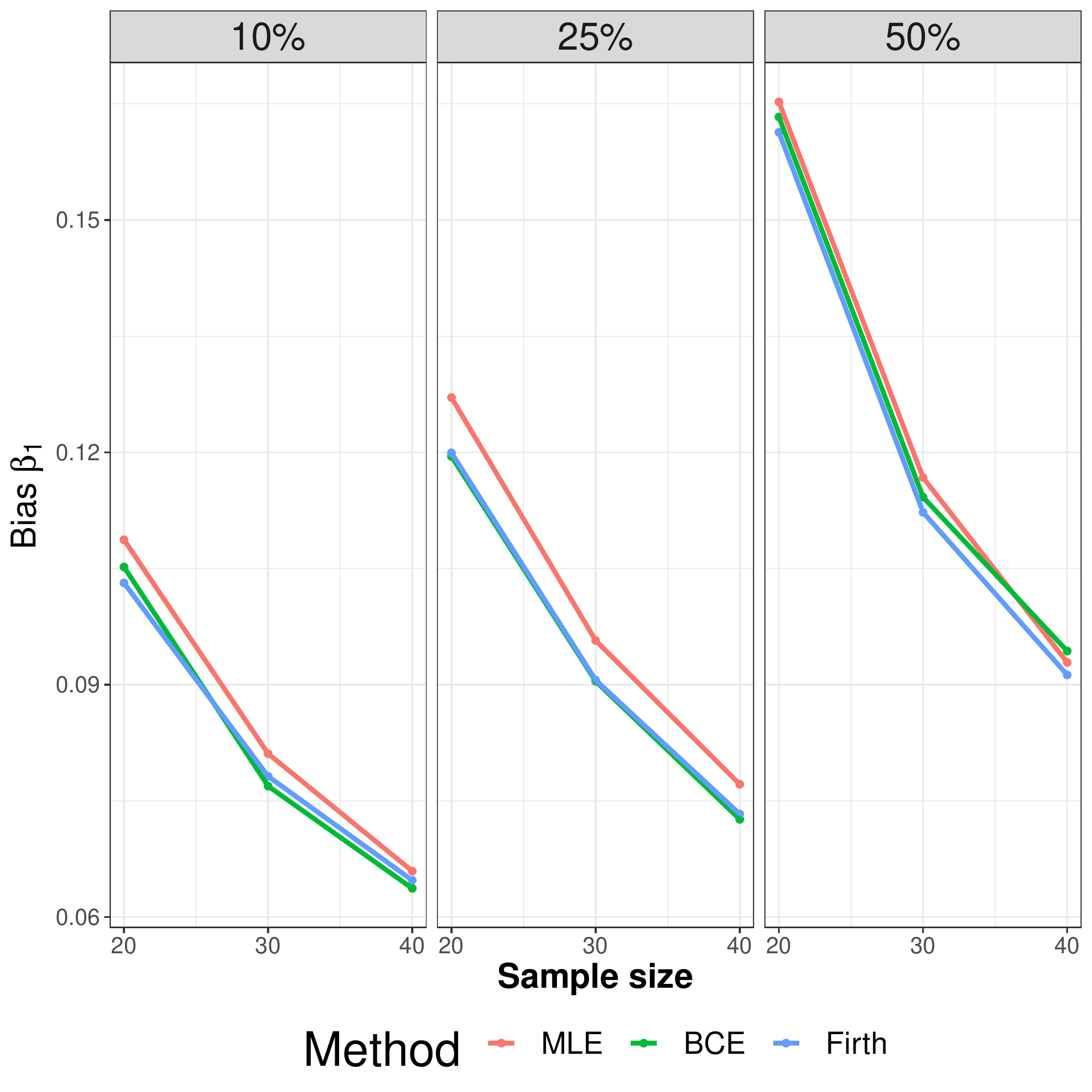}} \end{minipage} &
  \begin{minipage}{.28\textwidth}{\includegraphics[width=1\textwidth]{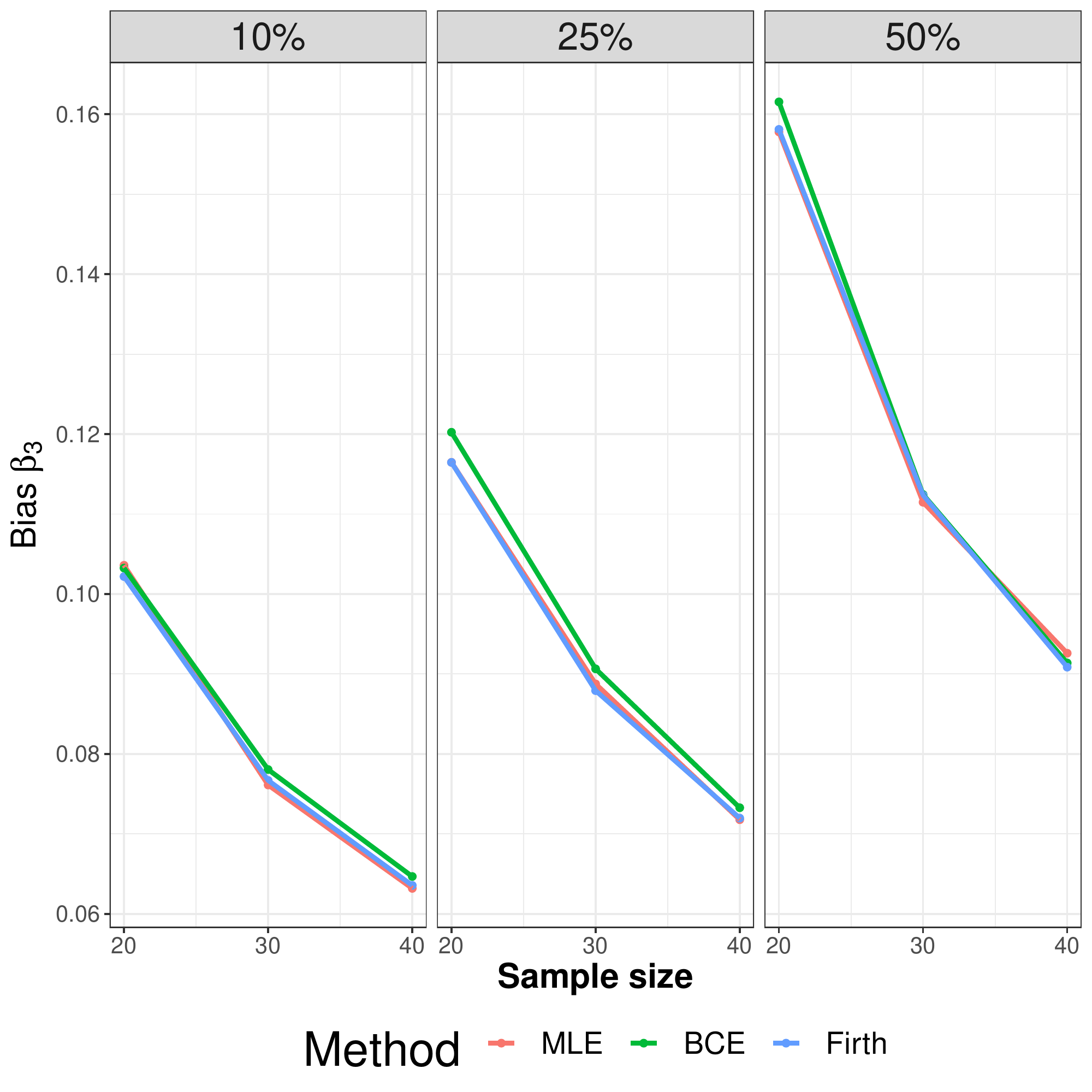}} \end{minipage} \\
  $p=7$ &
  \begin{minipage}{.28\textwidth}{\includegraphics[width=1\textwidth]{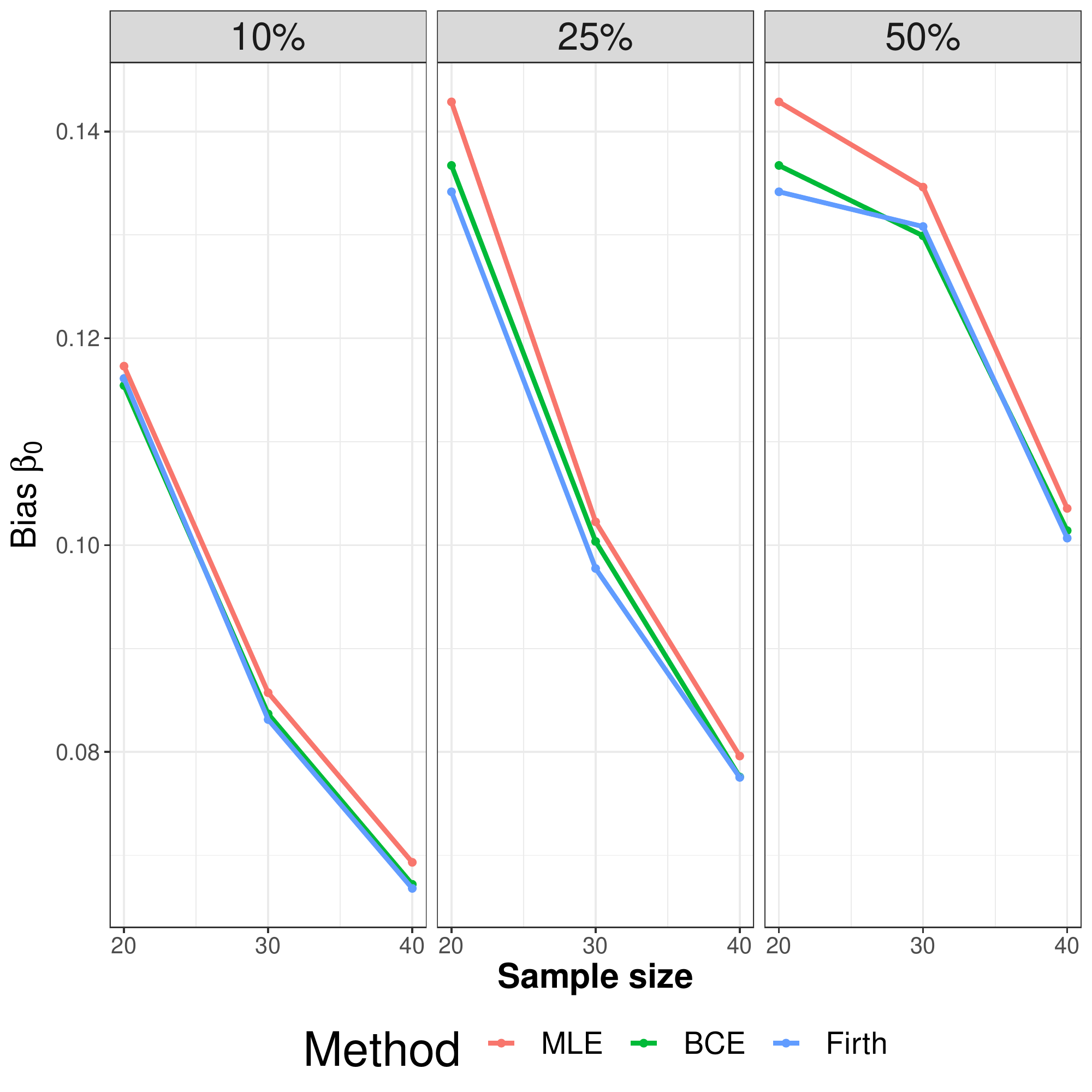}} \end{minipage} &
  \begin{minipage}{.28\textwidth}{\includegraphics[width=1\textwidth]{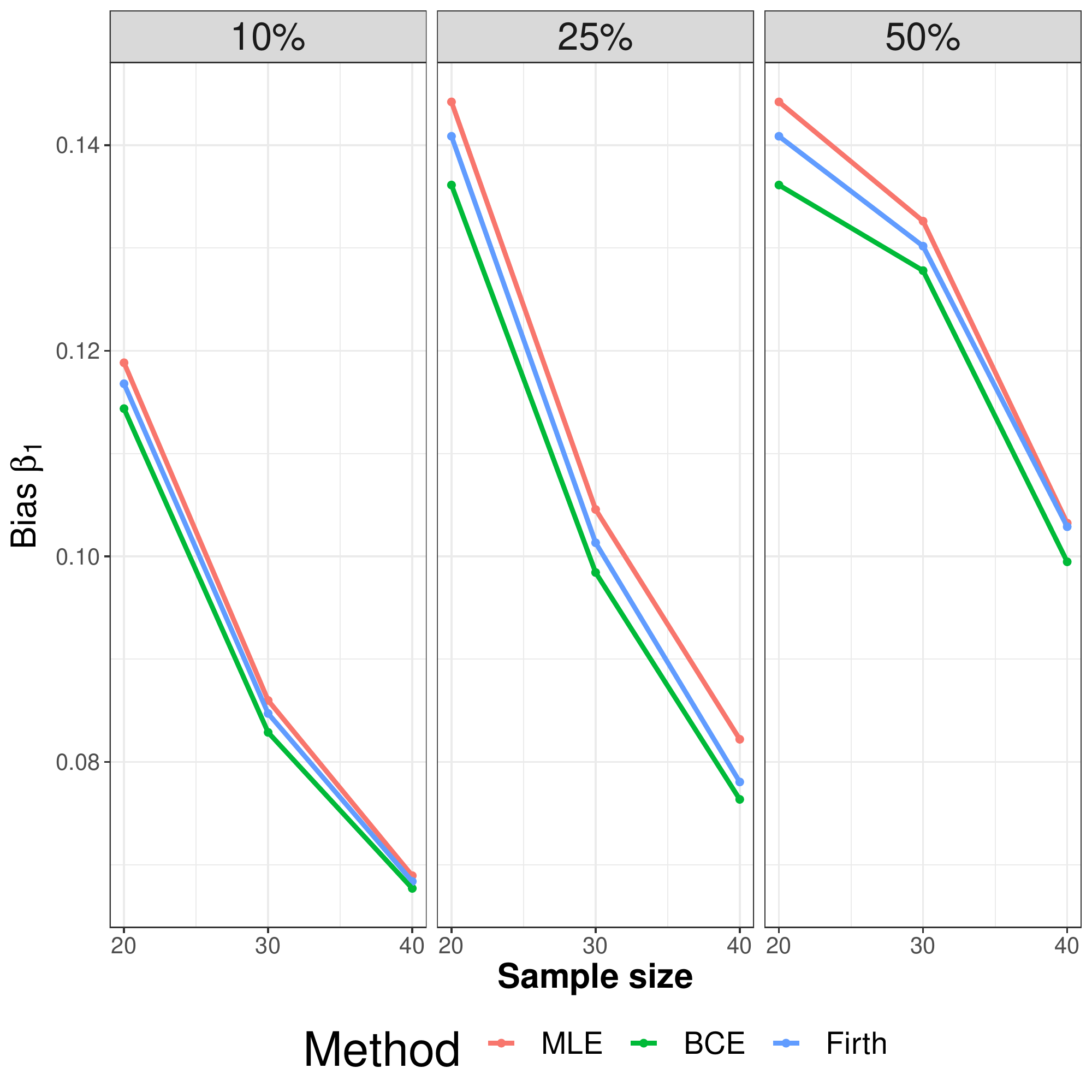}} \end{minipage} &
  \begin{minipage}{.28\textwidth}{\includegraphics[width=1\textwidth]{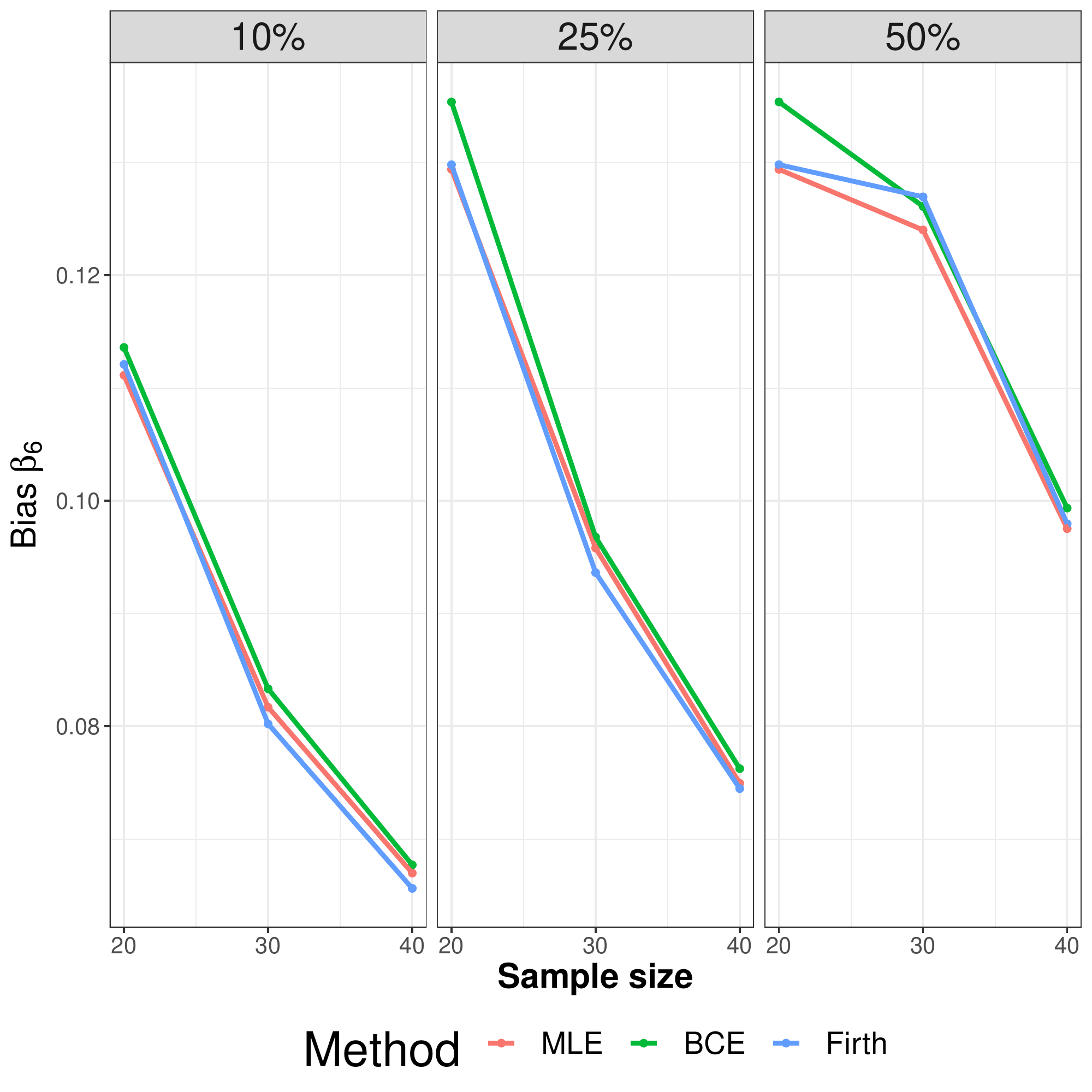}} \end{minipage} \\
\end{tabular}
}
\caption{Empirical bias for different estimators. Case $\sigma=0.5$.}
\label{fig:s1}
\end{center}
\end{figure}

\begin{figure}[!h]
\begin{center}
\resizebox{\linewidth}{!}{
\begin{tabular}{cccc}
  $p=3$ &
  \begin{minipage}{.28\textwidth}{\includegraphics[width=1\textwidth]{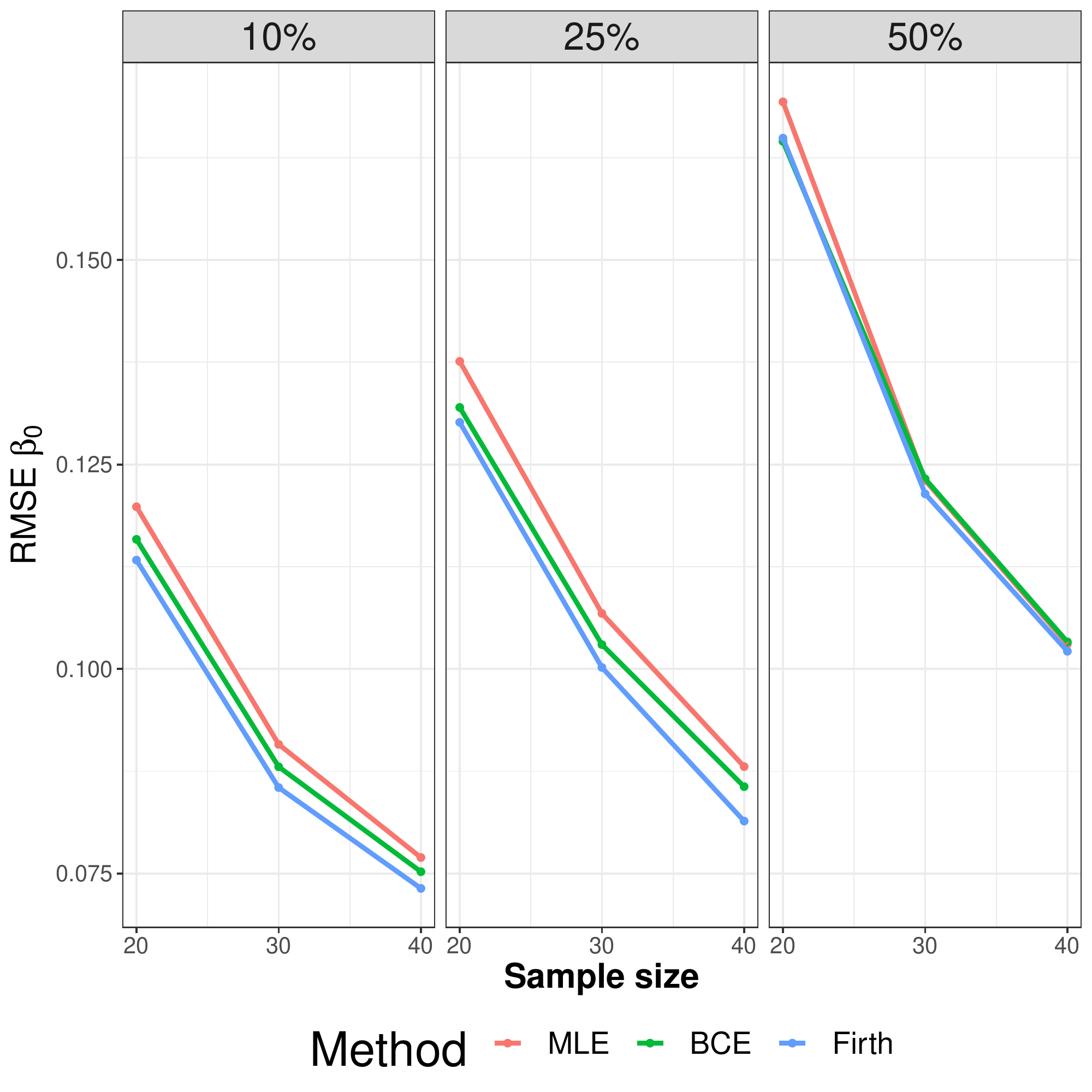}} \end{minipage} &
  \begin{minipage}{.28\textwidth}{\includegraphics[width=1\textwidth]{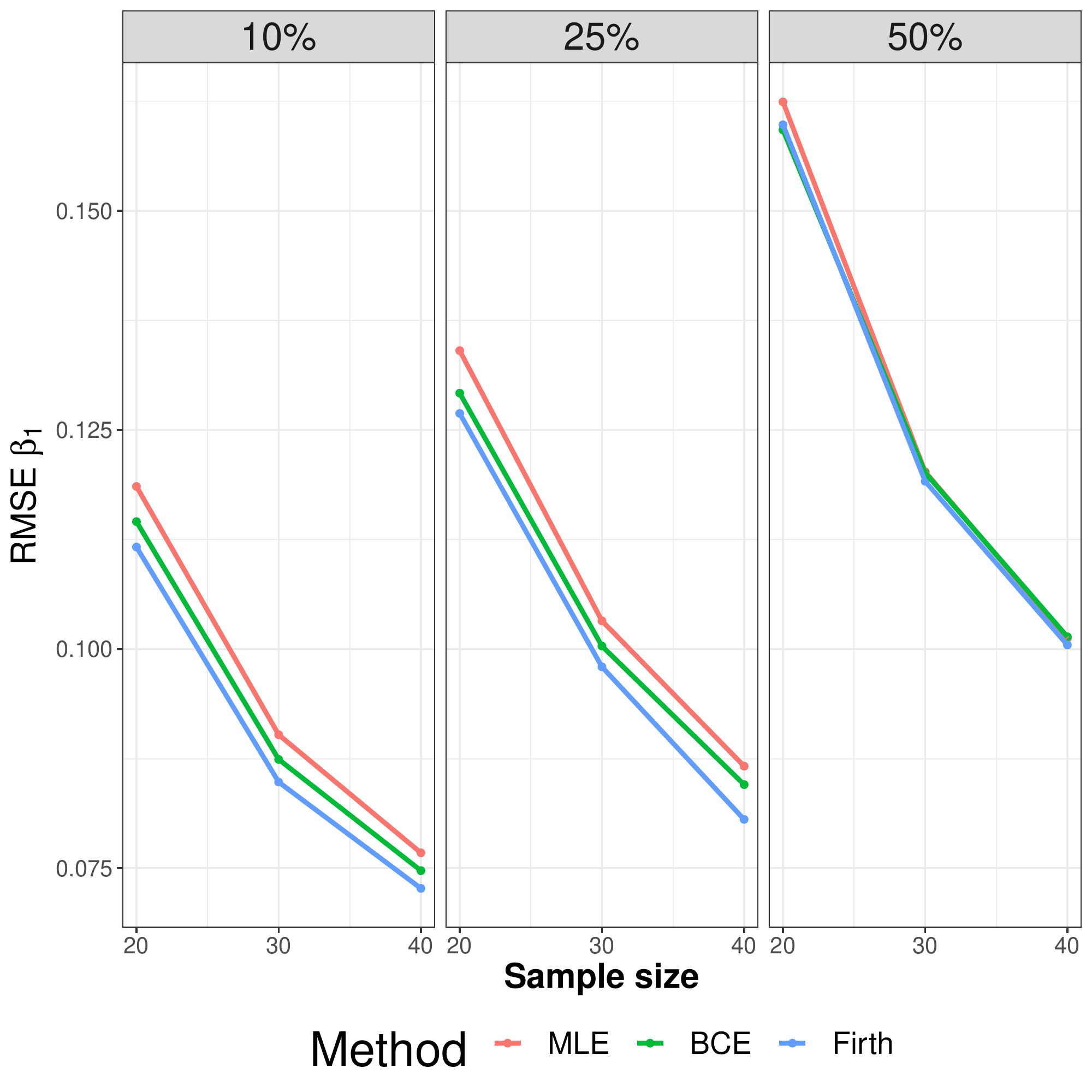}} \end{minipage} &
  \begin{minipage}{.28\textwidth}{\includegraphics[width=1\textwidth]{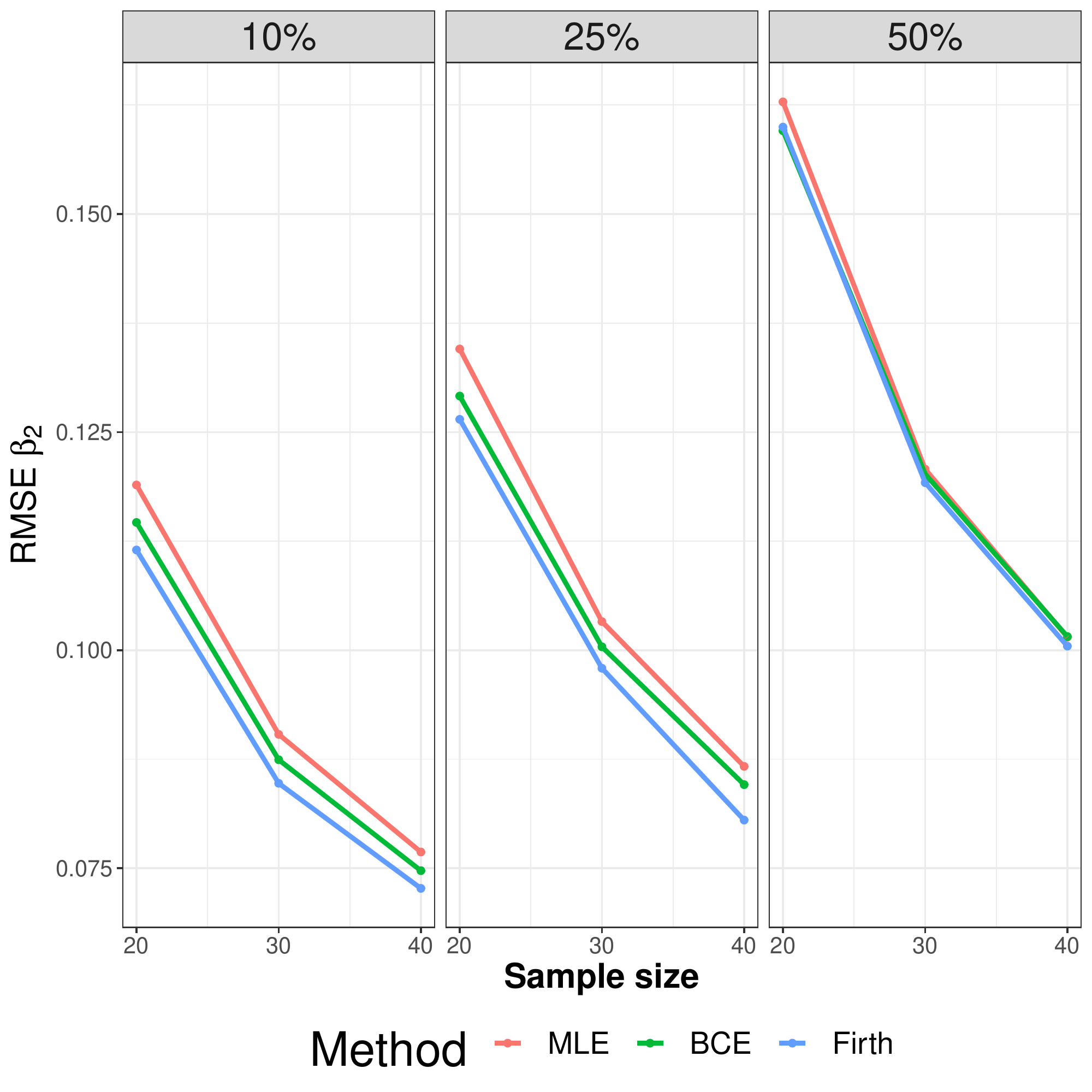}} \end{minipage} \\
  $p=5$ &
  \begin{minipage}{.28\textwidth}{\includegraphics[width=1\textwidth]{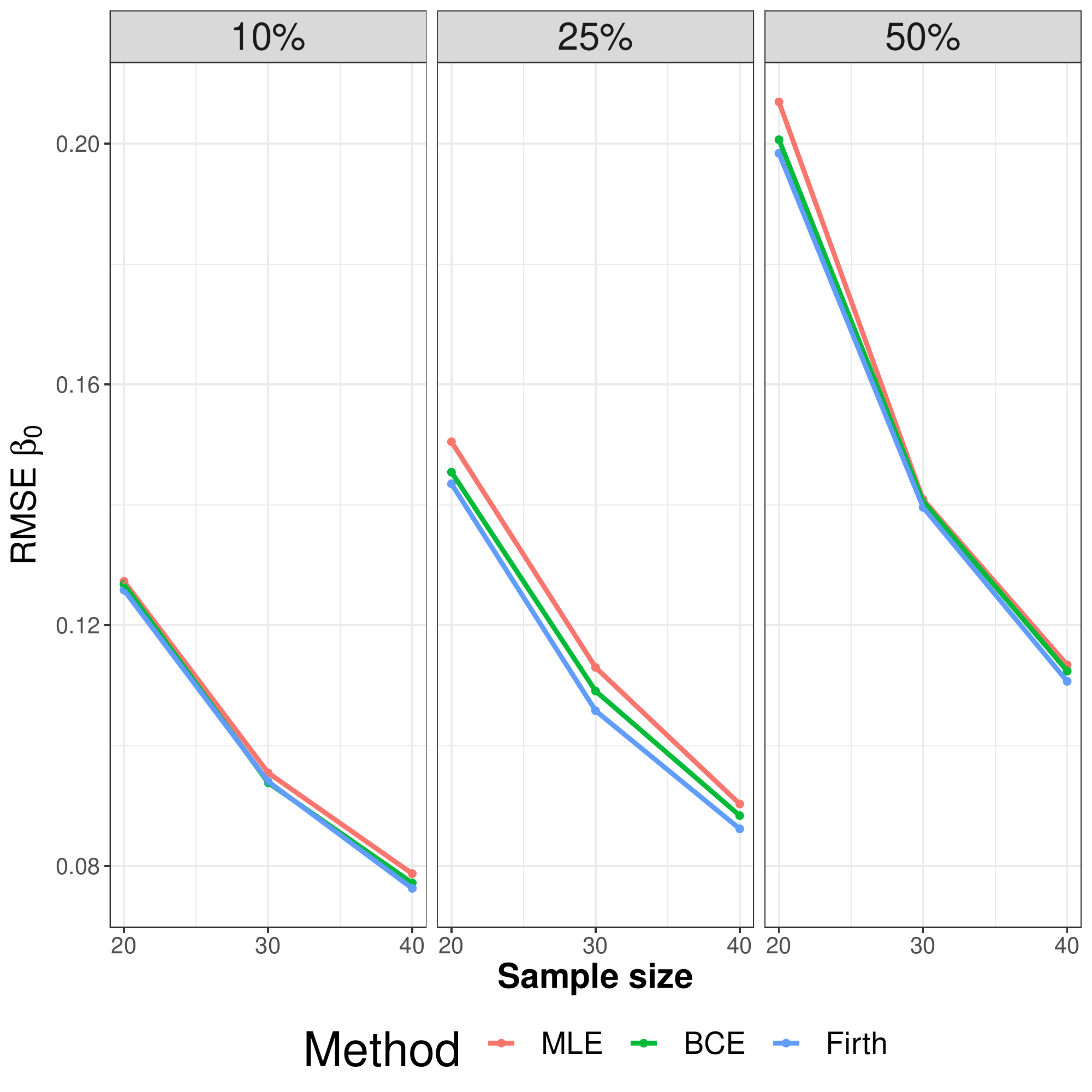}} \end{minipage} &
  \begin{minipage}{.28\textwidth}{\includegraphics[width=1\textwidth]{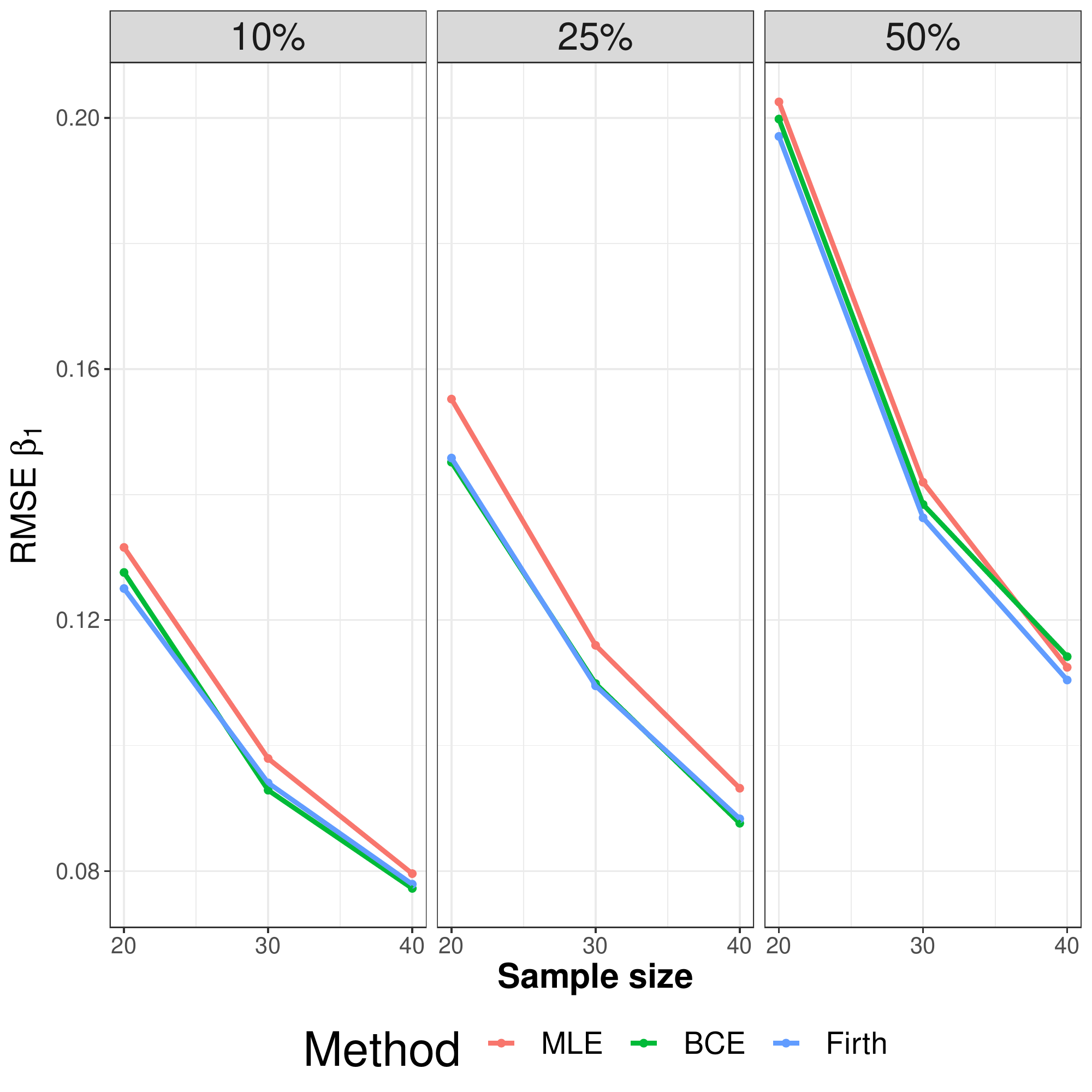}} \end{minipage} &
  \begin{minipage}{.28\textwidth}{\includegraphics[width=1\textwidth]{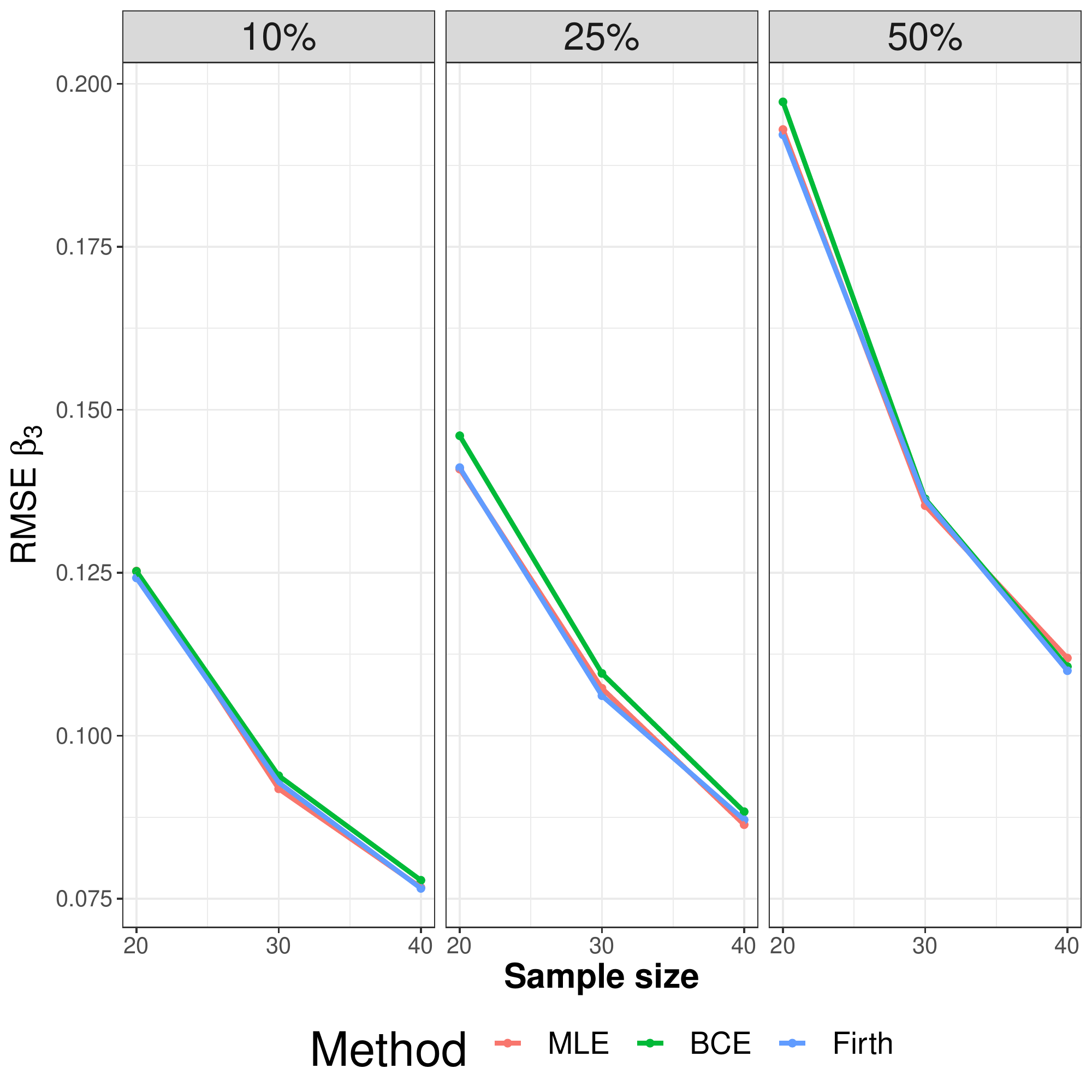}} \end{minipage} \\
  $p=7$ &
  \begin{minipage}{.28\textwidth}{\includegraphics[width=1\textwidth]{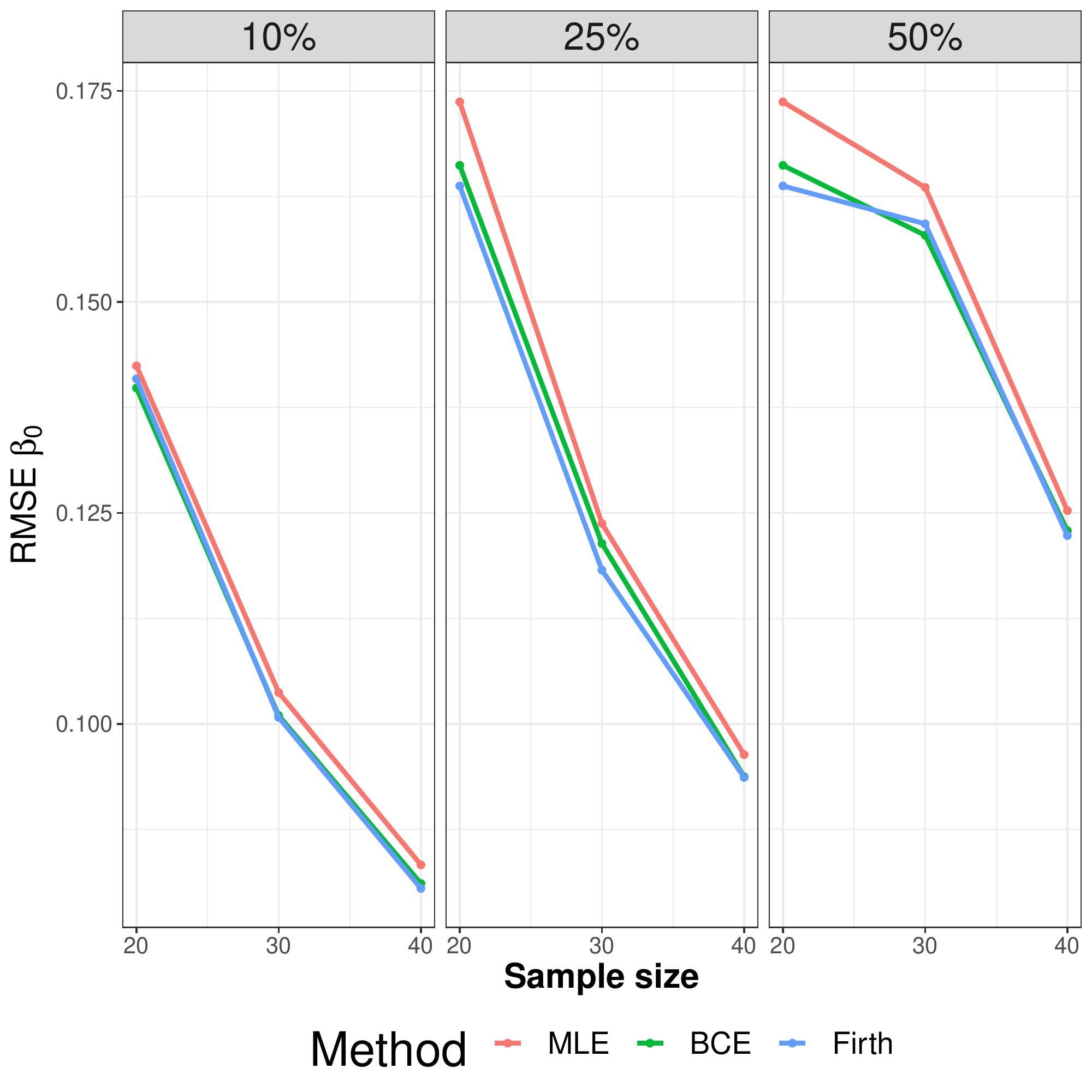}} \end{minipage} &
  \begin{minipage}{.28\textwidth}{\includegraphics[width=1\textwidth]{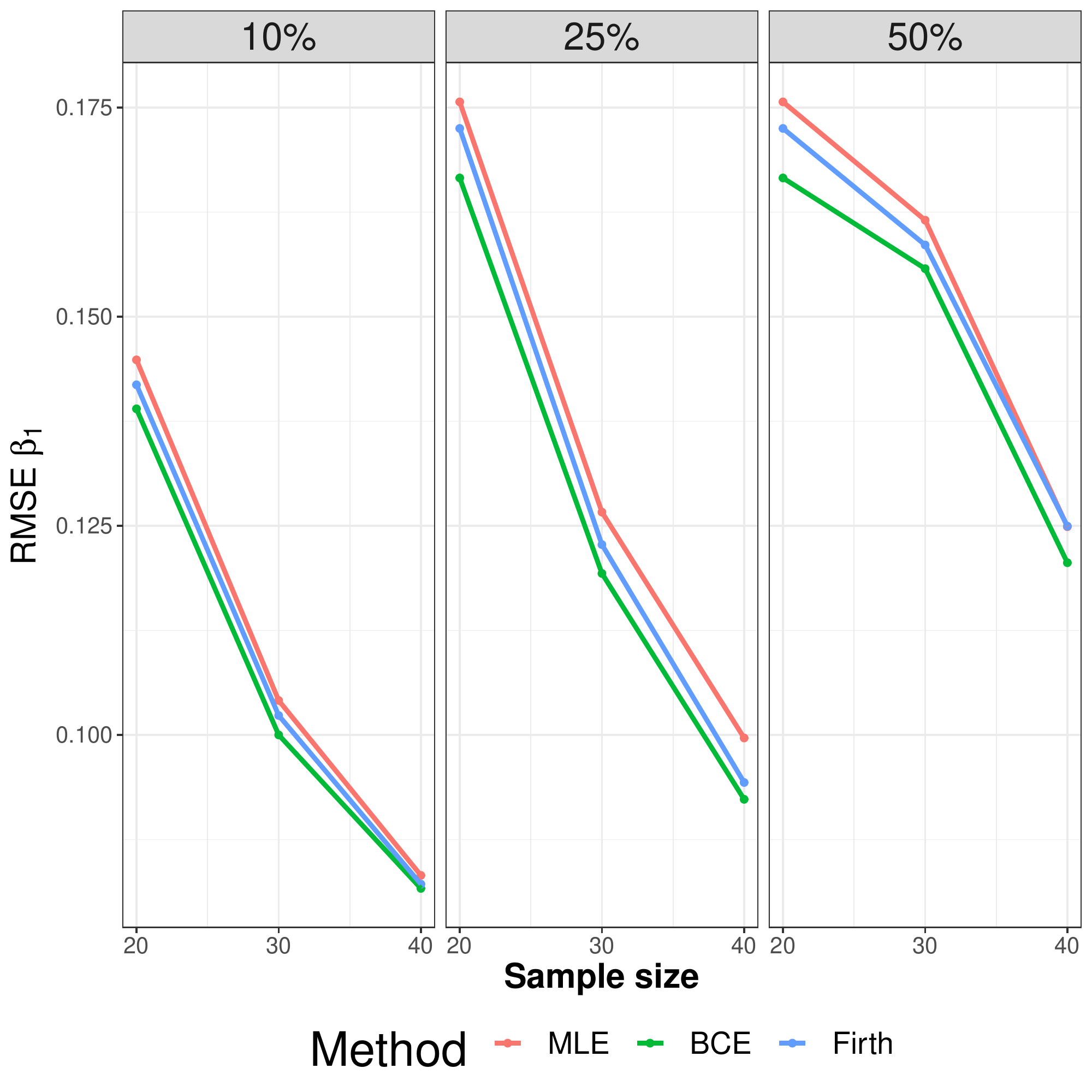}} \end{minipage} &
  \begin{minipage}{.28\textwidth}{\includegraphics[width=1\textwidth]{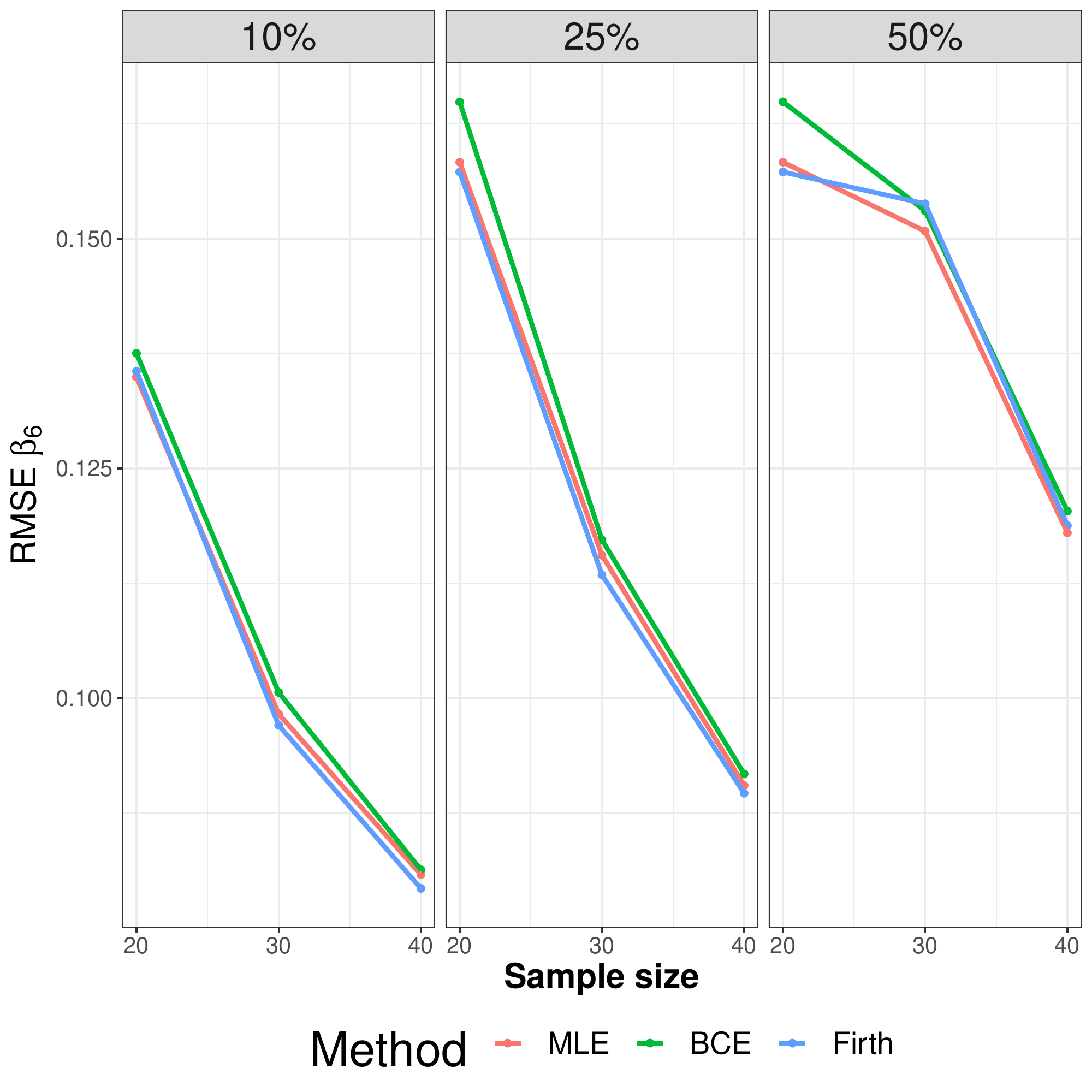}} \end{minipage} \\
\end{tabular}
}
\caption{Empirical RMSE for different estimators. Case $\sigma=0.5$.}
\label{fig:s2}
\end{center}
\end{figure}

\begin{figure}[!h]
\begin{center}
\resizebox{\linewidth}{!}{
\begin{tabular}{cccc}
  $p=3$ &
  \begin{minipage}{.28\textwidth}{\includegraphics[width=1\textwidth]{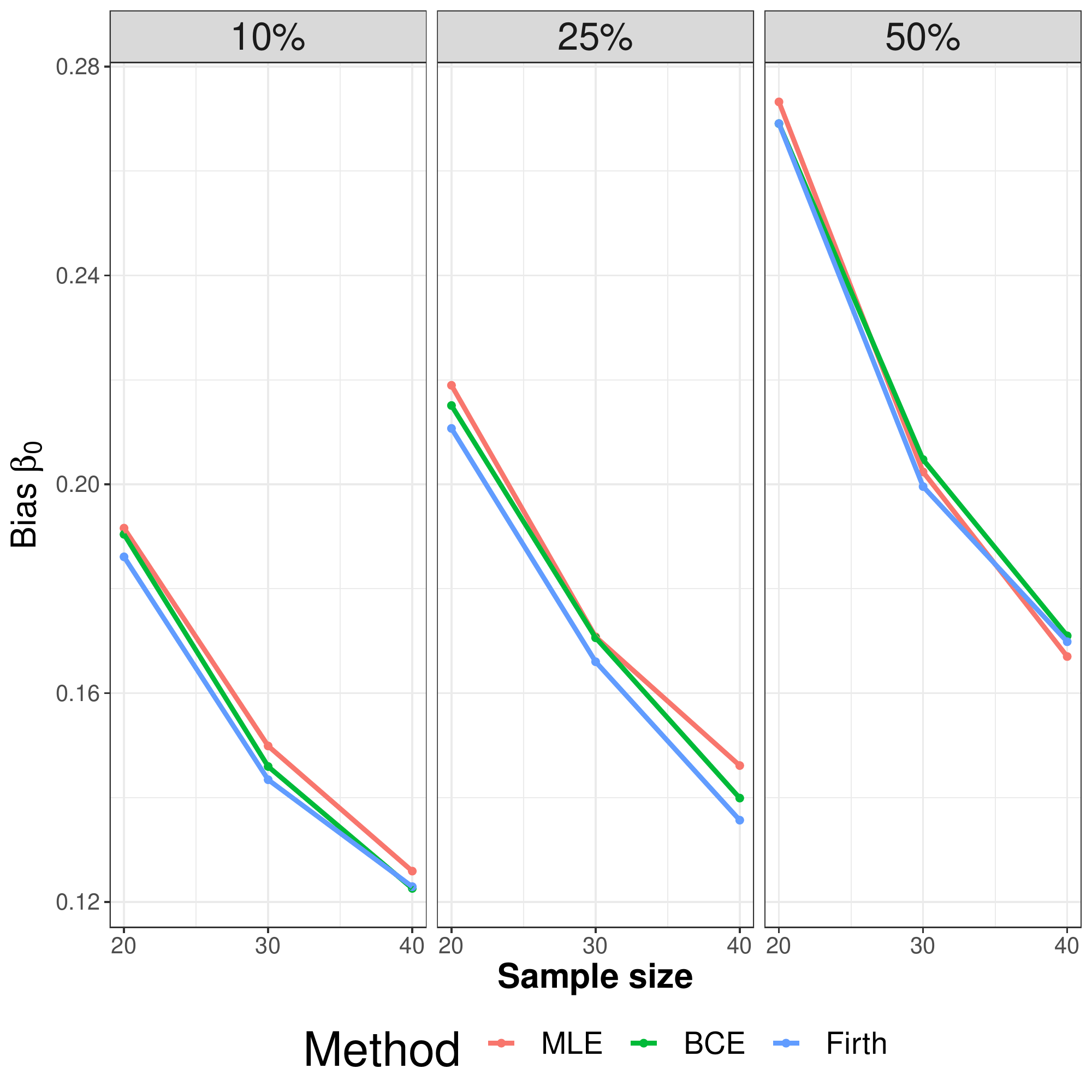}} \end{minipage} &
  \begin{minipage}{.28\textwidth}{\includegraphics[width=1\textwidth]{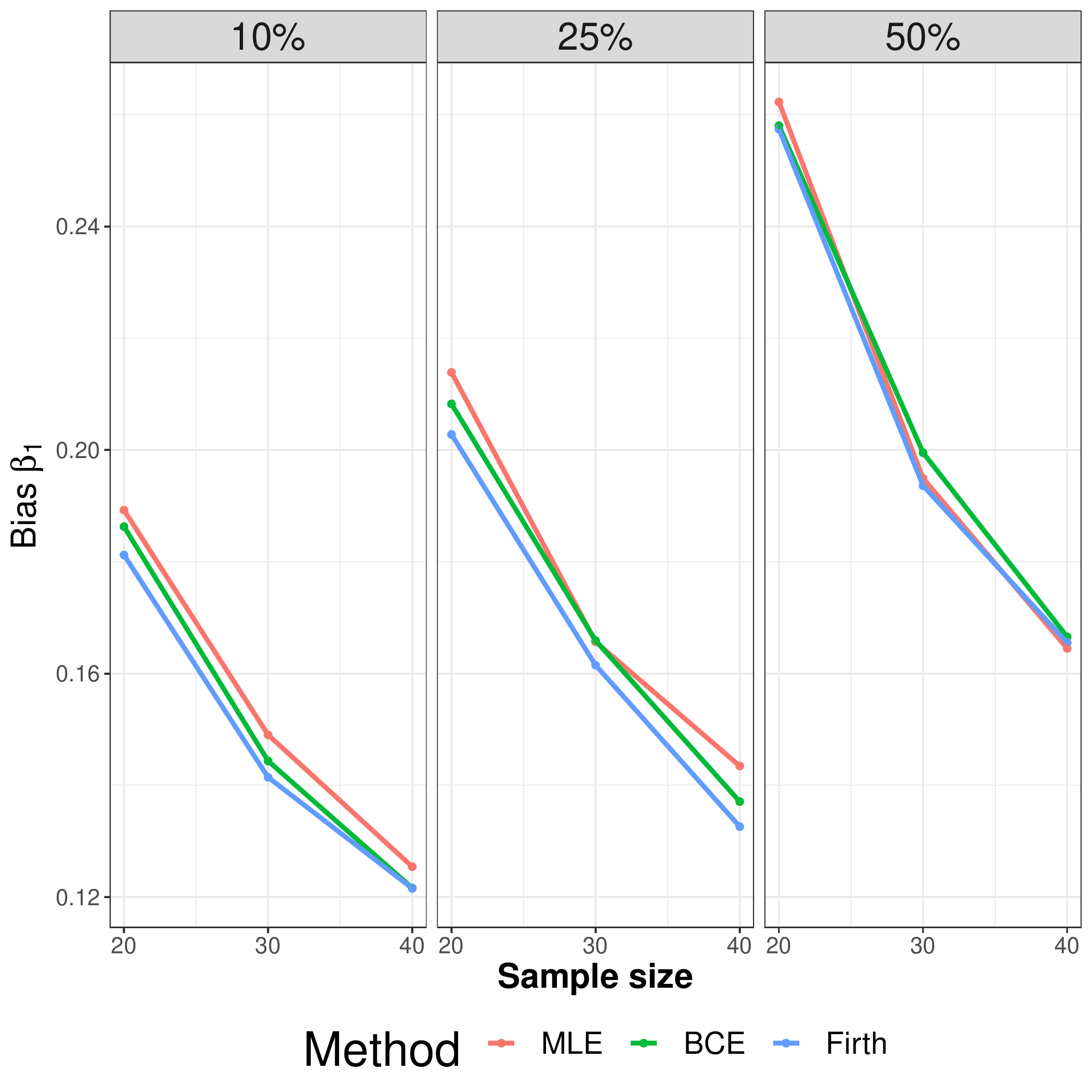}} \end{minipage} &
  \begin{minipage}{.28\textwidth}{\includegraphics[width=1\textwidth]{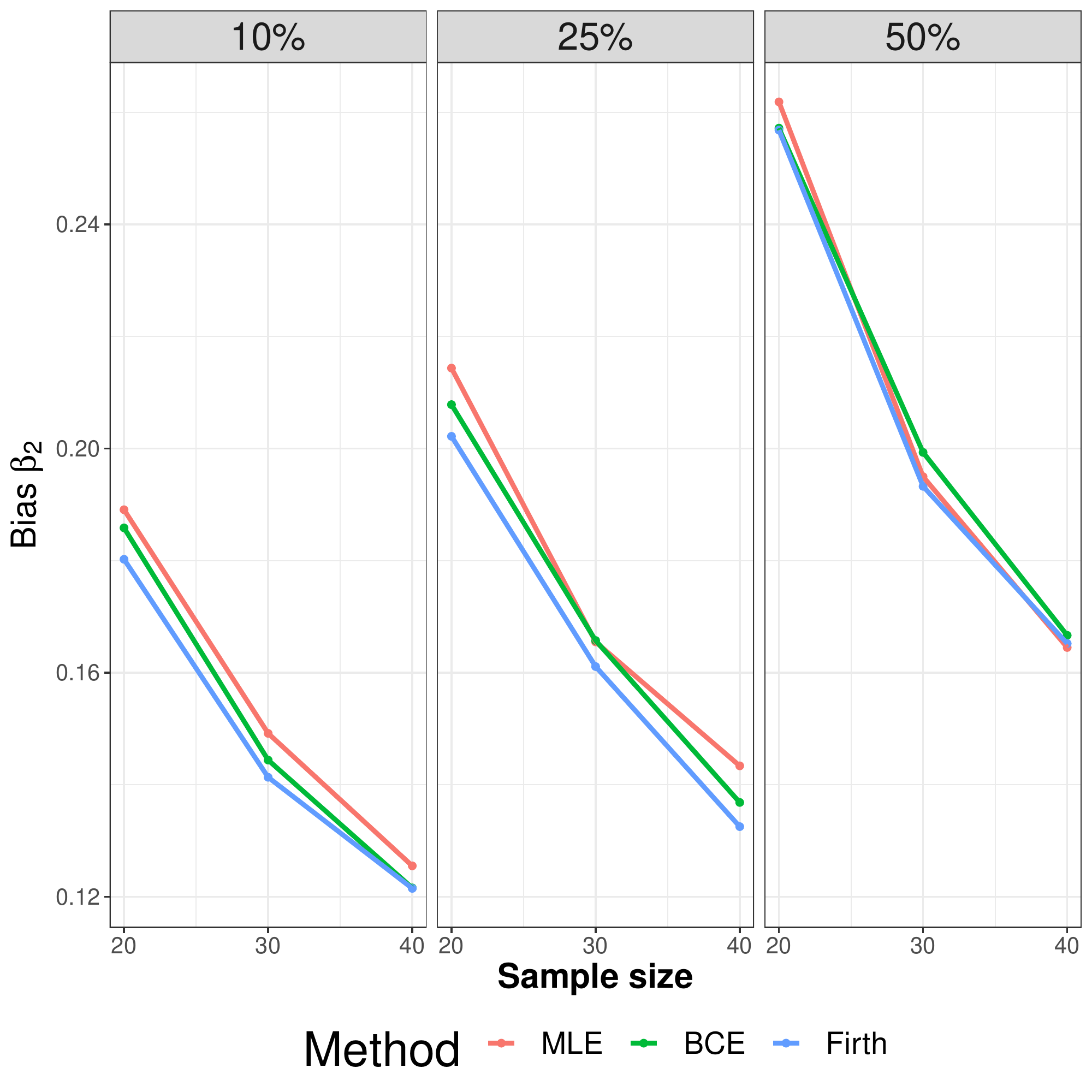}} \end{minipage} \\
  $p=5$ &
  \begin{minipage}{.28\textwidth}{\includegraphics[width=1\textwidth]{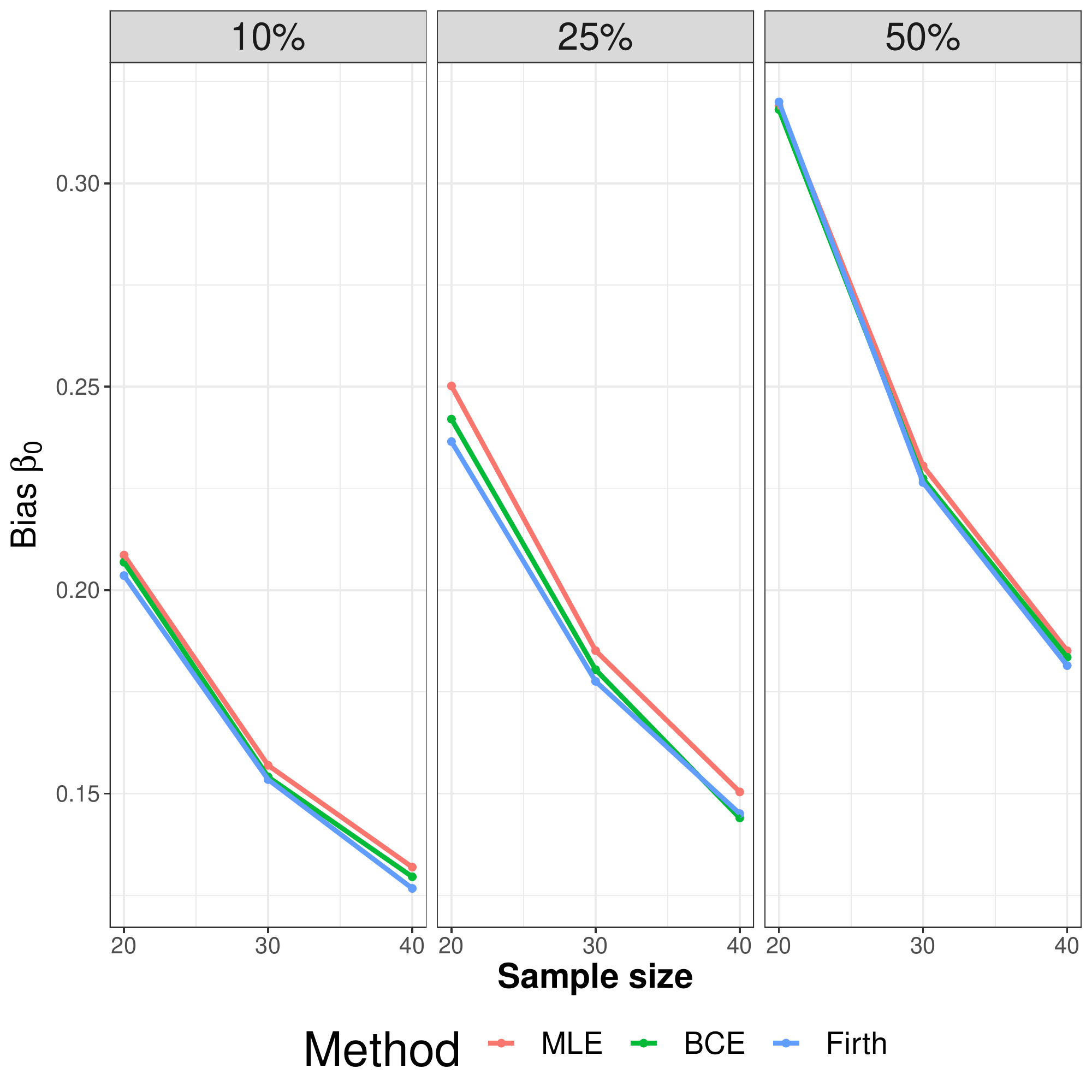}} \end{minipage} &
  \begin{minipage}{.28\textwidth}{\includegraphics[width=1\textwidth]{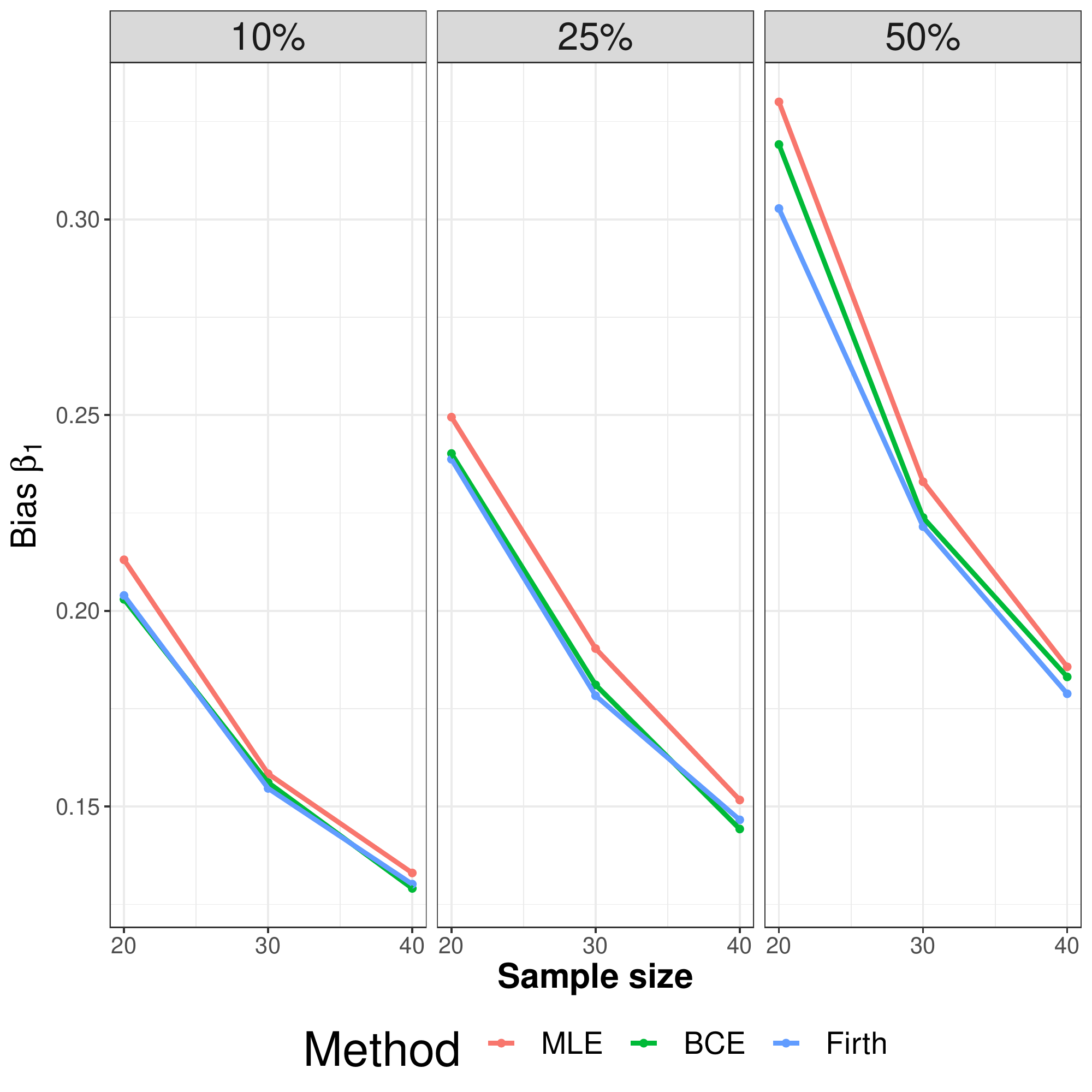}} \end{minipage} &
  \begin{minipage}{.28\textwidth}{\includegraphics[width=1\textwidth]{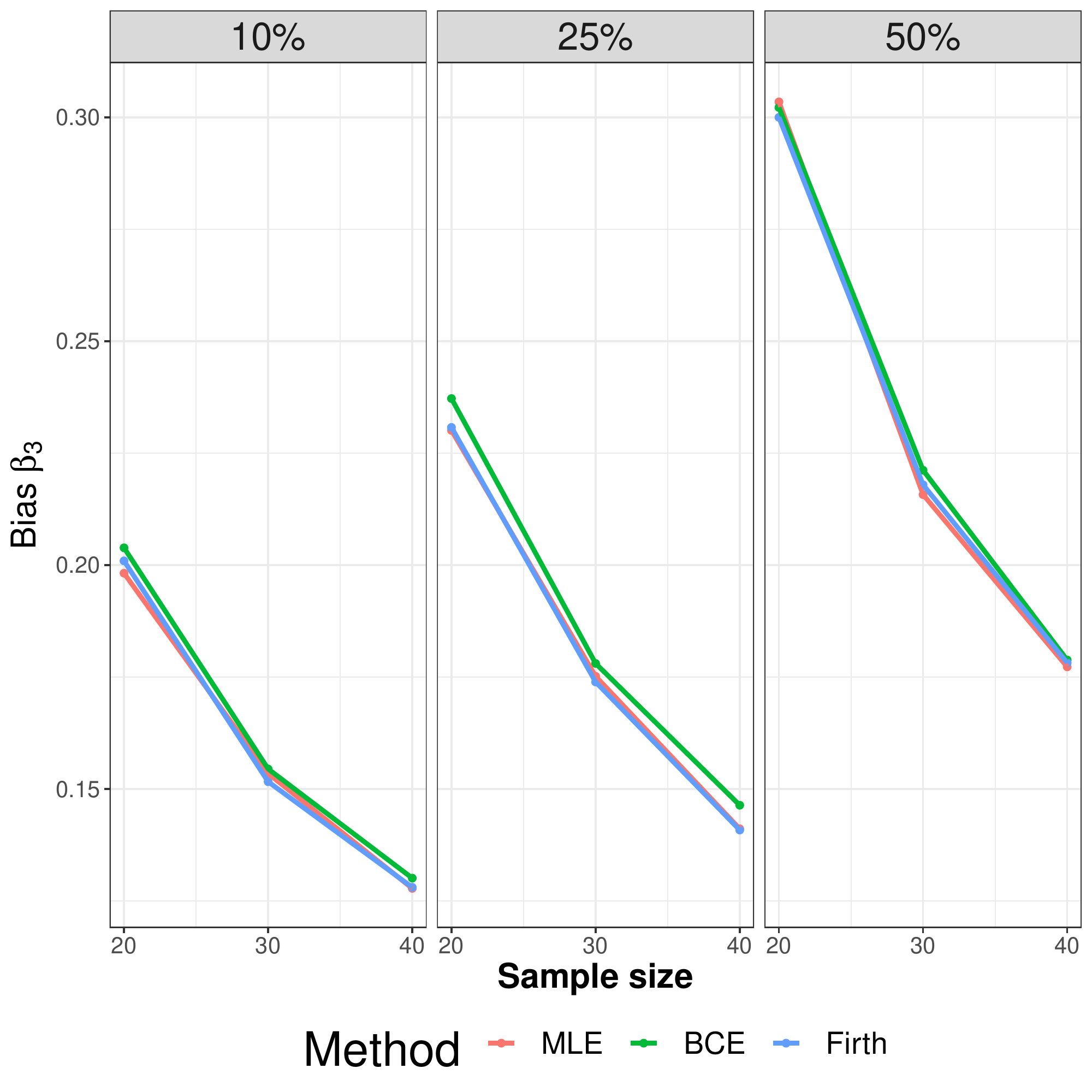}} \end{minipage} \\
  $p=7$ &
  \begin{minipage}{.28\textwidth}{\includegraphics[width=1\textwidth]{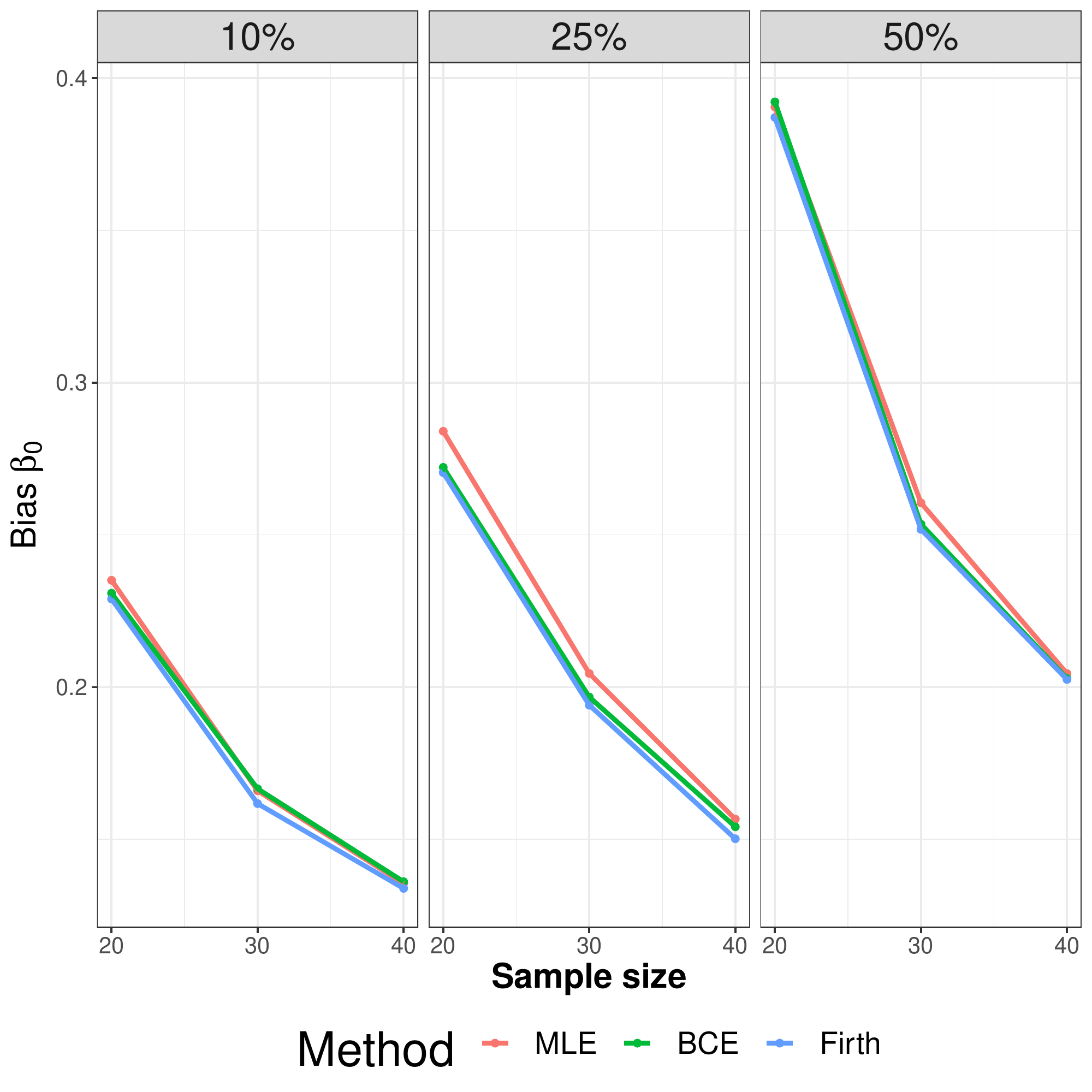}} \end{minipage} &
  \begin{minipage}{.28\textwidth}{\includegraphics[width=1\textwidth]{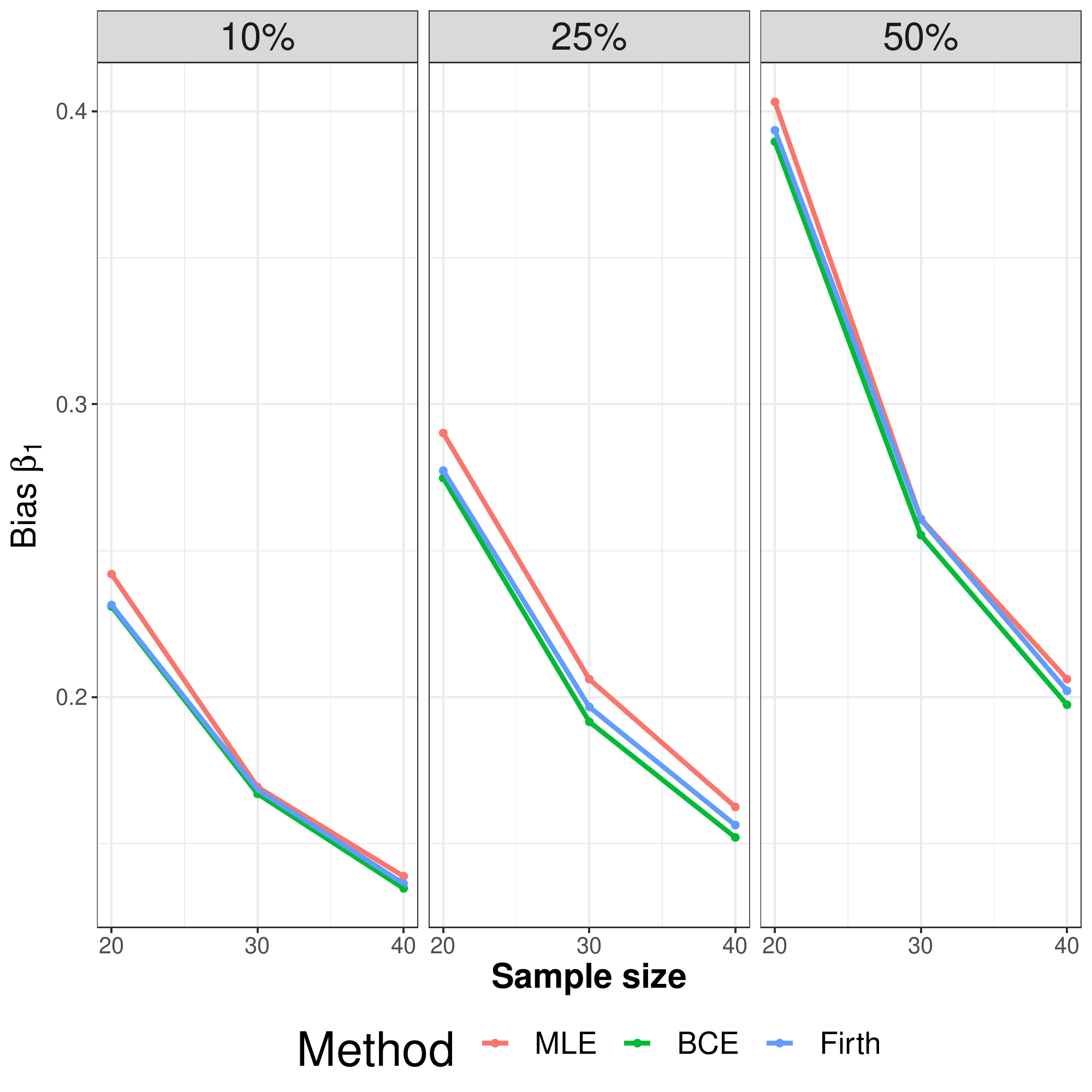}} \end{minipage} &
  \begin{minipage}{.28\textwidth}{\includegraphics[width=1\textwidth]{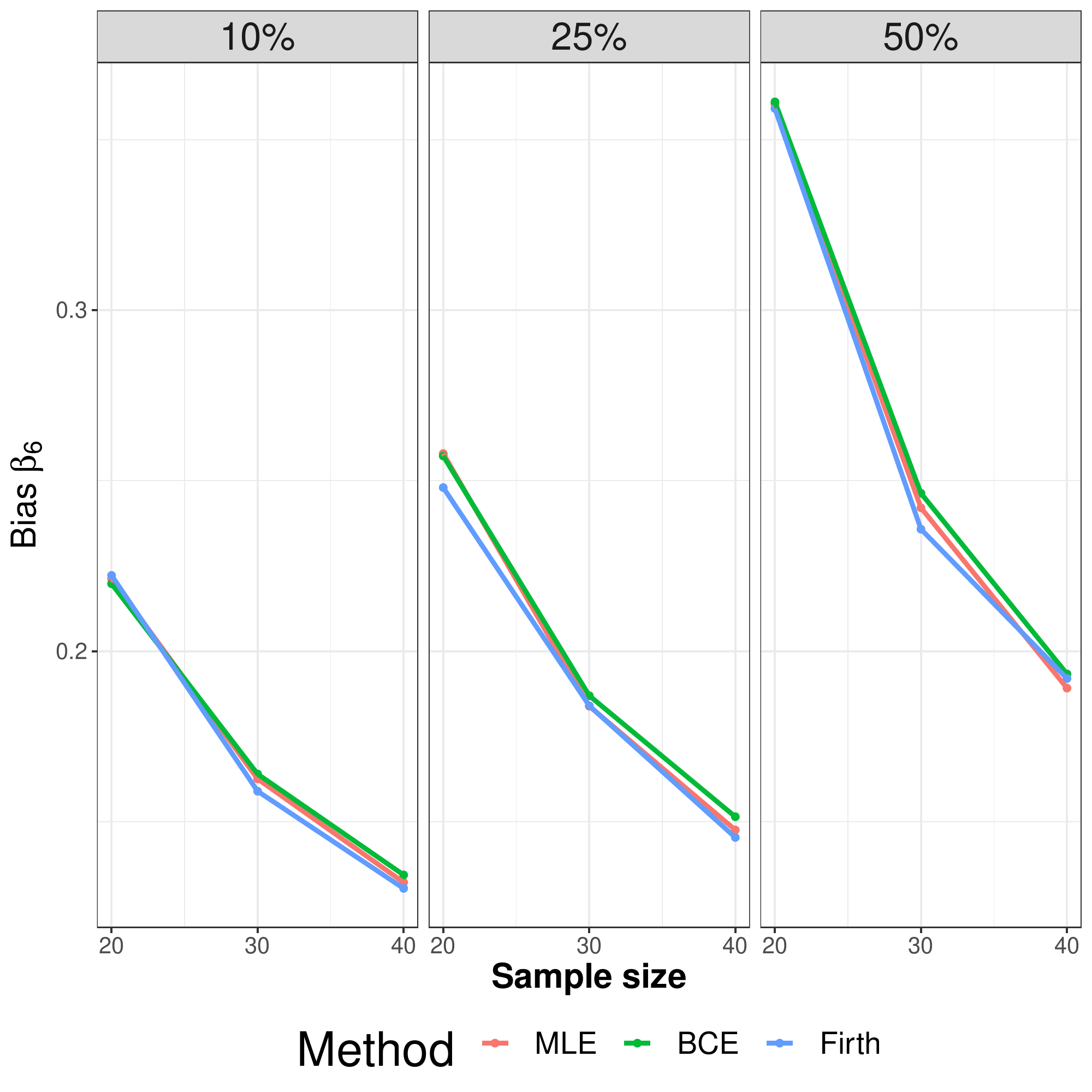}} \end{minipage} \\
\end{tabular}
}
\caption{Empirical bias for different estimators. Case $\sigma=1$.}
\label{fig:s3}
\end{center}
\end{figure}

\begin{figure}[!h]
\begin{center}
\resizebox{\linewidth}{!}{
\begin{tabular}{cccc}
  $p=3$ &
  \begin{minipage}{.28\textwidth}{\includegraphics[width=1\textwidth]{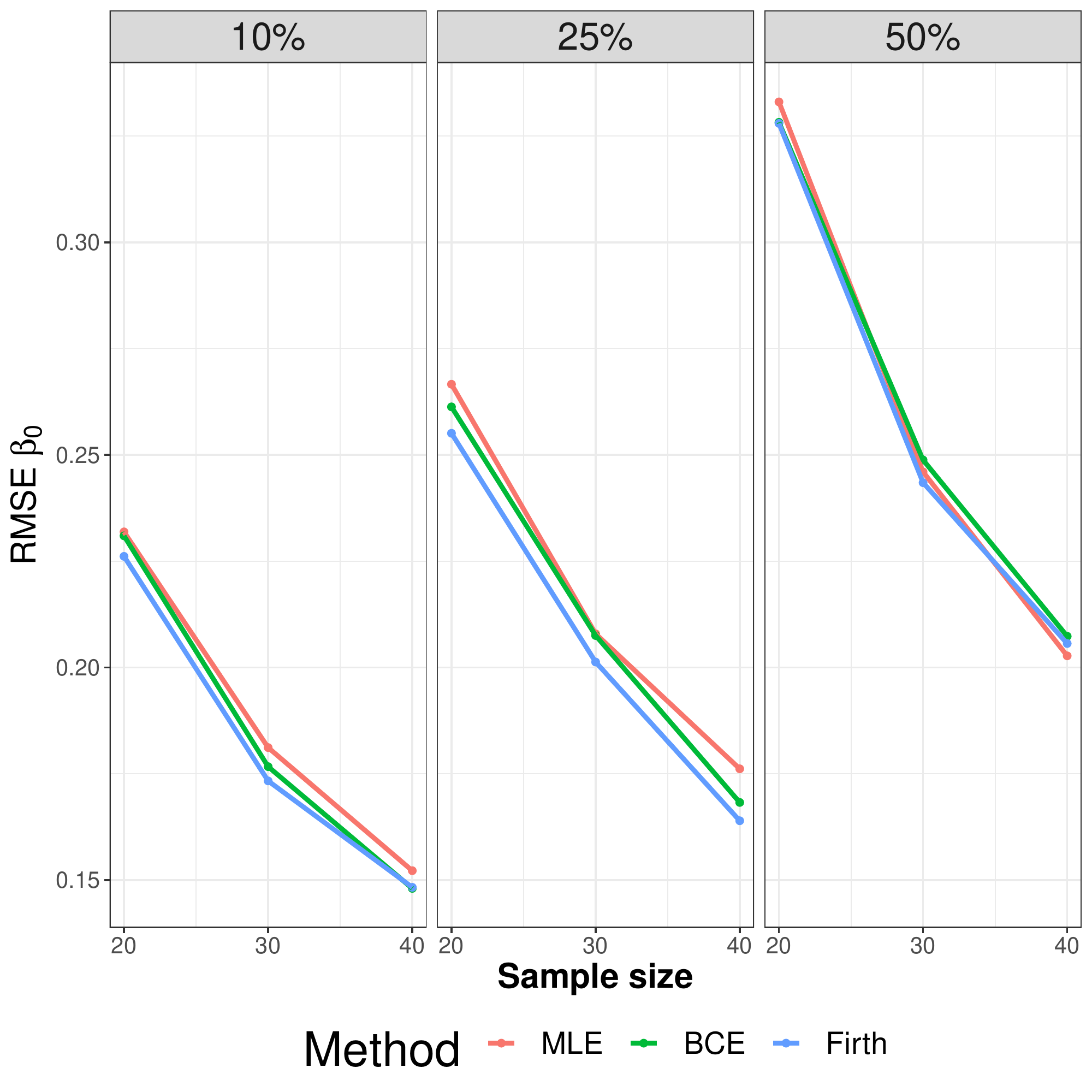}} \end{minipage} &
  \begin{minipage}{.28\textwidth}{\includegraphics[width=1\textwidth]{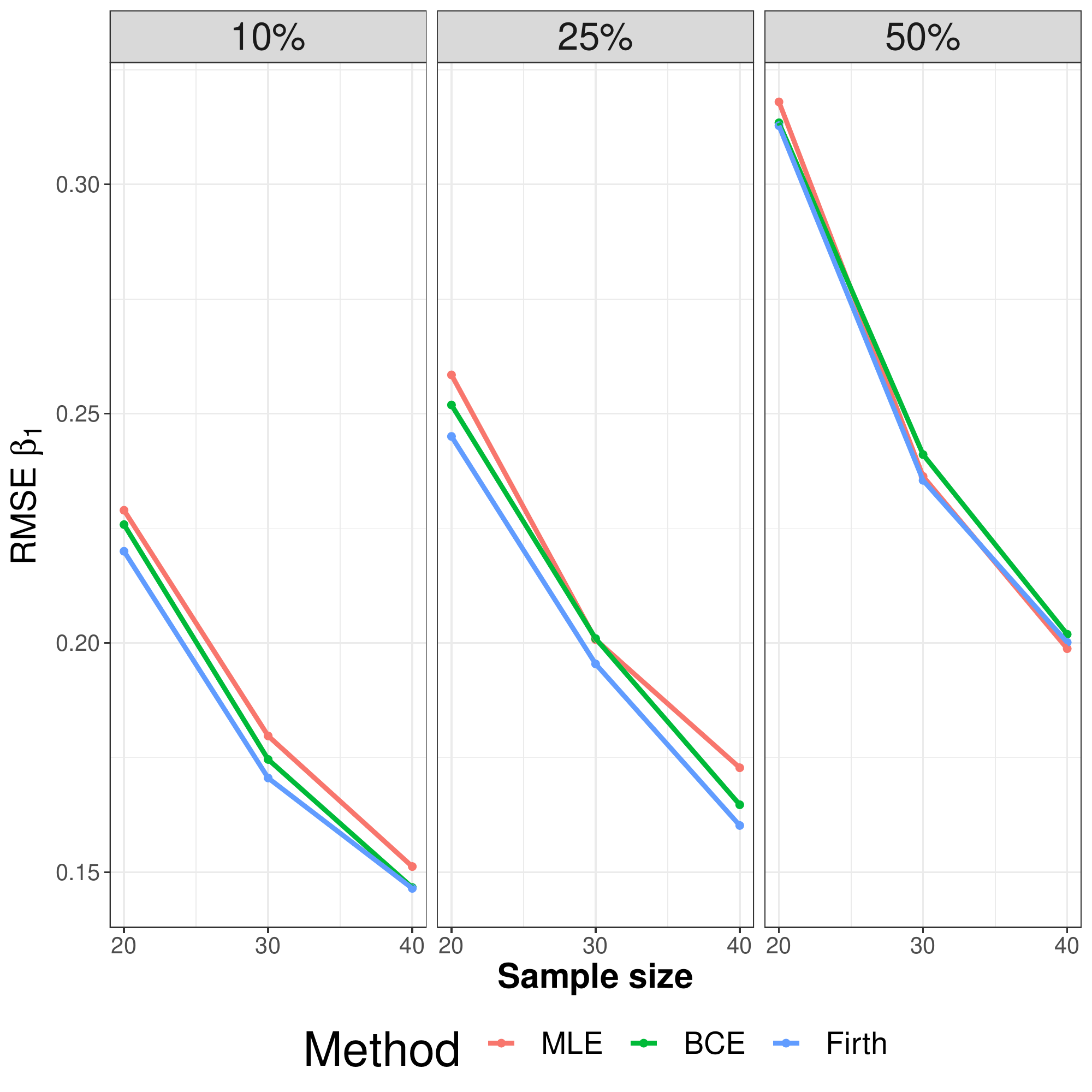}} \end{minipage} &
  \begin{minipage}{.28\textwidth}{\includegraphics[width=1\textwidth]{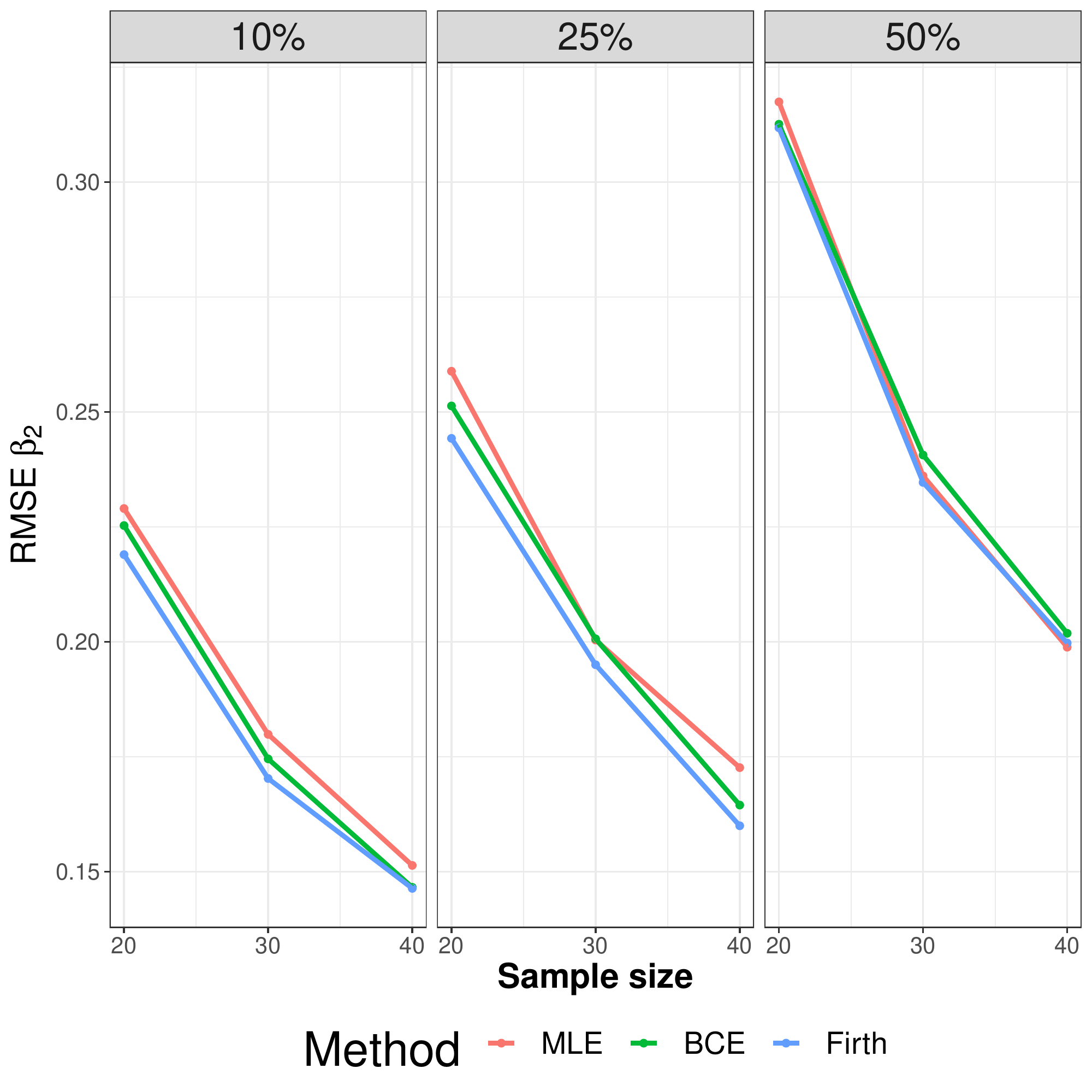}} \end{minipage} \\
  $p=5$ &
  \begin{minipage}{.28\textwidth}{\includegraphics[width=1\textwidth]{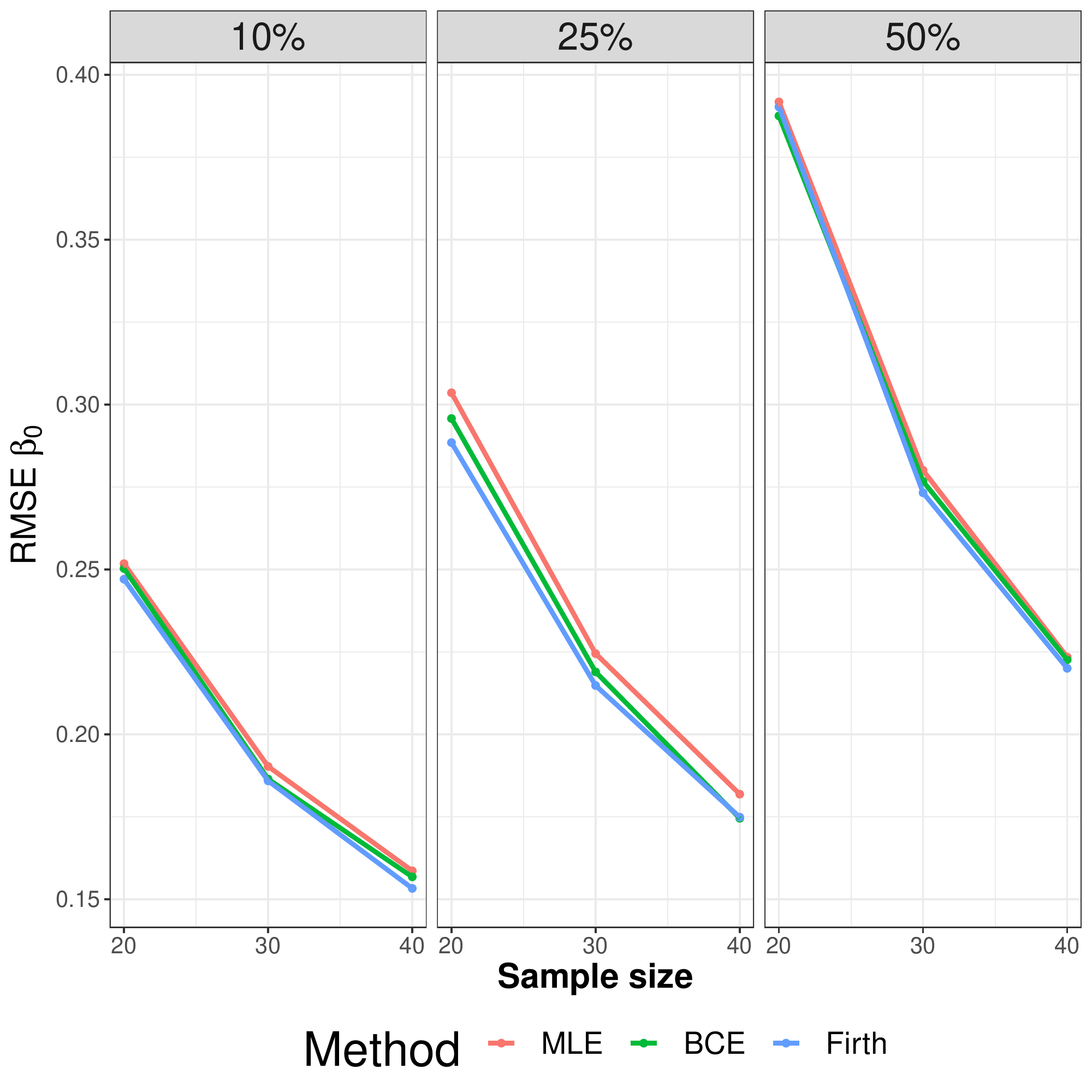}} \end{minipage} &
  \begin{minipage}{.28\textwidth}{\includegraphics[width=1\textwidth]{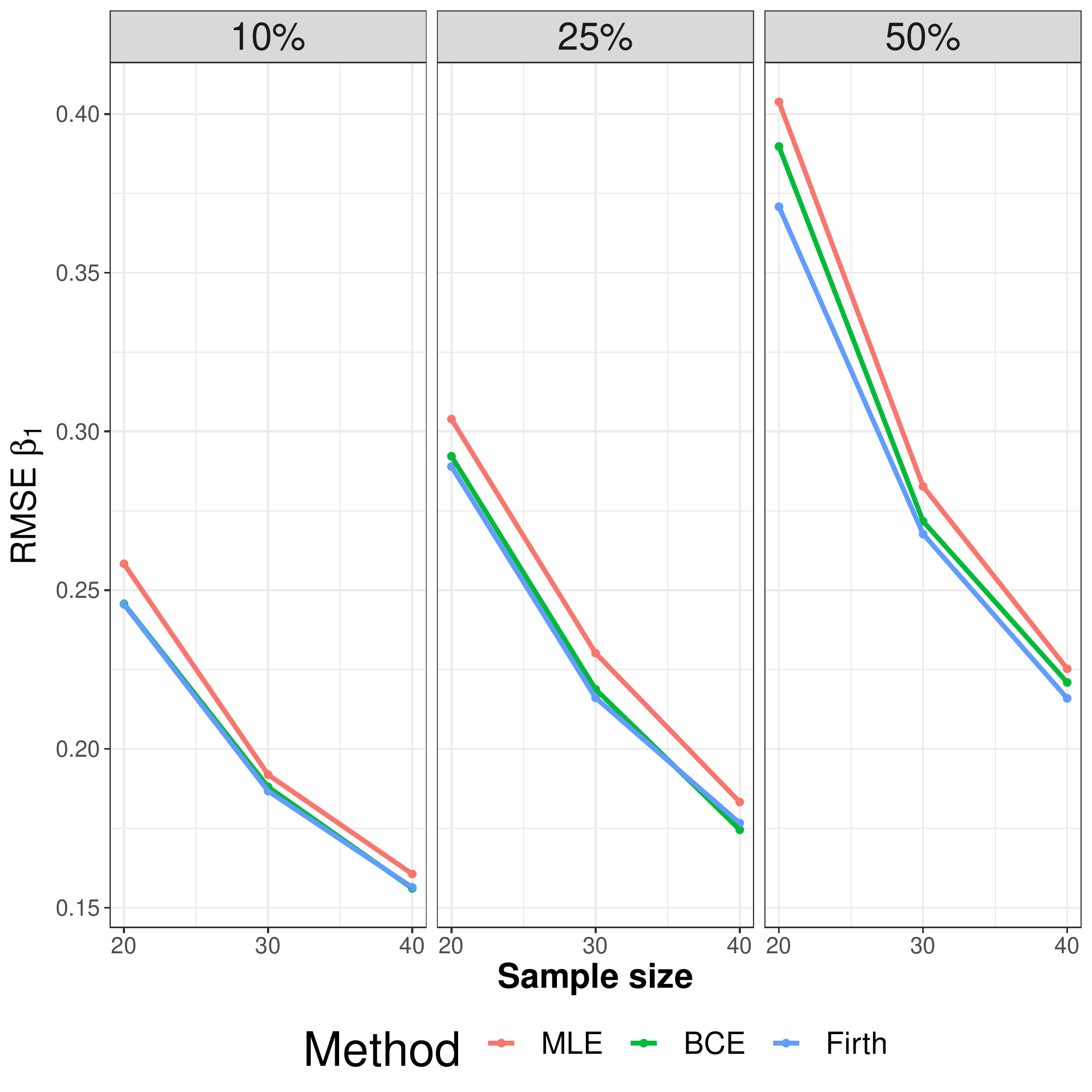}} \end{minipage} &
  \begin{minipage}{.28\textwidth}{\includegraphics[width=1\textwidth]{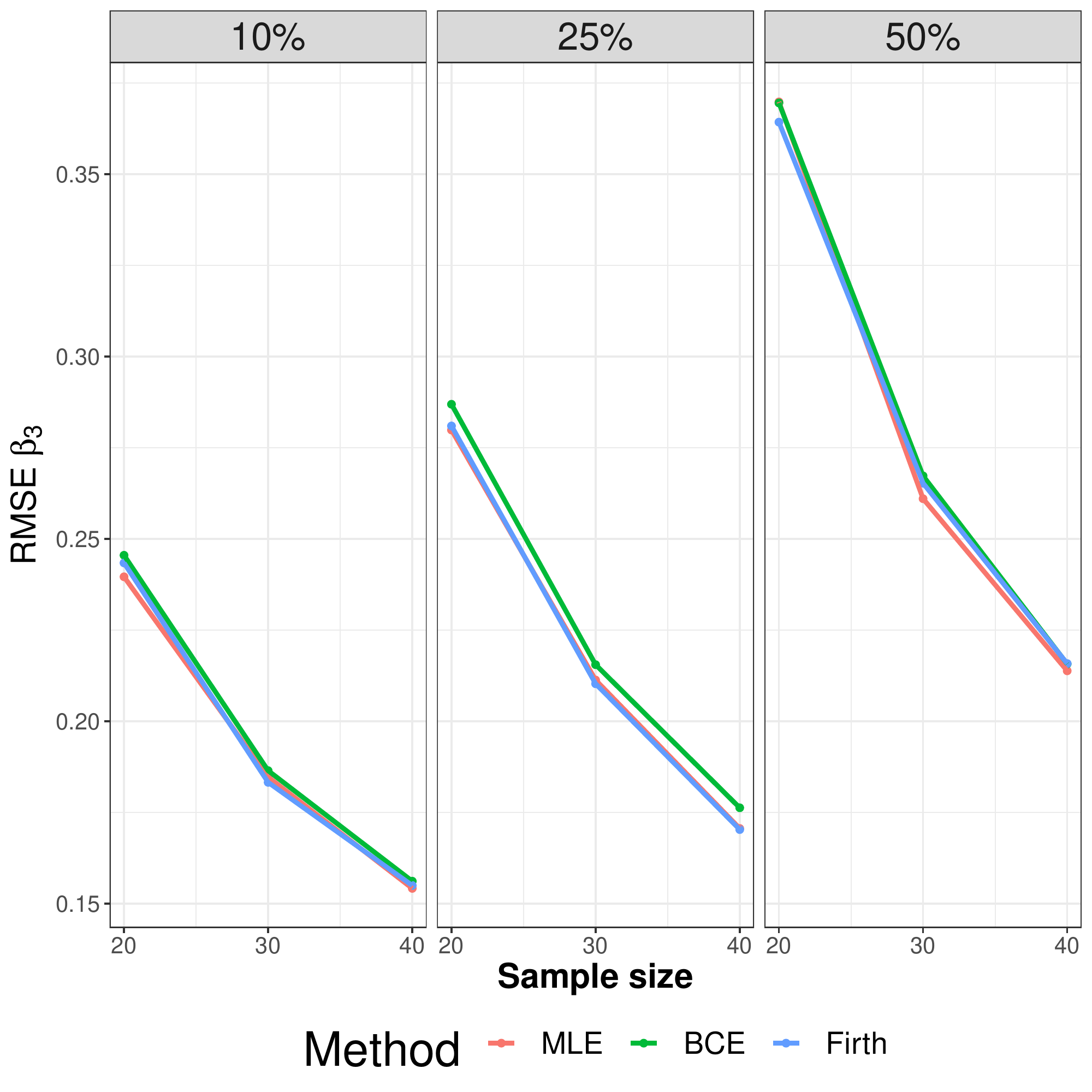}} \end{minipage} \\
  $p=7$ &
  \begin{minipage}{.28\textwidth}{\includegraphics[width=1\textwidth]{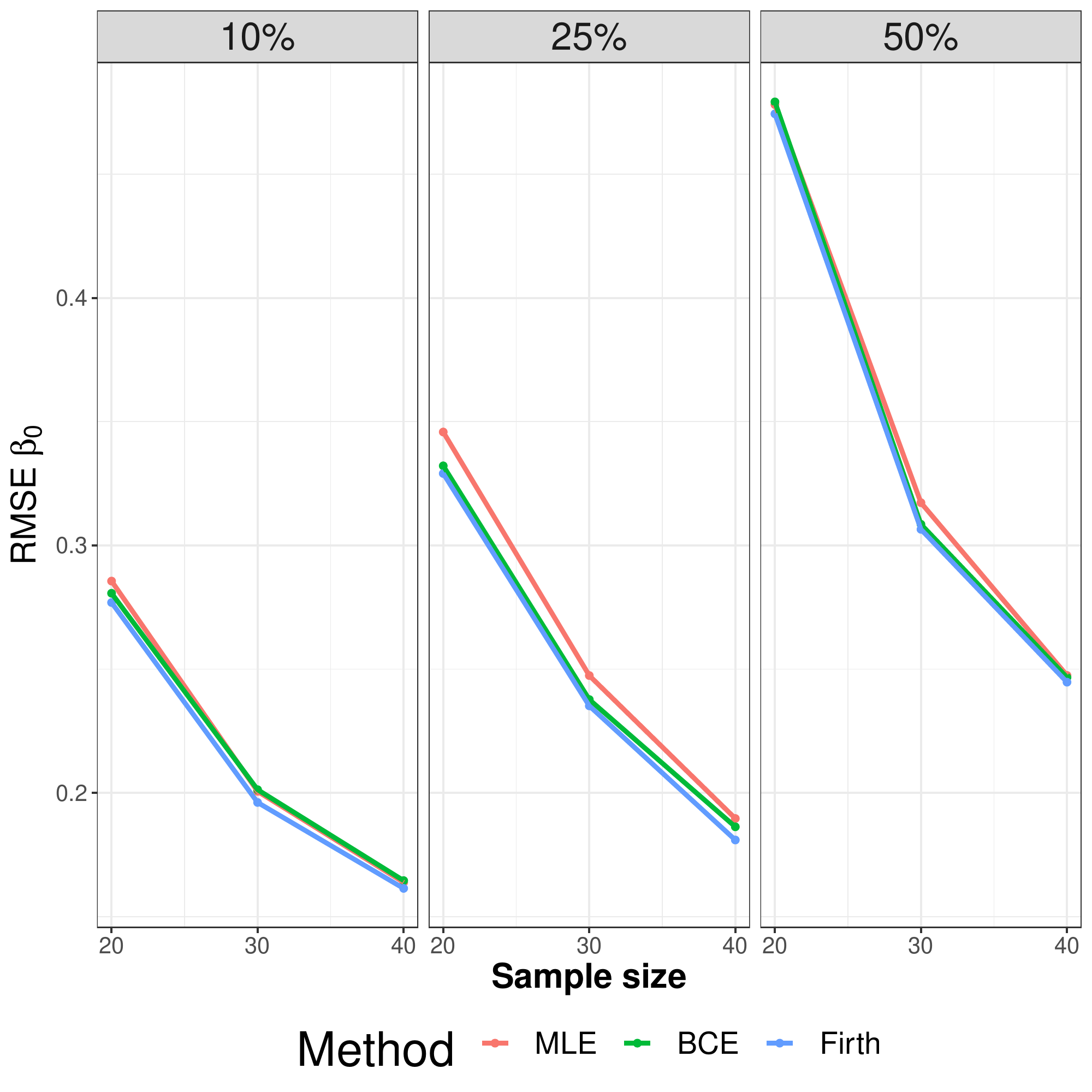}} \end{minipage} &
  \begin{minipage}{.28\textwidth}{\includegraphics[width=1\textwidth]{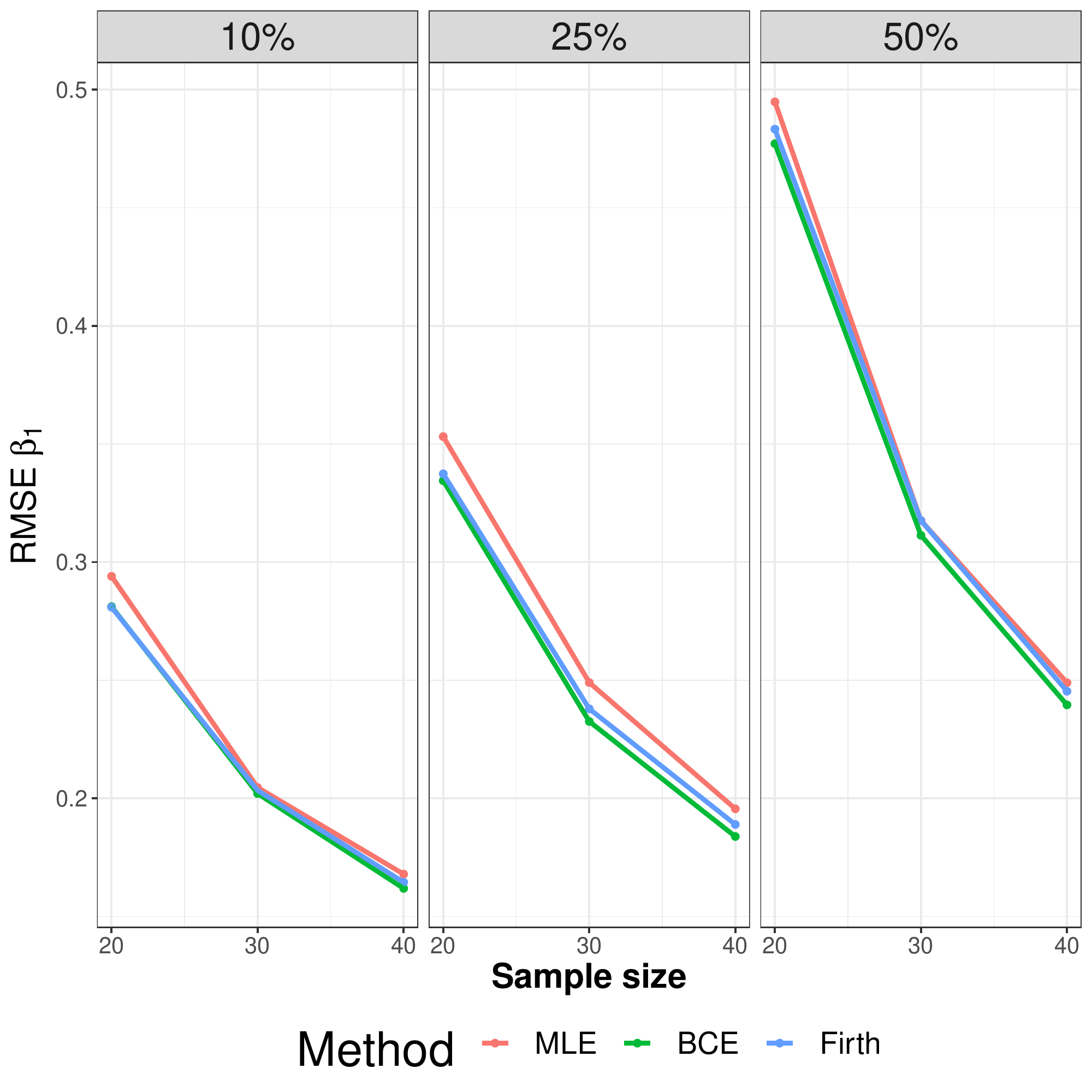}} \end{minipage} &
  \begin{minipage}{.28\textwidth}{\includegraphics[width=1\textwidth]{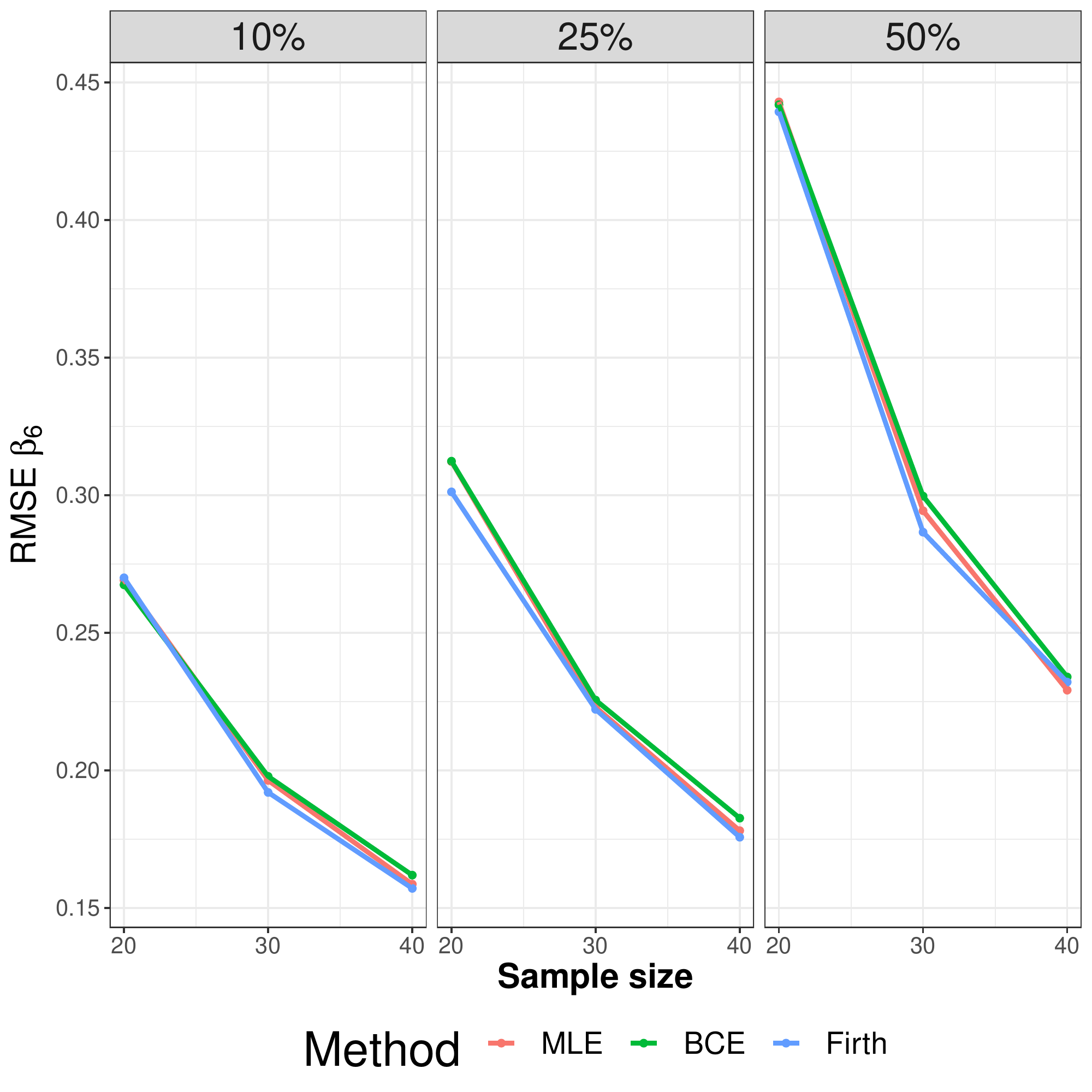}} \end{minipage} \\
\end{tabular}
}
\caption{Empirical RMSE for different estimators. Case $\sigma=1$.}
\label{fig:s4}
\end{center}
\end{figure}

\begin{figure}[!h]
\begin{center}
\resizebox{\linewidth}{!}{
\begin{tabular}{cccc}
  $p=3$ &
  \begin{minipage}{.28\textwidth}{\includegraphics[width=1\textwidth]{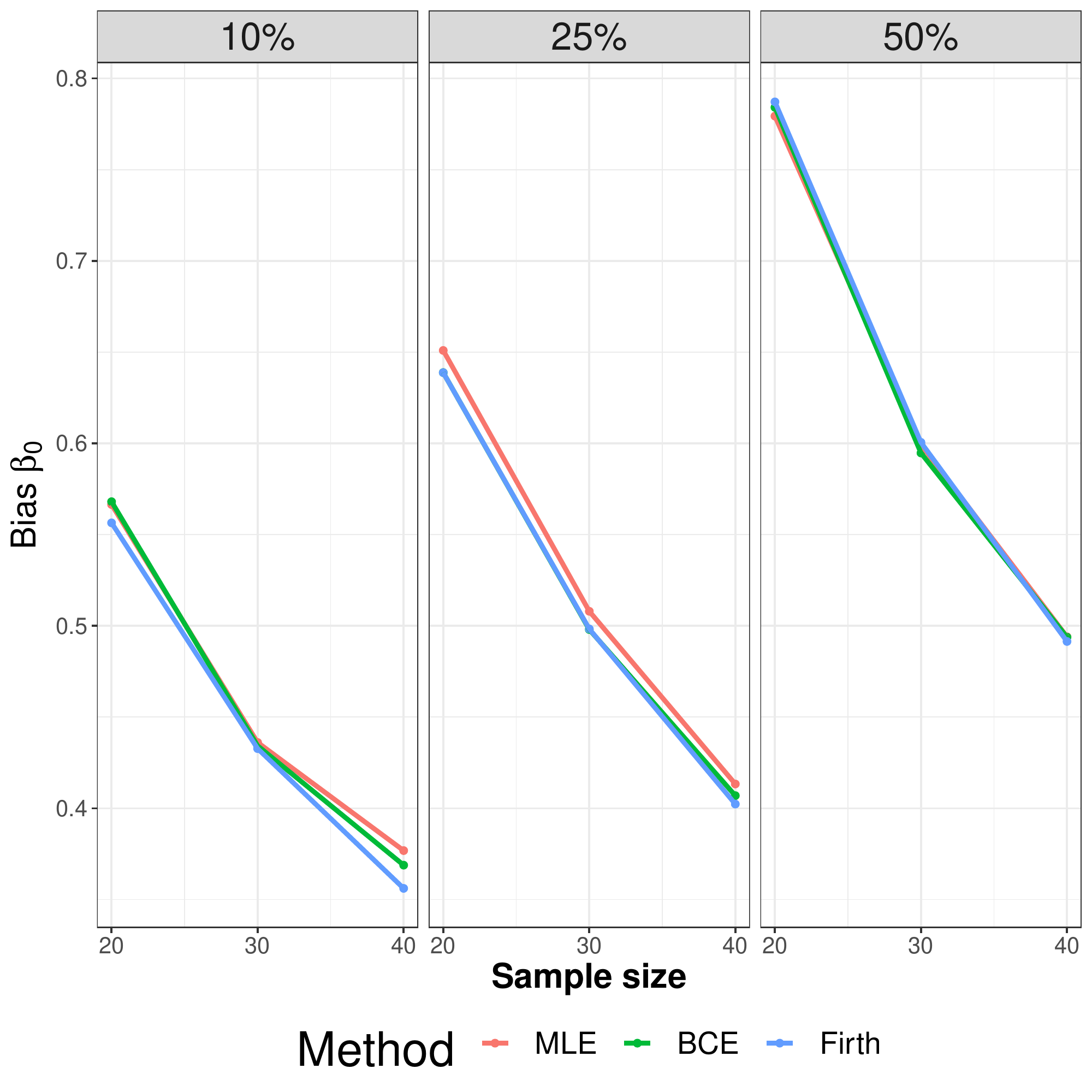}} \end{minipage} &
  \begin{minipage}{.28\textwidth}{\includegraphics[width=1\textwidth]{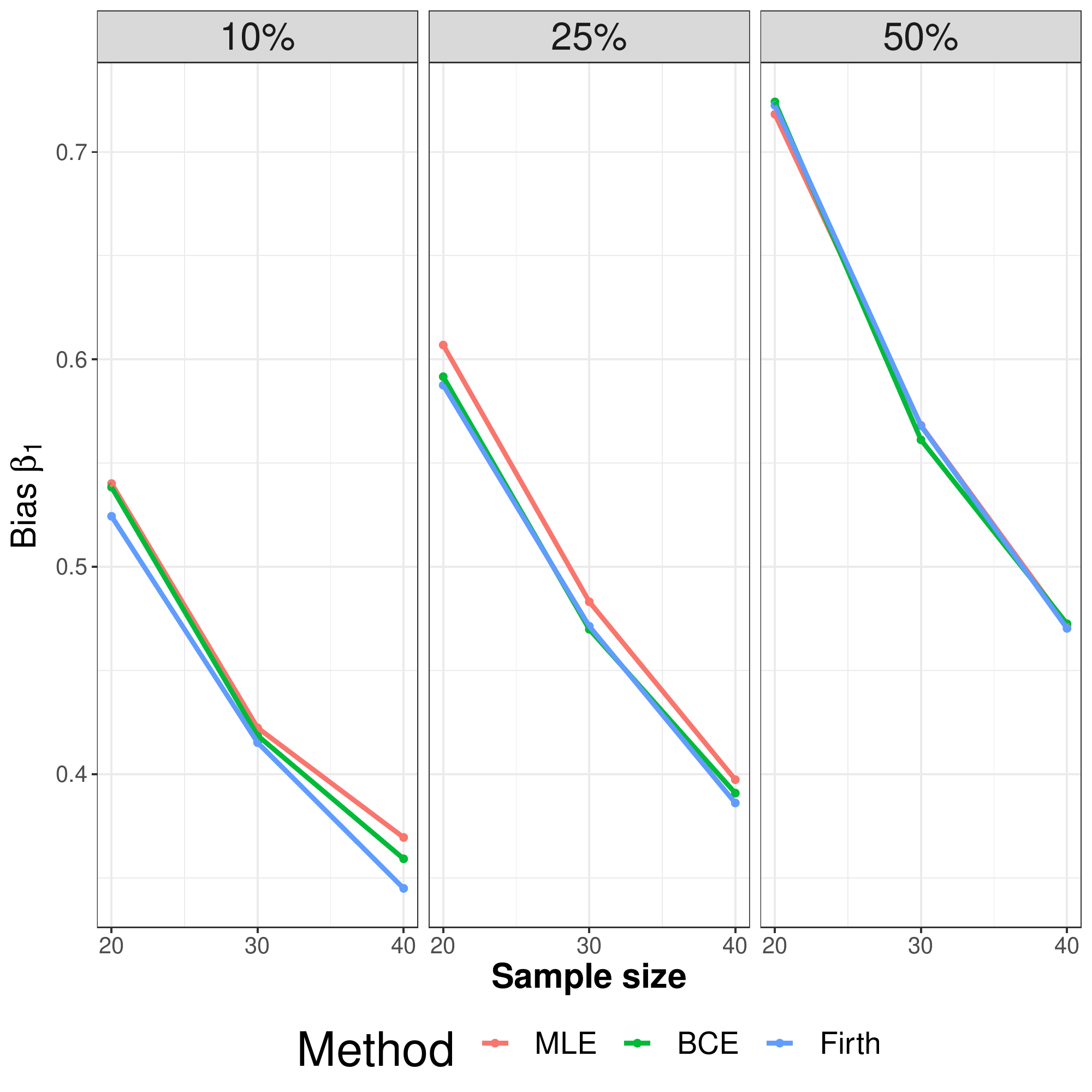}} \end{minipage} &
  \begin{minipage}{.28\textwidth}{\includegraphics[width=1\textwidth]{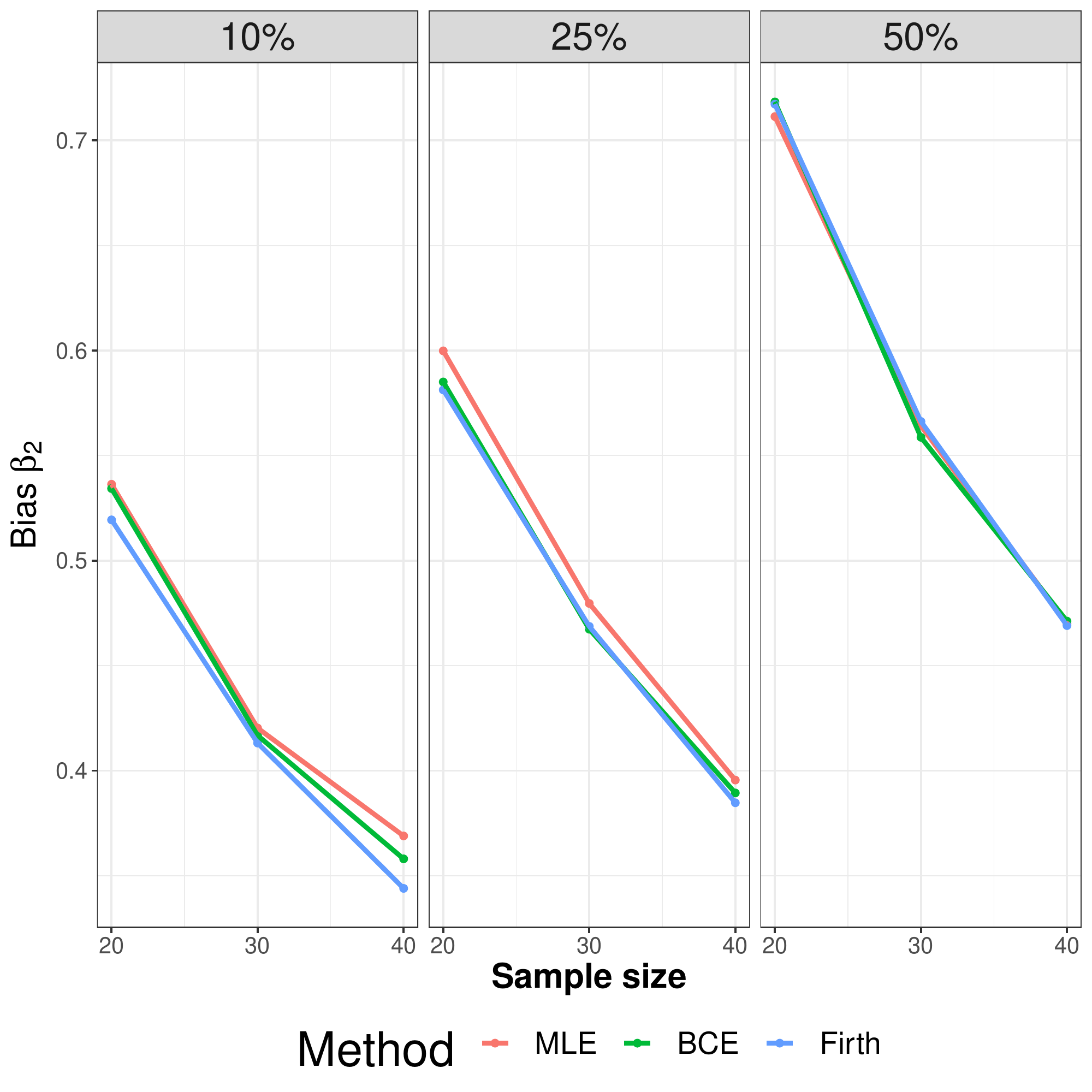}} \end{minipage} \\
  $p=5$ &
  \begin{minipage}{.28\textwidth}{\includegraphics[width=1\textwidth]{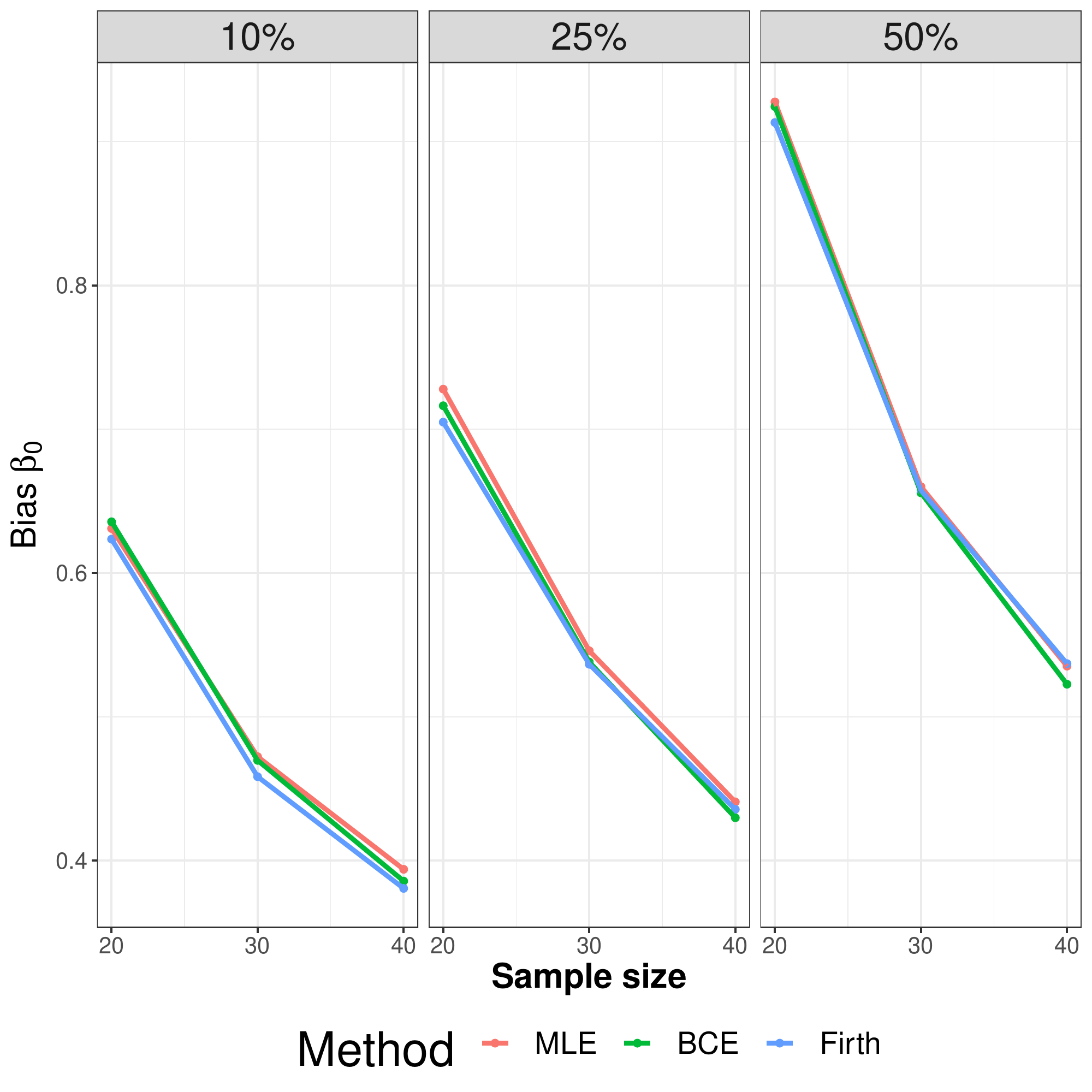}} \end{minipage} &
  \begin{minipage}{.28\textwidth}{\includegraphics[width=1\textwidth]{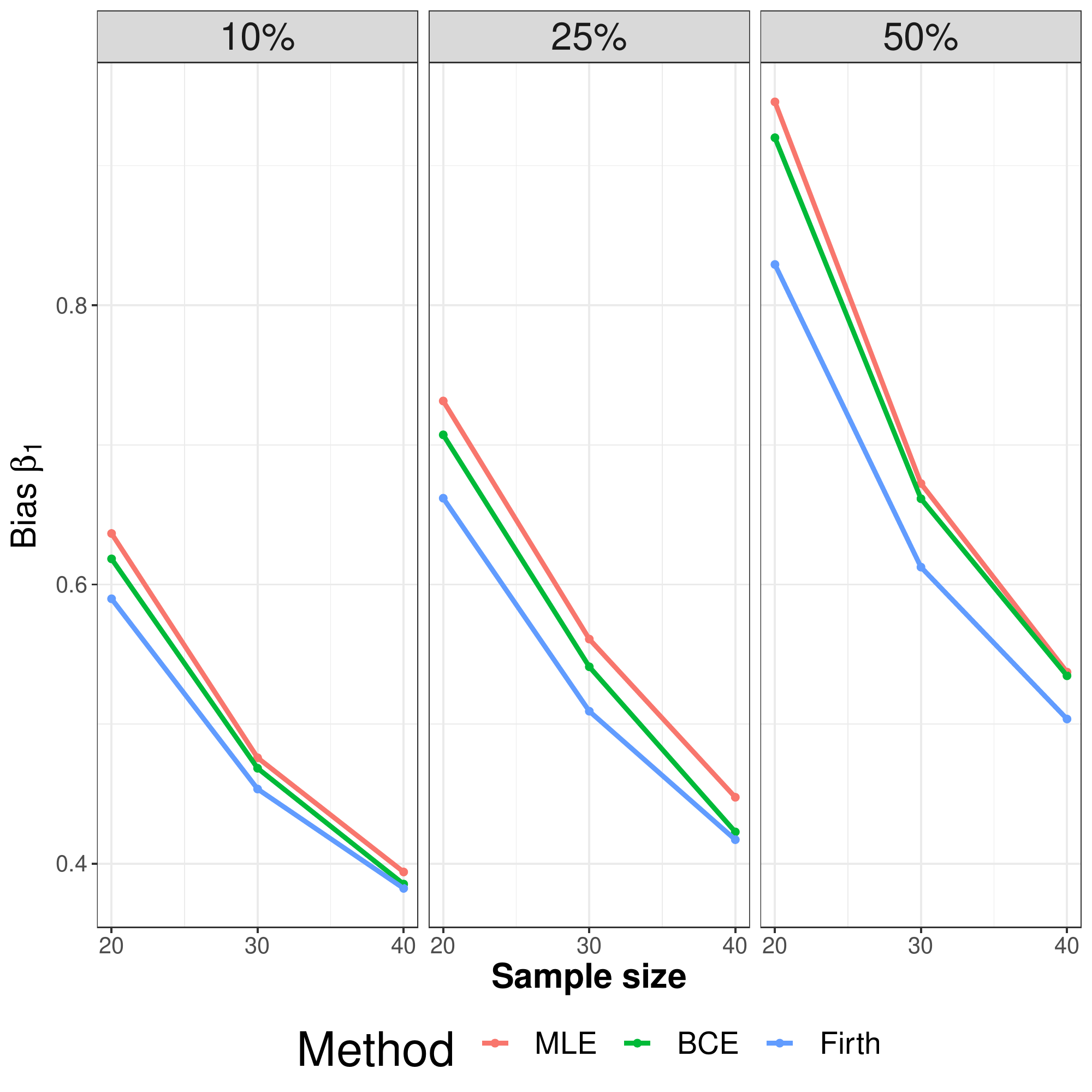}} \end{minipage} &
  \begin{minipage}{.28\textwidth}{\includegraphics[width=1\textwidth]{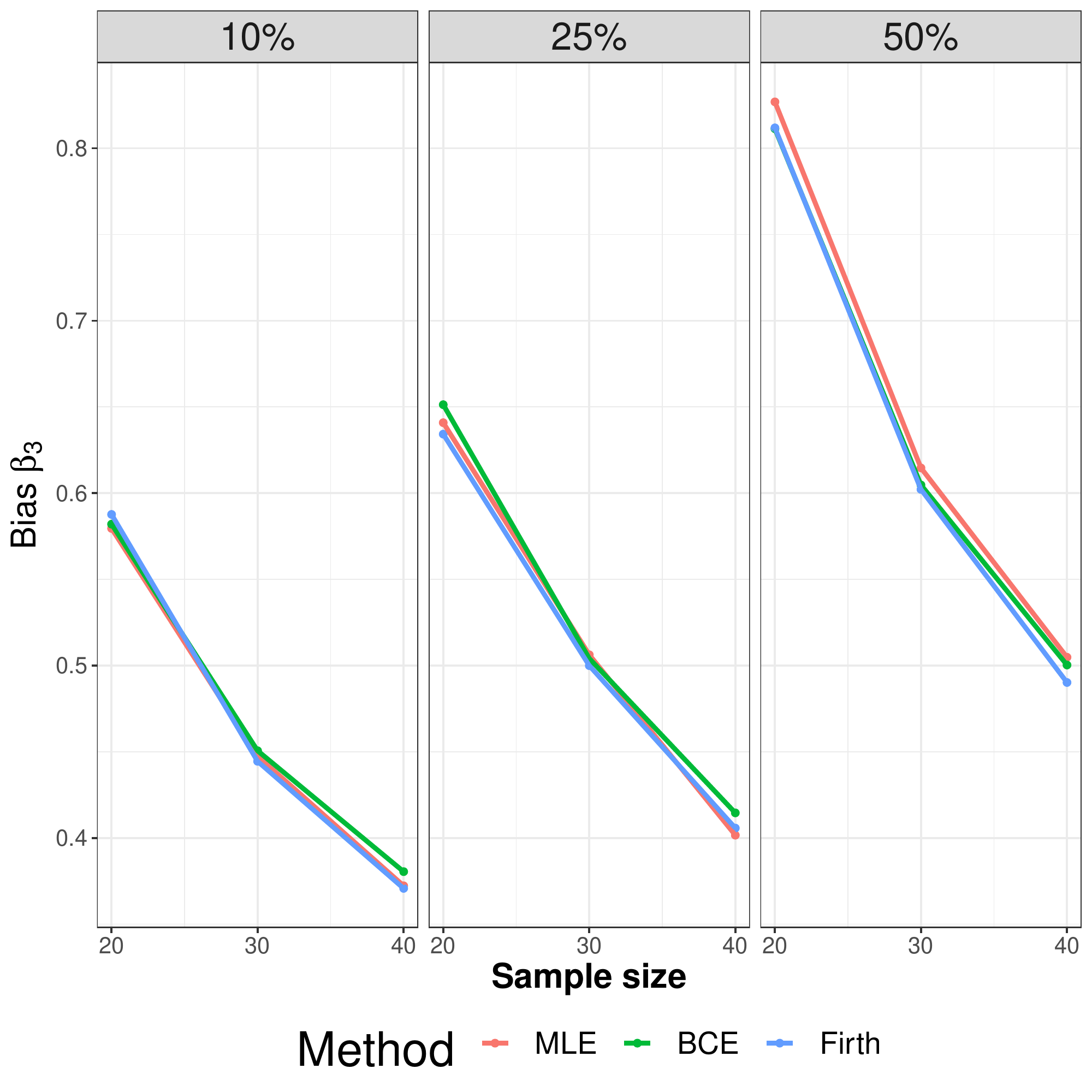}} \end{minipage} \\
  $p=7$ &
  \begin{minipage}{.28\textwidth}{\includegraphics[width=1\textwidth]{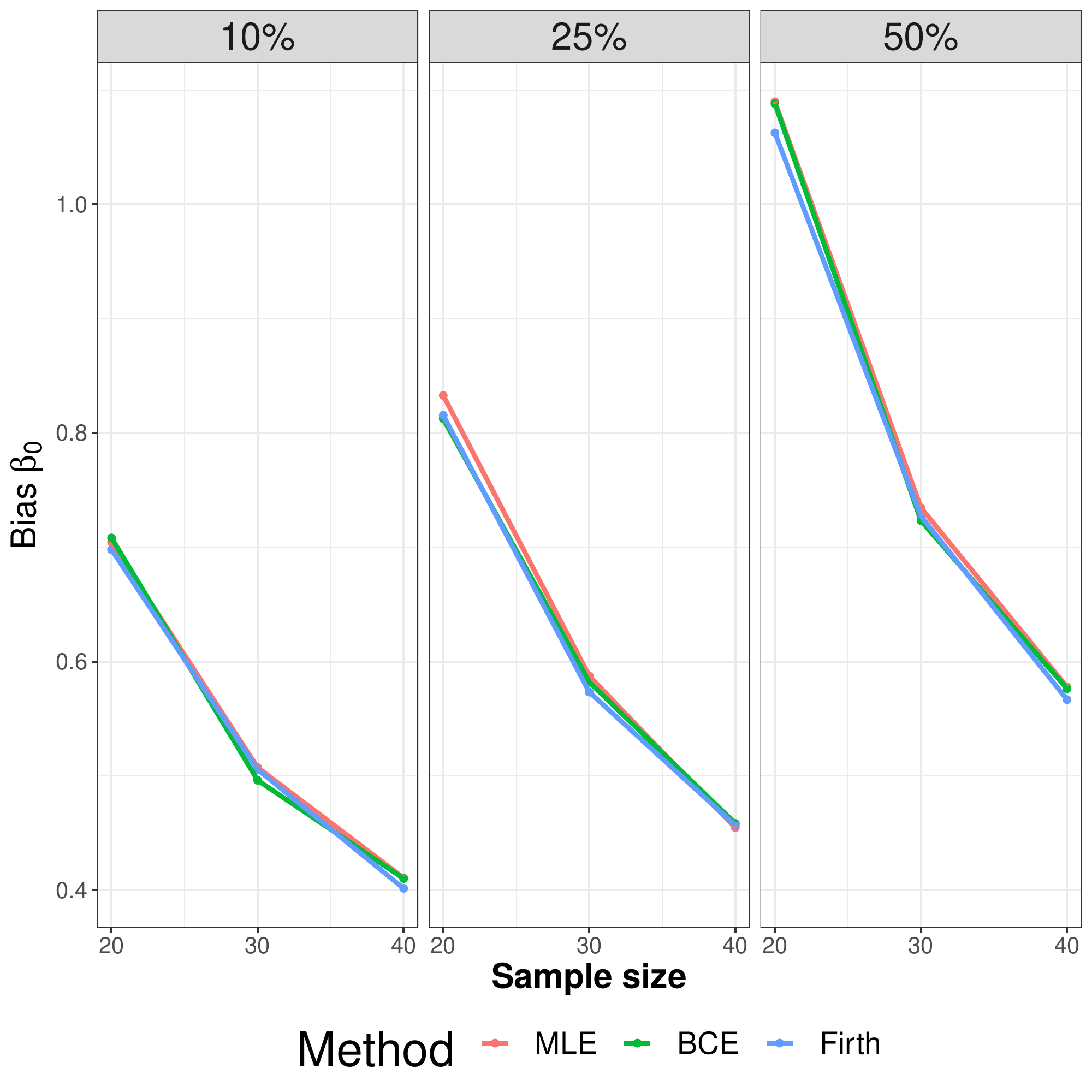}} \end{minipage} &
  \begin{minipage}{.28\textwidth}{\includegraphics[width=1\textwidth]{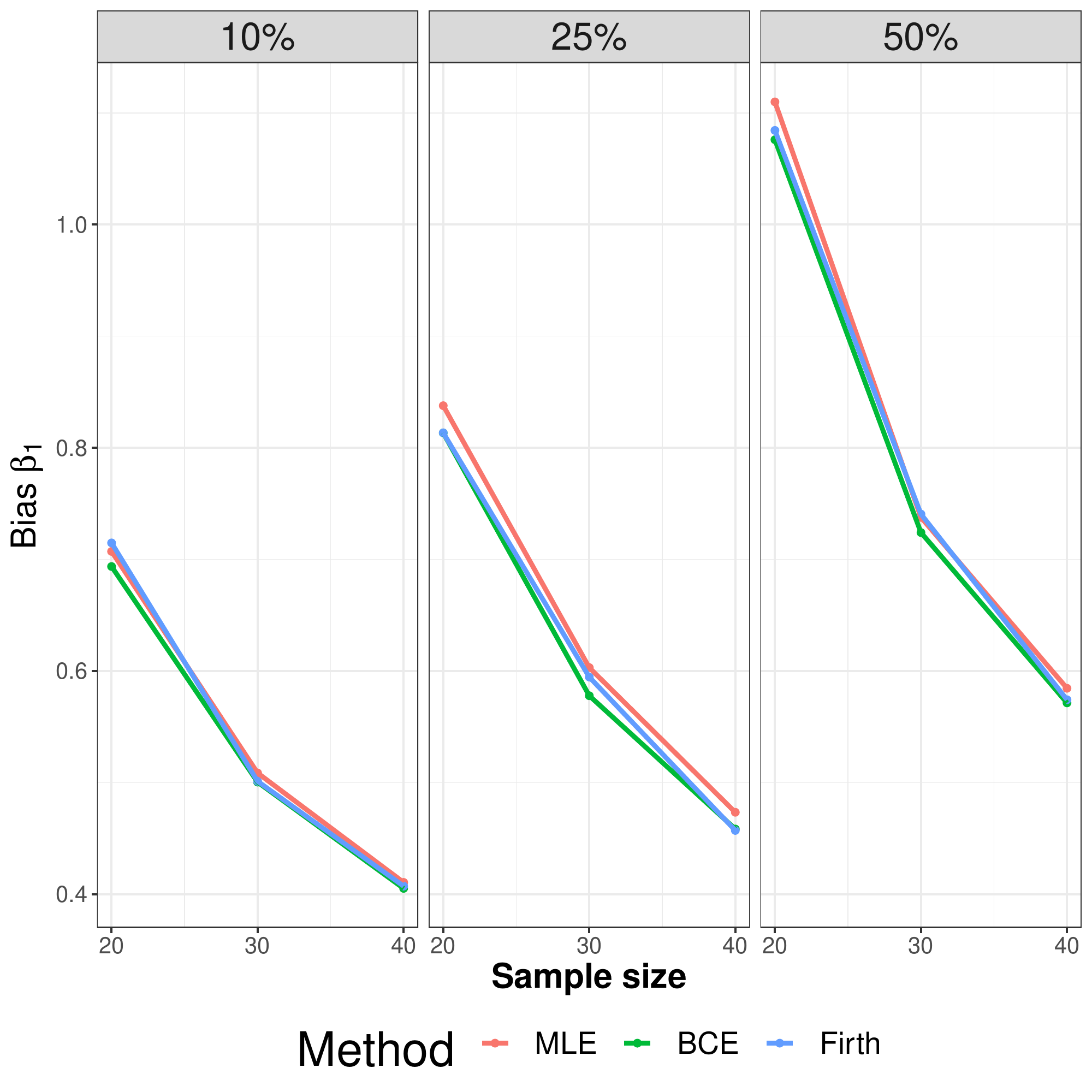}} \end{minipage} &
  \begin{minipage}{.28\textwidth}{\includegraphics[width=1\textwidth]{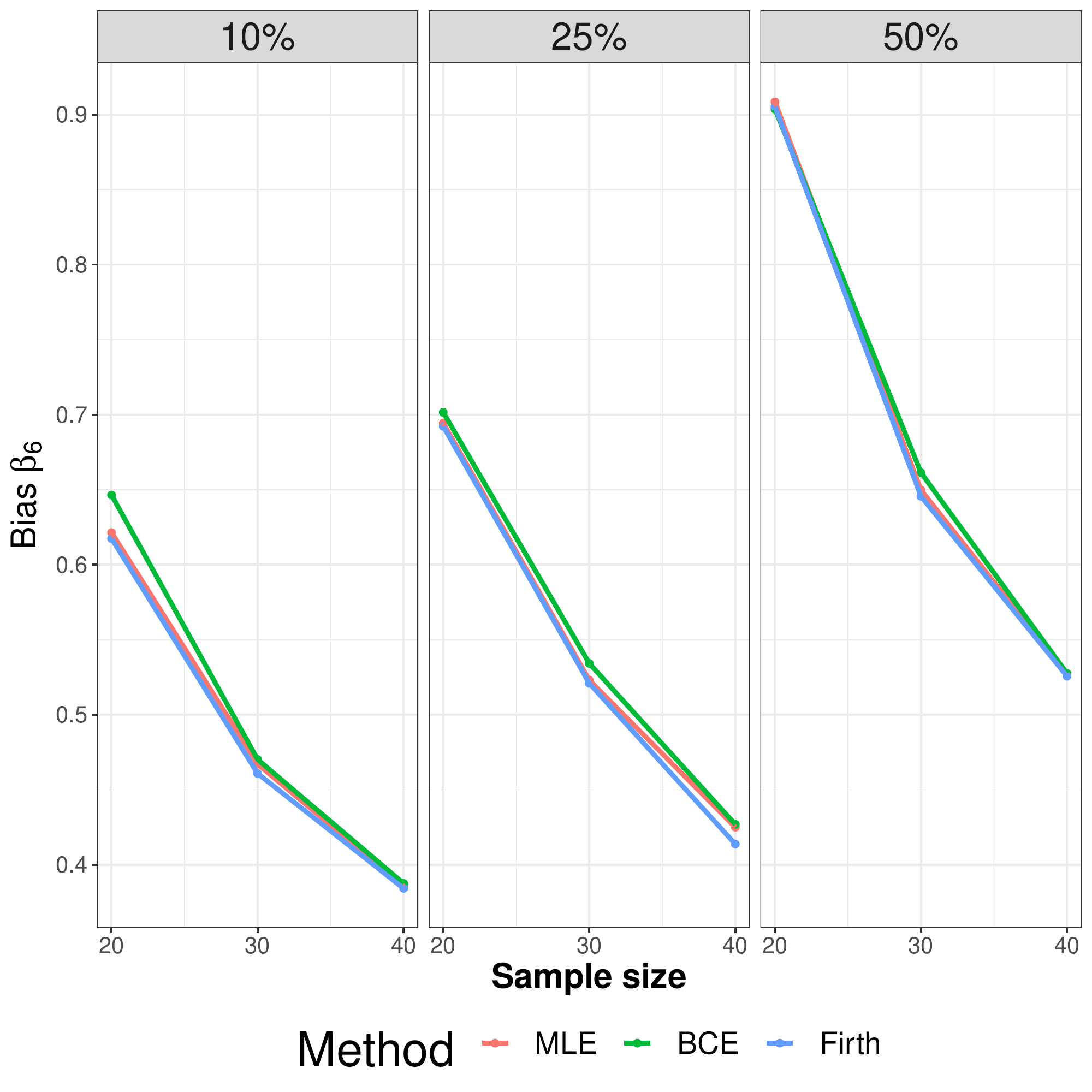}} \end{minipage} \\
\end{tabular}
}
\caption{Empirical bias for different estimators. Case $\sigma=3$.}
\label{fig:s5}
\end{center}
\end{figure}

\begin{figure}[!h]
\begin{center}
\resizebox{\linewidth}{!}{
\begin{tabular}{cccc}
  $p=3$ &
  \begin{minipage}{.28\textwidth}{\includegraphics[width=1\textwidth]{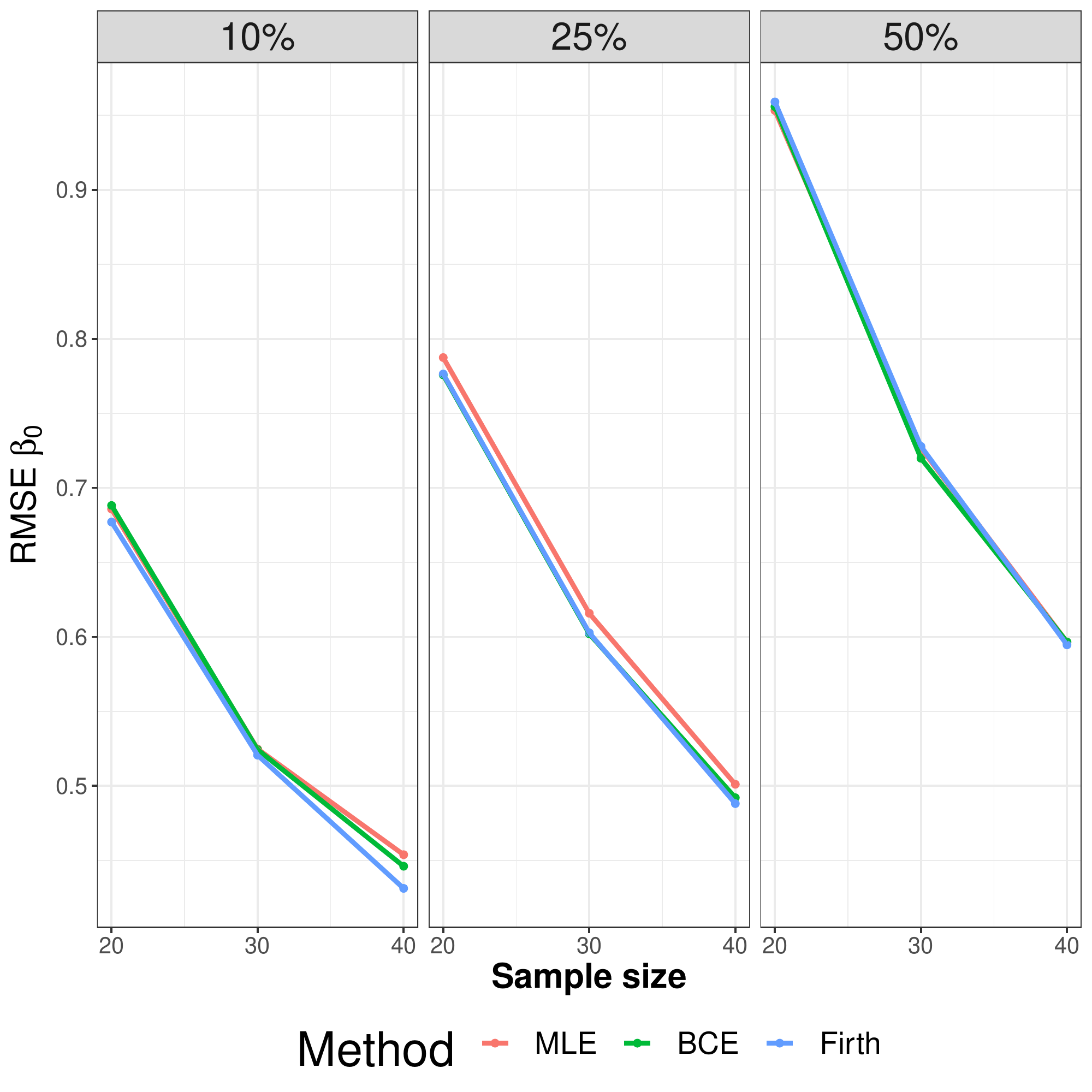}} \end{minipage} &
  \begin{minipage}{.28\textwidth}{\includegraphics[width=1\textwidth]{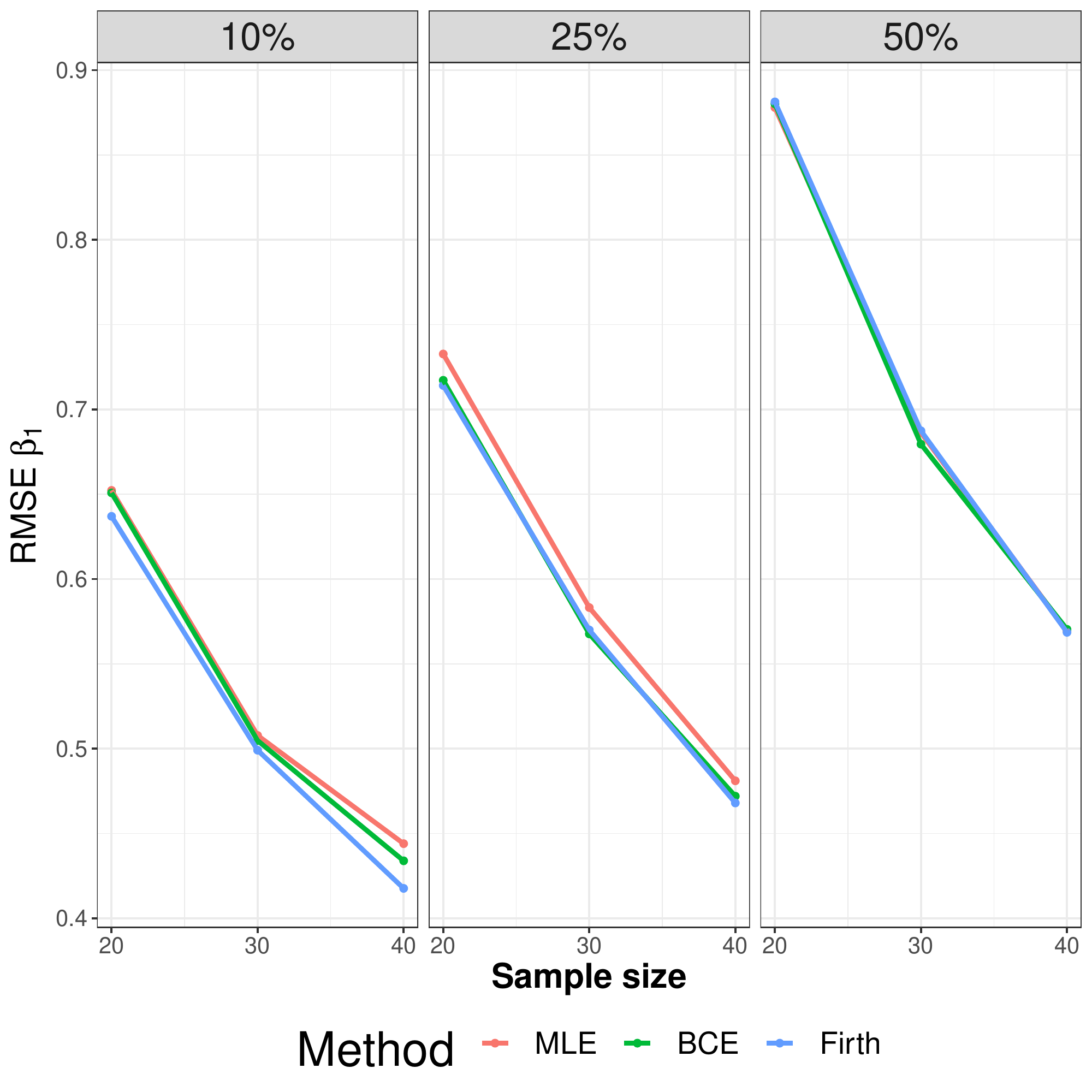}} \end{minipage} &
  \begin{minipage}{.28\textwidth}{\includegraphics[width=1\textwidth]{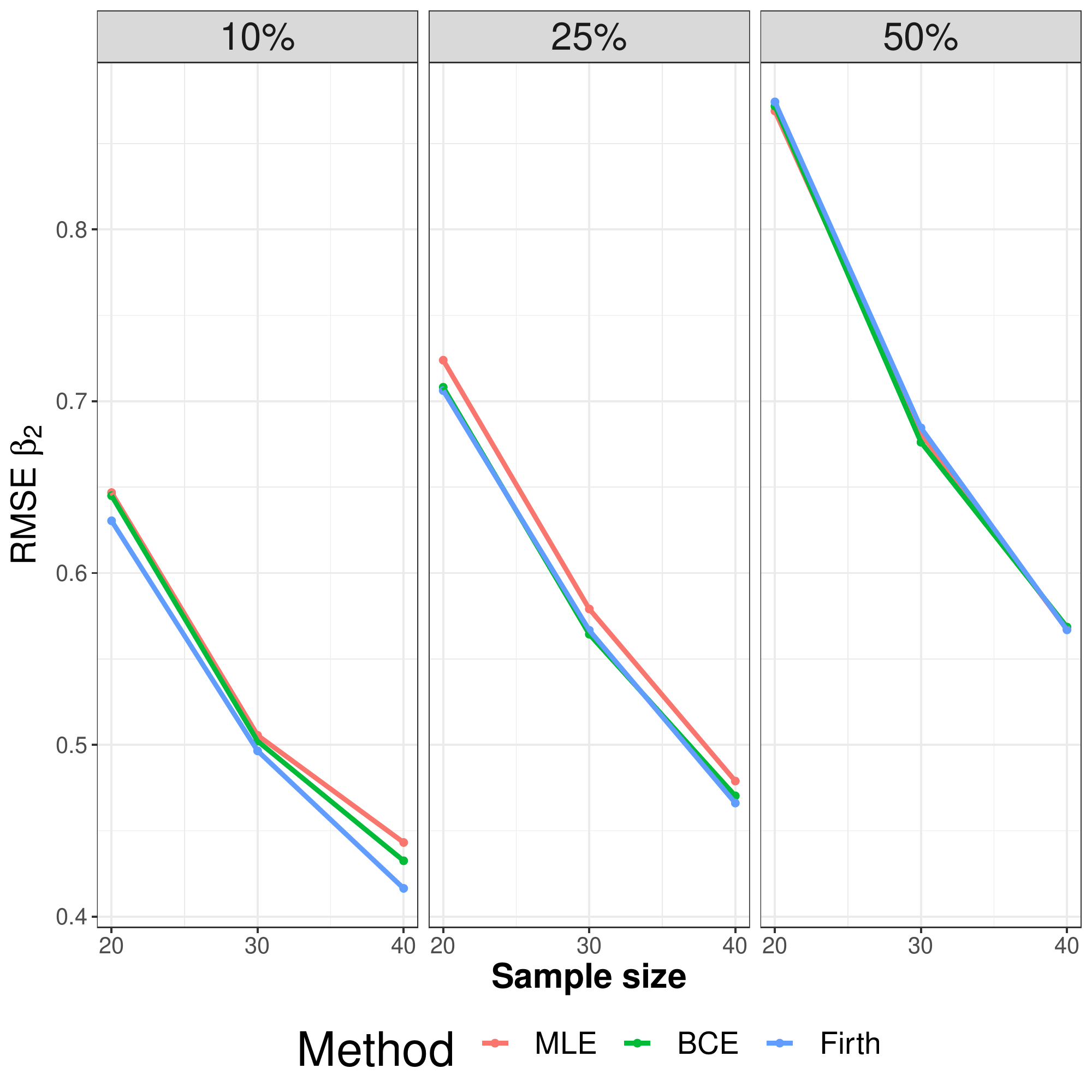}} \end{minipage} \\
  $p=5$ &
  \begin{minipage}{.28\textwidth}{\includegraphics[width=1\textwidth]{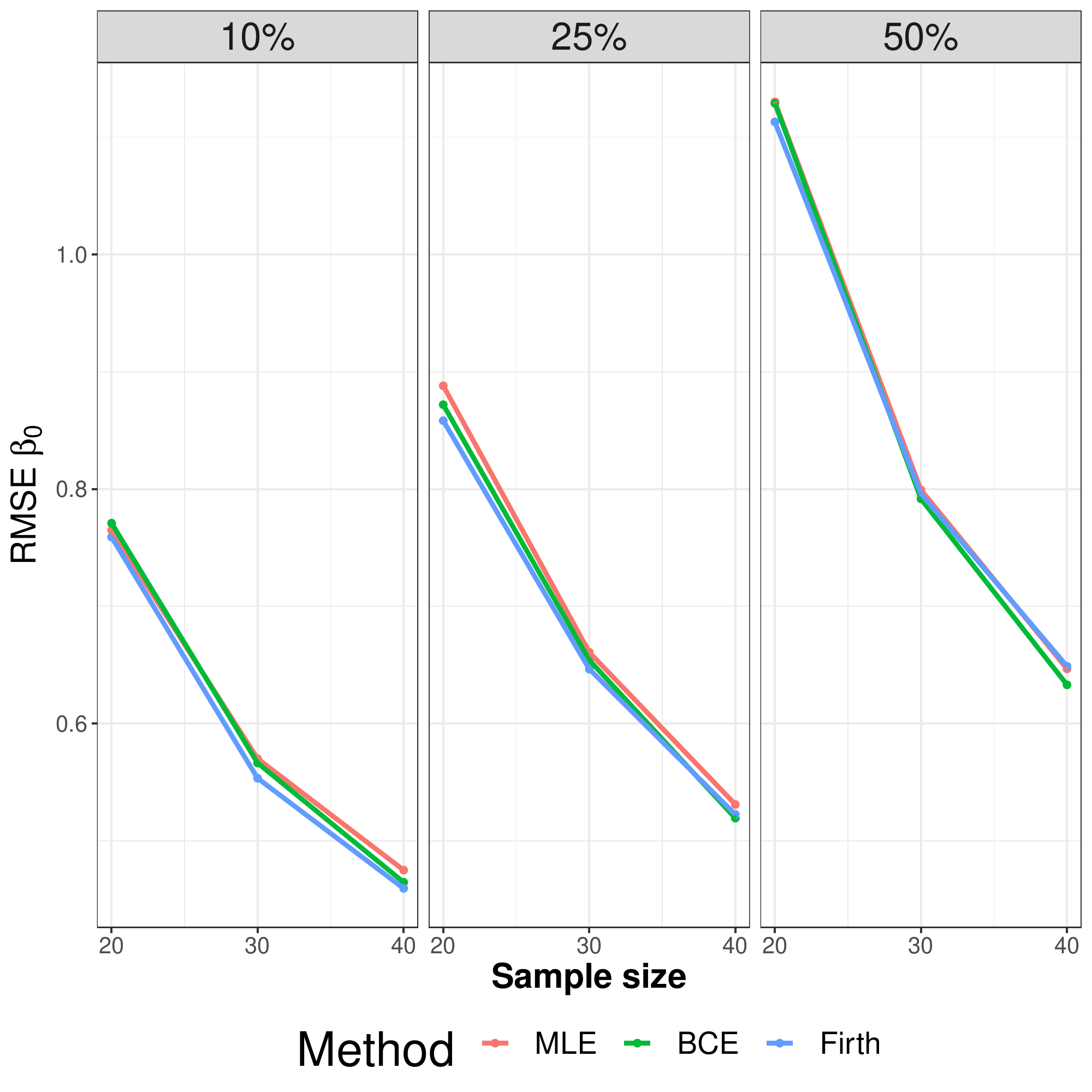}} \end{minipage} &
  \begin{minipage}{.28\textwidth}{\includegraphics[width=1\textwidth]{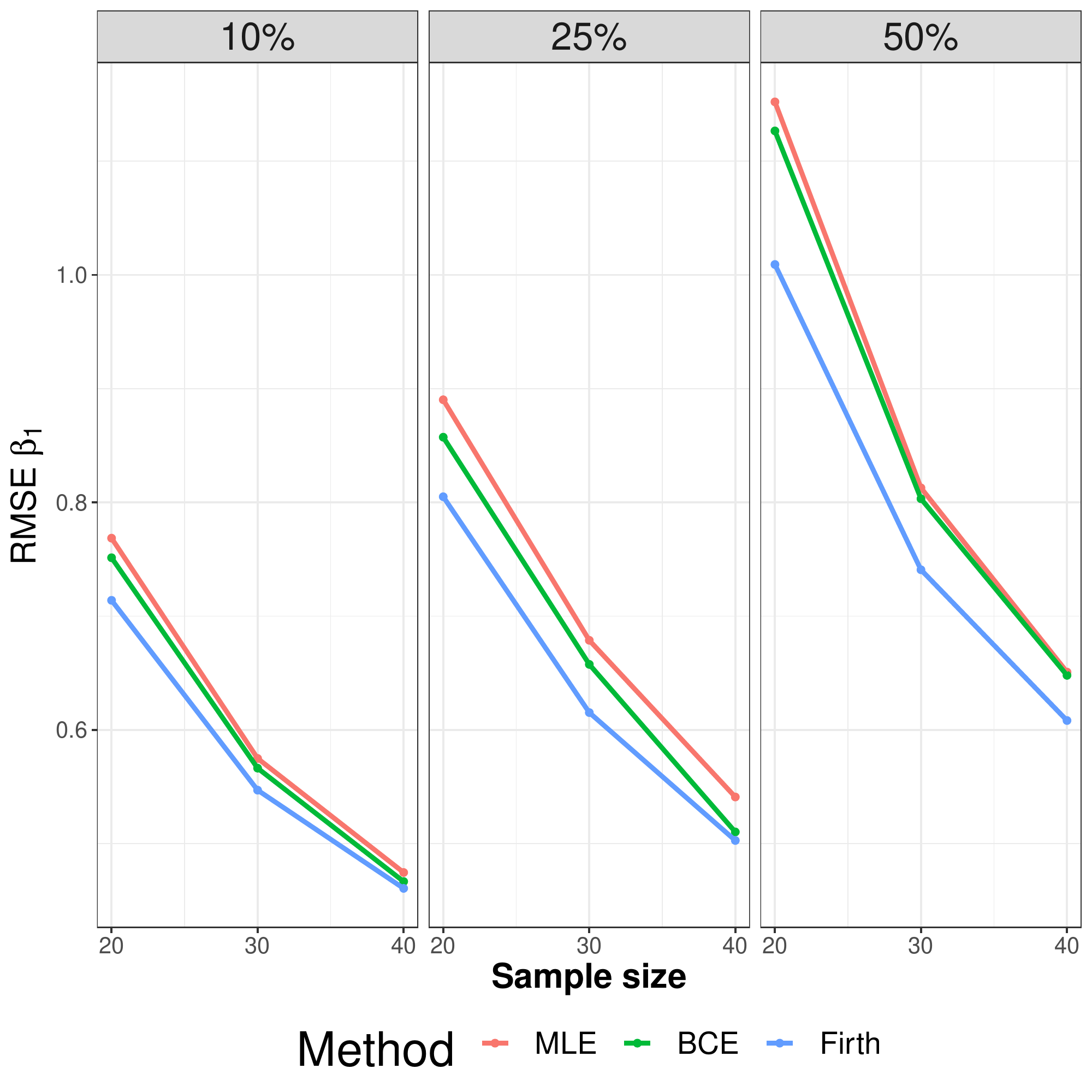}} \end{minipage} &
  \begin{minipage}{.28\textwidth}{\includegraphics[width=1\textwidth]{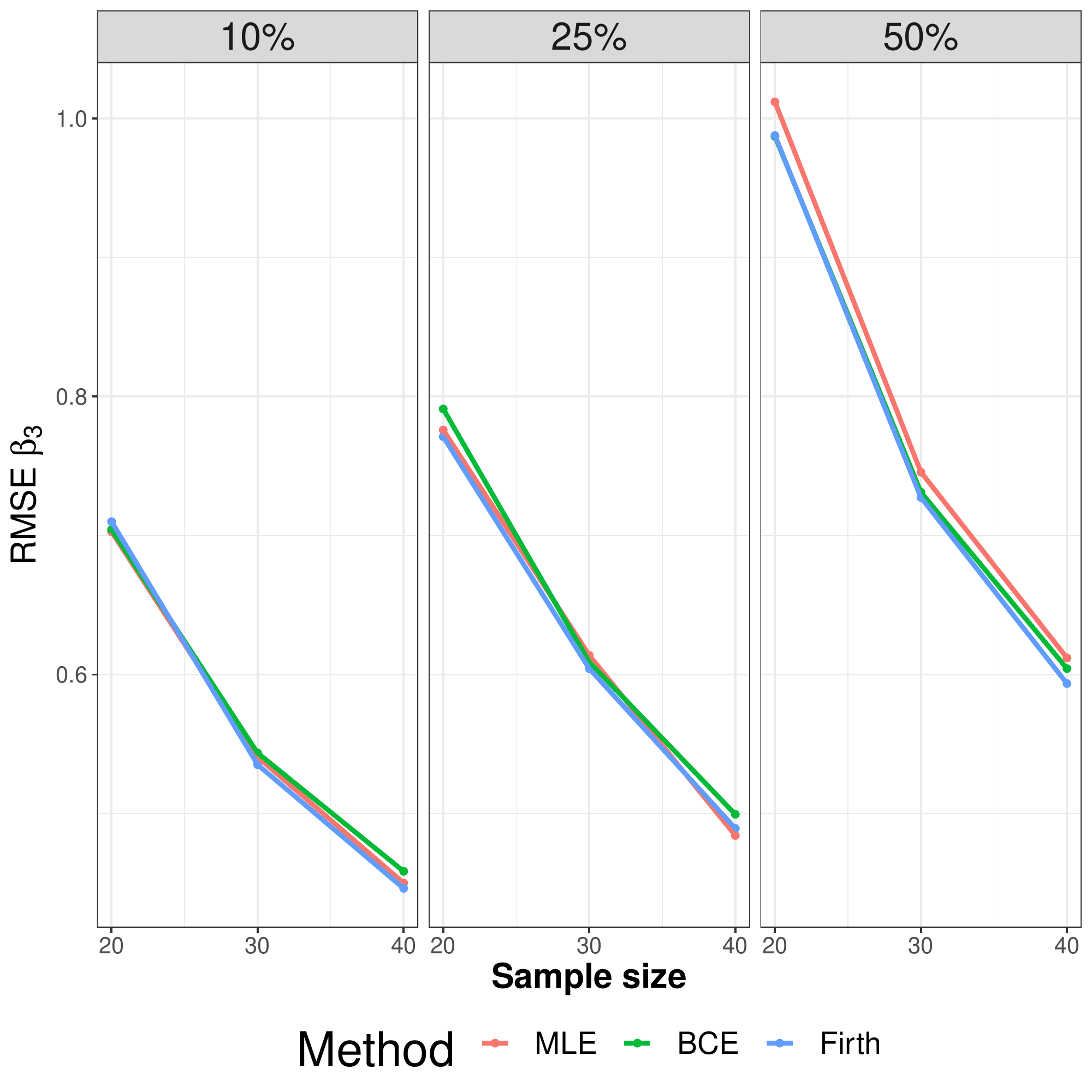}} \end{minipage} \\
  $p=7$ &
  \begin{minipage}{.28\textwidth}{\includegraphics[width=1\textwidth]{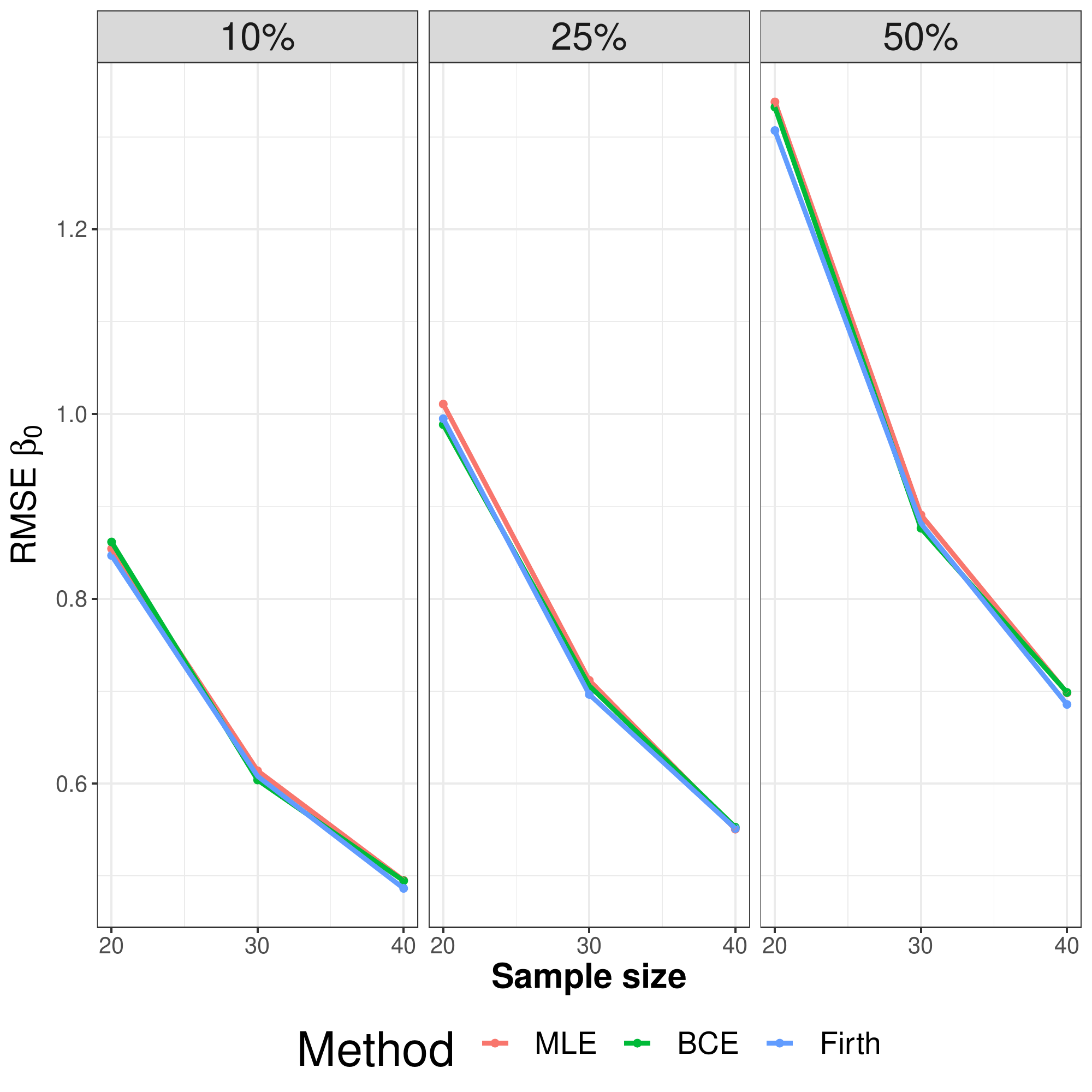}} \end{minipage} &
  \begin{minipage}{.28\textwidth}{\includegraphics[width=1\textwidth]{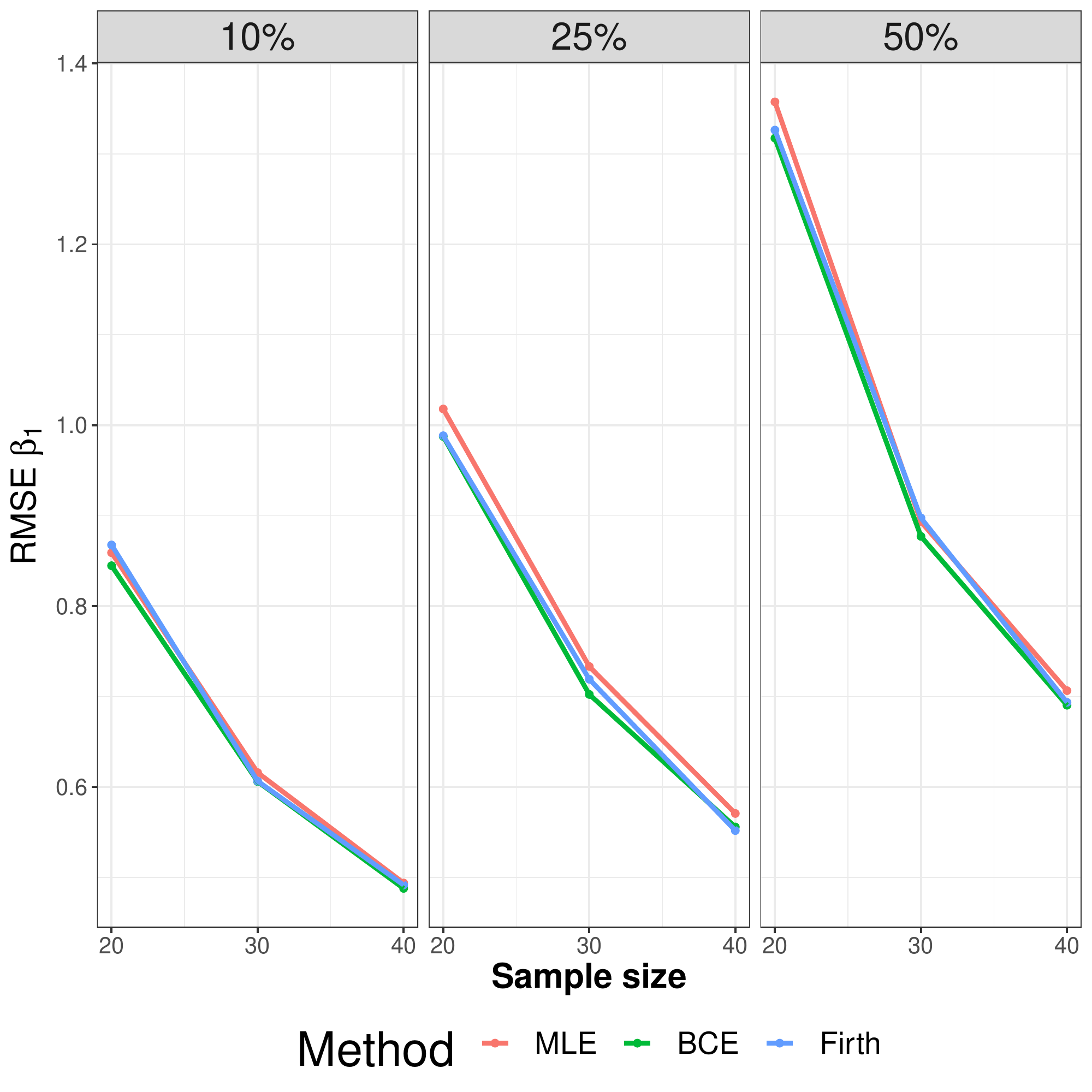}} \end{minipage} &
  \begin{minipage}{.28\textwidth}{\includegraphics[width=1\textwidth]{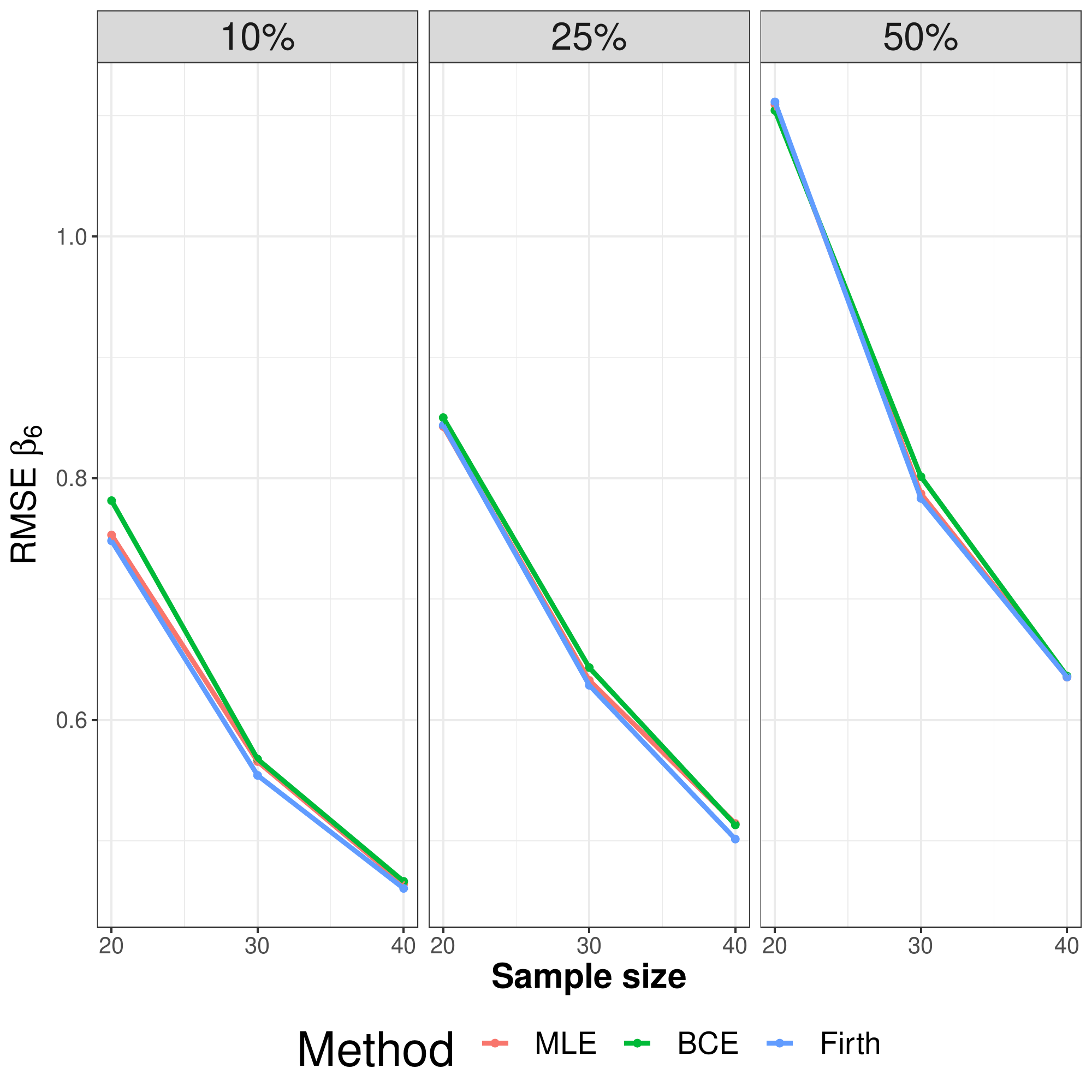}} \end{minipage} \\
\end{tabular}
}
\caption{Empirical RMSE for different estimators. Case $\sigma=3$.}
\label{fig:s6}
\end{center}
\end{figure}

\FloatBarrier

\newpage

\subsection{Assessing the variance estimation}

{\small\tabcolsep=0.5pt 
\begin{longtable}{ 
p{.05\textwidth}p{.05\textwidth}p{.025\textwidth}p{.025\textwidth} cccccccccccc}
\caption{Distance between different estimated covariance matrices.} 
\label{tab:s_cov} \\
\hline
           &            &            &            &                         \multicolumn{ 4}{c}{$n=20$} &                         \multicolumn{ 4}{c}{$n=30$} &                         \multicolumn{ 4}{c}{$n=40$} \\

           &            &            &            & \multicolumn{ 2}{c}{Cov($\widehat{\beta}$) and} & \multicolumn{ 2}{c}{Cov$(\widetilde{\beta})$ and} & \multicolumn{ 2}{c}{Cov($\widehat{\beta}$) and} & \multicolumn{ 2}{c}{Cov$(\widetilde{\beta})$ and} & \multicolumn{ 2}{c}{Cov($\widehat{\beta}$) and} & \multicolumn{ 2}{c}{Cov$(\widetilde{\beta})$ and} \\

 c &      $\sigma$ &          $p$ &   d &          $\widehat{K}$ & Cov$_2(\widehat{\beta})$ &          $\widetilde{K}$ & Cov$_2(\widetilde{\beta})$ &          $\widehat{K}$ & Cov$_2(\widehat{\beta})$ &          $\widetilde{K}$ & Cov$_2(\widetilde{\beta})$ &          $\widehat{K}$ & Cov$_2(\widehat{\beta})$ &          $\widetilde{K}$ & Cov$_2(\widetilde{\beta})$ \\
\hline
\endfirsthead
\hline
           &            &            &            &                         \multicolumn{ 4}{c}{$n=20$} &                         \multicolumn{ 4}{c}{$n=30$} &                         \multicolumn{ 4}{c}{$n=40$} \\

           &            &            &            & \multicolumn{ 2}{c}{Cov($\widehat{\beta}$) and} & \multicolumn{ 2}{c}{Cov$(\widetilde{\beta})$ and} & \multicolumn{ 2}{c}{Cov($\widehat{\beta}$) and} & \multicolumn{ 2}{c}{Cov$(\widetilde{\beta})$ and} & \multicolumn{ 2}{c}{Cov($\widehat{\beta}$) and} & \multicolumn{ 2}{c}{Cov$(\widetilde{\beta})$ and} \\

 c &      $\sigma$ &          $p$ &   d &          $\widehat{K}$ & Cov$_2(\widehat{\beta})$ &          $\widetilde{K}$ & Cov$_2(\widetilde{\beta})$ &          $\widehat{K}$ & Cov$_2(\widehat{\beta})$ &          $\widetilde{K}$ & Cov$_2(\widetilde{\beta})$ &          $\widehat{K}$ & Cov$_2(\widehat{\beta})$ &          $\widetilde{K}$ & Cov$_2(\widetilde{\beta})$ \\
\hline
\endhead
      10\% &        0.5 &          3 &         $d_1$ &     0.0634 &     0.0262 &     0.0593 &     0.0175 &     0.0431 &     0.0275 &     0.0403 &     0.0243 &     0.0320 &     0.0151 &     0.0297 &     0.0116 \\

           &            &            &         $d_2$ &     0.0070 &     0.0010 &     0.0059 &     0.0005 &     0.0026 &     0.0012 &     0.0022 &     0.0009 &     0.0017 &     0.0003 &     0.0015 &     0.0003 \\

           &            &            &         $d_3$ &     0.0129 &     0.0025 &     0.0115 &     0.0014 &     0.0053 &     0.0029 &     0.0044 &     0.0022 &     0.0033 &     0.0009 &     0.0031 &     0.0006 \\
\cline{3-16}
           &            &          5 &         $d_1$ &     0.1046 &     0.0657 &     0.0988 &     0.0640 &     0.0616 &     0.0338 &     0.0588 &     0.0323 &     0.0485 &     0.0278 &     0.0468 &     0.0280 \\

           &            &            &         $d_2$ &     0.0232 &     0.0095 &     0.0207 &     0.0088 &     0.0085 &     0.0031 &     0.0078 &     0.0030 &     0.0051 &     0.0023 &     0.0048 &     0.0023 \\

           &            &            &         $d_3$ &     0.0642 &     0.0343 &     0.0590 &     0.0328 &     0.0257 &     0.0132 &     0.0241 &     0.0127 &     0.0168 &     0.0100 &     0.0161 &     0.0097 \\
\cline{3-16}

           &            &          7 &         $d_1$ &     0.2691 &     0.2520 &     0.2694 &     0.2545 &     0.1790 &     0.1684 &     0.1783 &     0.1689 &     0.1470 &     0.1397 &     0.1463 &     0.1398 \\

           &            &            &         $d_2$ &     0.1799 &     0.1562 &     0.1786 &     0.1581 &     0.0867 &     0.0777 &     0.0863 &     0.0783 &     0.0583 &     0.0537 &     0.0579 &     0.0538 \\
\hline
\end{longtable}
}

\subsection{Assessing the type I error}

\begin{longtable}{cccccccccc}
\caption{Simulated rejection rates for $\mathcal{H}: {\bm \beta}_q = \textbf{0}_q$, with different combinations of $\sigma$, censoring, $n$, $p$ and $q$.}
\label{tab:s_typeI} \\
\hline 
     $\sigma$ &  censoring &          $n$ &          $p$ &          $q$ &       MLE &       MLE2 &       BCE &       BCE2 &      Firth\\
\hline
\endfirsthead
\hline
     $\sigma$ &  censoring &          $n$ &          $p$ &          $q$ &       MLE &       MLE2 &       BCE &       BCE2 &      Firth\\
\hline
\endhead
       0.5 &       10\% &         20 &          3 &          1 &     0.0723 &     0.0463 &     0.0723 &     0.0590 &     0.0722 \\

           &            &            &            &          2 &     0.0874 &     0.0474 &     0.0877 &     0.0627 &     0.0869 \\
\cline{4-10}
           &            &            &          5 &          1 &     0.0794 &     0.0415 &     0.0796 &     0.0580 &     0.0779 \\

           &            &            &            &          3 &     0.1131 &     0.0518 &     0.1132 &     0.0729 &     0.1121 \\
\cline{4-10}

           &            &            &          7 &          2 &     0.1223 &     0.0531 &     0.1226 &     0.0791 &     0.1190 \\

           &            &            &            &          4 &     0.1495 &     0.0580 &     0.1492 &     0.0853 &     0.1460 \\
\cline{3-10}

           &            &         30 &          3 &          1 &     0.0636 &     0.0422 &     0.0638 &     0.0522 &     0.0639 \\

           &            &            &            &          2 &     0.0770 &     0.0473 &     0.0769 &     0.0605 &     0.0767 \\
\hline
\end{longtable}

\subsection{Assessing the power of the tests}

\begin{longtable}{cccccccccc}
\caption{Simulated rejection rates for $\mathcal{H}: {\bm \beta}_q = \textbf{0}_q$, with different combinations of $\sigma$, censoring, $n$ and $q$ when the true parameter vector is ${\bm \beta}^\top=(\psi \textbf{1}_q^\top,\textbf{0}_{p-q}^\top)$}
\label{tab:s_power} \\
\hline
     $\sigma$ &  censoring &          $n$ &          $q$ &          $\psi$ &       MLE &       MLE2 &       BCE &       BCE2 &      Firth \\
\hline
\endfirsthead
\hline
     $\sigma$ &  censoring &          $n$ &          $q$ &          $\psi$ &       MLE&       MLE2 &       BCE &       BCE2 &      Firth \\
\hline
\endhead
       0.5 &       10\% &         20 &          1 &       0.05 &     0.1047 &     0.0583 &     0.1047 &     0.0799 &     0.1030 \\

           &            &            &            &       0.10 &     0.1599 &     0.0956 &     0.1605 &     0.1256 &     0.1575 \\

           &            &            &            &       0.25 &     0.4796 &     0.3537 &     0.4798 &     0.4231 &     0.4726 \\

           &            &            &            &       0.50 &     0.9141 &     0.8577 &     0.9143 &     0.8909 &     0.9092 \\

           &            &            &            &       1.00 &     0.9993 &     0.9982 &     0.9993 &     0.9989 &     0.9993 \\

           &            &            &            &       2.00 &     1.0000 &     1.0000 &     1.0000 &     1.0000 &     1.0000 \\
\cline{4-10}
           &            &            &          3 &       0.05 &     0.1453 &     0.0673 &     0.1446 &     0.0938 &     0.1426 \\

           &            &            &            &       0.10 &     0.2495 &     0.1337 &     0.2492 &     0.1763 &     0.2442 \\
\hline
\end{longtable}

\section{Cumulants}
\label{sec:cumulants}

Let $Y_1, \ldots, Y_n$ a random sample of (4), presented in the main article, the logarithm of the likelihood function is given by
%
%
\begin{eqnarray}
\label{eR:llf}
\ell({\bm \beta}) = \sum_{i=1}^{n} \left\{ \delta_i \left[- n \log\sigma + \frac{y_i - \mu_i}{\sigma}\right] - \exp\left( \frac{y_i - \mu_i}{\sigma}\right) \right\}.
\end{eqnarray}

The derivatives of the logarithm of the likelihood function $\ell({\bm \beta})$ concerning the unknown parameter $\beta$ are denoted by
%
%
\begin{align*}
	U_{r} = \partial l({\bm \beta})/\partial \beta_{r}, \ \
	U_{rs} = \partial^{2} l({\bm \beta})/\partial \beta_{r} \partial \beta_{s}, \ \
	U_{rst} = \partial^{3} l({\bm \beta})/\partial \beta_{r} \partial \beta_{s} \partial \beta_{t}, \ \
	\cdots
\end{align*}

\noindent where the indices $r$, $s$, $t$ denote the components of the ${\bm \beta}$ vector.

The first four derivatives of (\ref{eR:llf}) can be expressed, respectively, as
%
%
\begin{align*}
\frac{\partial\ell({\bm \beta})}{\partial\beta_r} &= \frac{1}{\sigma}\sum_{i=1}^{n} \left\{ - \delta_i + \exp\left( \frac{y_i - \mu_i}{\sigma}\right) \right\} x_{ri}; \\
\frac{\partial^2\ell({\bm \beta})}{\partial\beta_r \partial\beta_s} &= -\frac{1}{\sigma^2}\sum_{i=1}^{n} \exp\left( \frac{y_i - \mu_i}{\sigma}\right) x_{ri} x_{si}; \\
\frac{\partial^3\ell({\bm \beta})}{\partial\beta_r \partial\beta_s \partial\beta_t} &= \frac{1}{\sigma^3}\sum_{i=1}^{n} \exp\left( \frac{y_i - \mu_i}{\sigma}\right) x_{ri} x_{si} x_{ti}; \\
\frac{\partial^4\ell({\bm \beta})}{\partial\beta_r \partial\beta_s \partial\beta_t \partial\beta_u} &= - \frac{1}{\sigma^4}\sum_{i=1}^{n} \exp\left( \frac{y_i - \mu_i}{\sigma}\right) x_{ri} x_{si} x_{ti} x_{ui}.
\end{align*}

We use the notation introduced by Lawley\cite{Lawley:1956} to define the joint cumulants, and their derivatives, of the logarithm of the likelihood function:
%
%
\begin{align*}
	\kappa_{rs} = \mathds{E}(U_{rs}), \ \
	\kappa_{rst} = \mathds{E}(U_{rst}), \ \
	\kappa_{r,st} = \mathds{E}(U_{r} U_{st}), \ \
	\kappa_{st}^{(r)} = \partial \kappa_{st} / \partial \beta_{r}, \ \
 \cdots
\end{align*}

All $\kappa$'s refer to a total on the sample and are, in general, of  order $n$. The Fisher information matrix, ${\bm K}_{ {\bm \beta} {\bm \beta} }$, has elements $\kappa_{r,s} = -\kappa_{rs}$. Also, consider $\kappa^{r,s} = - \kappa^{rs}$ as the corresponding elements of its inverse, ${\bm K}_{ {\bm \beta} {\bm \beta} }^{-1}$.

The second- to forth-order cumulants are:
%
%
\begin{align*}
\kappa_{rs} &= -\frac{1}{\sigma^2}\sum_{i=1}^{n}w_i x_{ri} x_{si}; \ \
\kappa_{r,s} = - \kappa_{rs} = -\frac{1}{\sigma^2}\sum_{i=1}^{n}w_i x_{ri} x_{si};\\
\kappa_{rst} &= \frac{1}{\sigma^3}\sum_{i=1}^{n} w_i x_{ri} x_{si} x_{ti};  \ \
\kappa_{rs}^{(t)} = -\frac{1}{\sigma^2}\sum_{i=1}^{n}w_i^{\prime} x_{ri} x_{si} x_{ti}; \\
\kappa_{rs,t} &= -\frac{1}{\sigma^2}\sum_{i=1}^{n} \left(w_i^{\prime} + \frac{1}{\sigma} w_i\right) x_{ri} x_{si} x_{ti};  \ \
\kappa_{r,s,t} = \frac{1}{\sigma^2}\sum_{i=1}^{n} \left(3 w_i^{\prime} + \frac{2}{\sigma} w_i\right) x_{ri} x_{si} x_{ti};\\
\kappa_{rstu} &= - \frac{1}{\sigma^4}\sum_{i=1}^{n} w_i x_{ri} x_{si} x_{ti} x_{ui}; \\
\kappa_{rs}^{(tu)} &= -\frac{1}{\sigma^2}\sum_{i=1}^{n} w_i^{\prime\prime} x_{ri} x_{si} x_{ti} x_{ui};  \ \
\kappa_{rst}^{(u)} = \frac{1}{\sigma^3}\sum_{i=1}^{n} w_i^{\prime}  x_{ri} x_{si} x_{ti} x_{ui}; \\
\kappa_{r,stu} &= \frac{1}{\sigma^3} \sum_{i=1}^{n} \left( \frac{1}{\sigma} w_i + w_i^{\prime} \right) x_{ri} x_{si} x_{ti} x_{ui};  \ \
\kappa_{rs,tu} = \frac{1}{\sigma^4} \sum_{i=1}^{n} (2 w_i + 2 \sigma w_i^{\prime} - w_i^2) x_{ri} x_{si} x_{ti} x_{ui}; \\
\kappa_{r,s,tu} &= \frac{1}{\sigma^2} \sum_{i=1}^{n} \left\{ \frac{1}{\sigma^2} w_i^2 - \frac{3}{\sigma^2} w_i - \frac{4}{\sigma} w_i^{\prime} - w_i^{\prime\prime} \right\} x_{ri} x_{si} x_{ti} x_{ui}; \\
\kappa_{r,s,t,u} &= \frac{1}{\sigma^2} \sum_{i=1}^{n} \left\{ - \frac{3}{\sigma^2} w_i^2 + \frac{9}{\sigma^2} w_i + \frac{14}{\sigma} w_i^{\prime} + 6 w_i^{\prime\prime} \right\} x_{ri} x_{si} x_{ti} x_{ui};
\end{align*}

\noindent where \vspace{-10mm}
\begin{align*}
w_i &= 1 - \exp\left\{ -L_i^{1/\sigma} \exp(-\mu_i/\sigma) \right\}, \
w_i^{\prime} = - \frac{1}{\sigma} L_i^{1/\sigma} \exp\{ -L_i^{1/\sigma} \exp(-\mu_i/\sigma) - \mu_i/\sigma \}, \\
w_i^{\prime\prime} &= - \frac{1}{\sigma^2} L_i^{1/\sigma} \exp\{ -L_i^{1/\sigma} \exp(-\mu_i/\sigma) - \mu_i/\sigma \} \left[ L_i^{1/\sigma} \exp(-\mu_i/\sigma) - 1 \right].
\end{align*}

\noindent  It can be observed that $w_i^{\prime} = w_i^{\prime\prime} = 0$ for type II censoring.

\section{Details about corrections}

\subsection{Bias correction}

Regarding the cumulants presented in Section \ref{sec:cumulants}, we can write the $\mathcal{O}(n^{-1})$ biases of the MLE ${\bm \beta}$, Cox and Snell\cite{CoxSnell:1968}, as:
%
%
\begin{align}
\label{eq:biastensorial}
B(\widehat{\bm \beta}_a) &= \sum_{r,s,t = 1}^{p} \kappa^{ar} \kappa^{st} \left\{ \kappa_{rs}^{(t)} - \frac{1}{2} \kappa_{rst} \right\} \nonumber\\
&= -\frac{1}{2 \sigma^3} \sum_{r,s,t = 1}^{p} \sum_{i=1}^{n} \left( w_1 + 2 \sigma w_i^{\prime} \right) \kappa^{ar} \kappa^{st} x_{r i} x_{s i} x_{t i} \nonumber\\
&= -\frac{1}{2 \sigma^3} \sum_{i=1}^{n} \left( w_1 + 2 \sigma w_i^{\prime} \right) p_{a,i} z_{ii},
\end{align}

\noindent where $p_{a,i} = -\sum_{r = 1}^{p} \kappa^{ar} x_{r i}$ and $z_{ii} = -\sum_{s,t = 1}^{p} x_{s i}\kappa^{st} x_{t i}$. Eq. (11), in the main article, presents the matrix form to \eqref{eq:biastensorial}.

\subsection{Covariance correction}

We can write the $(a, b)$th element of the matrix ${\bm \Delta}$ (presented in the Eq. (13), in the main article) as $\delta_{ab}$, Magalh\~aes et al.\cite{MagalhaesBotterSandoval:2021}, with $\delta_{ab}  = -\frac{1}{2} \delta_{ab}^{(1)} + \frac{1}{4} \delta_{ab}^{(2)} + \frac{\tau_2}{2} \delta_{ab}^{(3)}$, where
%
%
\begin{align*}
	\delta_{ab}^{(1)} &= \sum_{c,d = 1}^{p} \kappa^{cd} \left\{ \tau_1 \left[ 2\kappa_{bc}^{(ad)} - \kappa_{bcd}^{(a)} \right] + \kappa_{ac,bd} \right\}, \\
	\delta_{ab}^{(2)} &= \sum_{r,c = 1}^{p} \sum_{s,t = 1}^{p} \kappa^{rs} \kappa^{ct}  \left\{ \kappa_{arc} [ 3\kappa_{bst} + 2\kappa_{b,st} + 8\kappa_{bs,t} ] + \ 2 \kappa_{ar,c} [ 2 \kappa_{b,st} + \kappa_{bt, s}] \right\}, \\
	\delta_{ab}^{(3)} &= \sum_{c = 1}^{p} \sum_{r,s,t = 1}^{p} \kappa^{rs} \kappa^{ct} \left\{ \kappa_{bc}^{(a)} \left( \kappa_{st}^{(r)} + \kappa_{r, st} \right) \right\}.
\end{align*}

\noindent with ${\bm \tau} = (\tau_1, \tau_2) = (1, 1)$ indicating the second-order covariance matrix of the MLE and ${\bm \tau} = (0, -1)$ indicating the second-order covariance matrix of the BCE. Also regarding the cumulants presented in Section \ref{sec:cumulants}, 
%
%
\begin{align}\label{eq:covtensorial}
	\delta_{ab}^{(1)} &= \frac{1}{\sigma^4} \sum_{i = 1}^{n} w_{i}^{\star} z_{ii} x_{ai} x_{bi}, \nonumber\\
	\delta_{ab}^{(2)} &= - \frac{1}{\sigma^6} \sum_{i = 1}^{n} \sum_{j = 1}^{n} \left( w_i w_j - 2 \sigma w_i w_j^{\prime} - 6 \sigma^2 w_i^{\prime} w_j^{\prime} \right) z_{ij} z_{ij} x_{ai} x_{bj}, \\
	\delta_{ab}^{(3)} &= \frac{1}{\sigma^5} \sum_{i = 1}^{n} w_i^{\prime} w_i^{\star\star} z_{ij} z_{jj} x_{ai} x_{bi}, \nonumber
\end{align}

\noindent where $z_{ij} = -\sum_{c,t = 1}^{p} x_{c i} \kappa^{ct} x_{t j}$, $z_{jj} = -\sum_{r,s = 1}^{p} x_{r j} \kappa^{rs} x_{s j}$, $w_i^{\star} = w_i (w_i - 2) - 2 \sigma w_i^{\prime} + \sigma \tau_1 (w_i^{\prime} + 2 \sigma w_i^{\prime \prime})$ and $w_{i}^{\star\star} = \sum_{j = 1}^{n} (w_j + 2 \sigma w_j^{\prime}) z_{ij} z_{jj}$. In the main article, Eq. (13) presents the matrix form of \eqref{eq:covtensorial}.

\bibliographystyle{ama}
\bibliography{Wcdsocm_bib}
